\documentclass[twocolumn, tighten, times, trackchanges]{aastex62_rev}

\usepackage{amsmath,amssymb}
\usepackage{aas_macros}
\usepackage{graphicx}
\usepackage{CJKutf8}
\usepackage{apjfonts}
\usepackage{gensymb}
\usepackage{enumitem}
\usepackage{longtable}
\usepackage{auto-pst-pdf}

\PassOptionsToPackage{hyphens}{url}
\usepackage{hyperref}

\usepackage[all]{hypcap} 

\newcommand{\mum}{\ifmmode{\rm \mu m}\else{$\mu$m}\fi}

\newcommand{\sevenrm}{\rm\scriptsize}

\newcommand{\cosmology}{$\Omega _m = 0.27$,~$\Omega_{\Lambda}=0.73$~and~$H_0=71$km~s$^{-1}$Mpc$^{-1}$}

\newcommand{\OIV}{[O{\sevenrm\,IV}]}

\submitjournal{ApJ}

\received{07-31-2019}
\revised{09-22-2019}
\accepted{09-24-2019}
\published{11-15-2019}

\shorttitle{Quasar Mid-IR Variability and Dust Reverberation}
\shortauthors{Lyu, Rieke \& Smith}

\begin{document}

\title{\bf\Large  
    Mid-IR Variability and Dust Reverberation Mapping of Low-$z$ Quasars \\
    I. Data, Methods and Basic Results
  }

\correspondingauthor{Jianwei Lyu}
\email{jianwei@email.arizona.edu}

\author[0000-0002-6221-1829]{Jianwei Lyu (\begin{CJK}{UTF8}{gkai}吕建伟\end{CJK})}
\affiliation{ Steward Observatory, University of Arizona, 
933 North Cherry Avenue, Tucson, AZ 85721, USA}
\author[0000-0003-2303-6519]{George H. Rieke}
\affiliation{ Steward Observatory, University of Arizona, 
933 North Cherry Avenue, Tucson, AZ 85721, USA}

\author[0000-0002-5083-3663]{Paul S. Smith}
\affiliation{ Steward Observatory, University of Arizona, 
933 North Cherry Avenue, Tucson, AZ 85721, USA}
 
\begin{abstract}
	The continued operation of the {\it Wide-field Infrared Survey Explorer} ({\it
	WISE}) combined with several ground-based optical transient
	surveys (e.g., CRTS, ASAS-SN and PTF) offer an unprecedented
	opportunity to explore the dust structures in luminous AGNs. We use
	these data for a mid-IR dust reverberation mapping (RM) study of 87
	archetypal Palomar--Green quasars at $z\lesssim0.5$. To cope with
	various contaminations of the photometry data and the sparse time
	sampling of the light curves, procedures to combine these datasets and
	retrieve the dust RM signals have been developed. We find that $\sim$70\%
	of the sample (with a completeness correction,  up to 95\%) has
	convincing mid-IR time lags in the {\it WISE} $W1$ ($\sim3.4~\mum$) and
	$W2$ ($\sim4.5~\mum$) bands and they are proportional to the square root
	of the AGN luminosity.  Combined with previous $K$-band ($\sim2.2~\mum$)
	RM results in the literature, the inferred dust emission size ratios
	are $R_{K}:R_{W1}:R_{W2}=0.6:1:1.2$. Under simple assumptions, we put
	preliminary constraints on the projected dust surface density at these
	bands and reveal the possibly different torus structures among
	hot-dust-deficient, warm-dust-deficient and normal quasars from the
	reverberation signals. With multi-epoch {\it Spitzer} data and later
	{\it WISE} photometry, we also explore AGN IR variability at
	10--24~$\mum$ over a 5 yr time-scale. Except for blazars and
	flat-spectrum radio sources, the majority of AGNs have typical
	variation amplitudes at 24~$\mum$ of no more than 10\% of that in the
	$W1$ band, indicating that the dust reverberation signals damp out
	quickly at longer wavelengths. In particular, steep-spectrum radio
	quasars also lack strong 24~$\mum$ variability, consistent with the
	unification picture of radio-loud AGNs.
\end{abstract}

\keywords{galaxies:active --- galaxies:Seyfert --- quasars:general ---
infrared:galaxies --- dust, extinction}

\section{Introduction}

The circumnuclear dusty structures in active galactic nuclei (AGNs) bridge the
gap between the black hole (BH) accretion disk and the host galaxy interstellar
medium (ISM) and lay the foundation for AGN unification proposals
\citep{Antonucci1993a, Urry1995, Netzer2015}. However, due to their complex
geometry and small sizes, it is very challenging to characterize torus
structures as well as the properties of the constituent dust grains.

Infrard (IR) reverberation mapping (RM) opens a window to peek inside the
so-called AGN dusty torus. When the UV/optical emission of the BH accretion
disk changes, the varying signal travels at the speed of light to the torus,
causing it to react as reprocessed emission in the IR but with a time lag. We
can analyze the optical and IR light curves of AGNs to retrieve spatial
information reflected in the response of the torus to changes in irradiation
from the central engine.  The procedure is similar to the RM
originally proposed by \cite{Blandford1982} and widely applied to study the
broad-line regions (BLRs) and accretion disks \citep[e.g.,][]{Peterson2004}. The
time lag of the IR emission was originally reported for the Seyfert 1
galaxy Fairall 9 by \citet{Clavel1989} and soon after analyzed in terms of RM
to constrain the AGN dusty circumnuclear structure by \citet{Barvainis1992}.
The behavior was subsequently observed for a number of AGNs; the most
noteworthy work is the systematic study of time lags between the near-IR
($K$-band) and the optical of $\sim$30 Seyfert-1 nuclei through ground-based
monitoring. The results follow an $R\propto L^{1/2}$ size-luminosity relation,
as expected if there is a similar dust sublimation temperature in all the
sources \citep[e.g.,][]{Oknyanskij2001, Suganuma2006, Lira2011, Koshida2014,
PozoNunez2014, Mandal2018, Ramolla2018}.

\cite{Glass2004} conducted a {\it JHKL} (1.25--3.45~$\mum$) long-term
monitoring program of 41 Seyfert nuclei and reported tentative time-lag
measurements between the $U$ ($\sim0.36~\mum$) and $L$ ($\sim3.45~\mum$) bands
for five objects. Given the stable and cool space environment, high-precision
photometry was made possible by the IRAC instrument on {\it Spitzer} and a
mid-IR dust reverberation study was carried out at 3.6 and 4.5~$\mum$ and
reported for the reddened type-1 AGN in NGC 6418 (\citealt{Vazquez2015}).
However, due to the limitations in observing at 3--5$~\mum$ from the ground and
the difficulty and cost of carrying out long-term coordinated ground- and
space-based targeted observations, no other detailed reverberation studies in
this wavelength range are available.  

Now there is a new possibility to conduct systematic dust RM at 3--5$~\mum$ for
bright AGNs with the continued operation of the {\it Wide-field Infrared Survey
Explorer} ({\it WISE}). Launched in 2009 December, this satellite performed a
mid-IR all-sky survey, was put into hibernation, and then reactivated as the
{\it Near-Earth Object WISE} (\textit{NEOWISE}) mission for asteroid hunting
\citep{neowise}.  The telescope completes an all-sky map roughly every 6 months
and had provided photometric data with 12--13 epochs that cover a time period
of 8 yr by the end of 2018.  The AGNs in Seyfert galaxies studied by
\citet{Koshida2014} had typical luminosities up to about $3 \times
10^{11}$~$L_\odot$ and reverberation time delays at this luminosity are
expected to be $\sim$ 100 days, i.e., too short to be sampled well at the
\textit{NEOWISE} cadence.  However, the cadence is satisfactory for more
luminous AGNs.  When combined with ground-based wide-field optical transient
surveys, these mid-IR multi-epoch data can be used to explore the dust
reverberation signals close to the spectral energy distribution (SED) peak of
the AGN hot dust emission at $\sim3~\mum$ for the first time.  Given the depth,
range, and sky coverage of these surveys, this approach has the potential to
draw general conclusions about the structure of a typical AGN and the
surrounding material in a statistically meaningful way. In this paper, we will
demonstrate how to make the best use of these public datasets and present the
results from an RM analysis at 3--5~$\mum$ of 87 $z<0.5$ Palomar-Green (PG)
quasars.

Despite the presence of terrestrial atmospheric windows at 10 and 20~$\mu$m,
few studies of AGN IR variability are available there because of the
sensitivity limitations of ground-based telescopes in the thermal IR.
Nonetheless, studies of individual blazars have found large-amplitude
coordinated variations from the visible through 10~$\mu$m
\citep[e.g.,][]{rieke1974}. \citet{Neugebauer1999} reported a broader-based
study based on coordinated data at $J  (1.27 \mu$m), $H (1.65 \mu$m), $K (2.23
\mu$m), $L' (3.69 \mu$m), and $N (10.6 \mu$m) of 25 PG quasars over several
decades. They concluded that the blazar 3C~273  (PG~1226+023) was the only
source that clearly varied at the $N$ band and another radio-quiet quasar,
PG~1535+547 might also be variable in this band, since its light curve mimicked
the pattern seen at the shorter wavelengths.  With the data collected for a
large sample of AGNs repeatedly observed by {\it Spitzer} and {\it WISE}, we
will provide an updated study of AGN variability behavior at 12~$\mum$ and
24~$\mum$ over a timescale of 4--5 yr and establish its relation with IR
variability at shorter wavelengths.

This paper is organized as follows. We describe the data and the compilation of
multiband light curves in Section~\ref{sec:data}.  To  retrieve the dust
reverberation signals rigorously, Section~\ref{sec:method} introduces a new
method for the cross-correlation analysis between the low-cadence optical and
mid-IR light curves. We present the results from the 3--5~$\mum$ dust RM
analysis, as well as the 10--24 $\mum$ variability study, in
Section~\ref{sec:result}.  Discussion about the origin of AGN mid-IR
variability, the circumnuclear dust structure, and its relation to the BLR can
be found in Section~\ref{sec:discussion}.  Section~\ref{sec:summary} is the
final summary.

Some technical details are left to appendices. Appendix~\ref{app:pho_stable}
evaluates the quality of the photometric survey data used.  We provide the
derivations that relate time lag to torus size in Appendix~\ref{app:model}. In
Appendix~\ref{app:koshida}, we reexamine the Seyfert 1
sample whose variability in the $K$ band was studied by \cite{Koshida2014}.

Throughout this paper, we adopt the cosmology \cosmology.

\section{Photometric Data and Light Curves}\label{sec:data}

\subsection{Data Ensemble}

\begin{figure}[htp]
    \begin{center}
	\includegraphics[width=1.0\hsize]{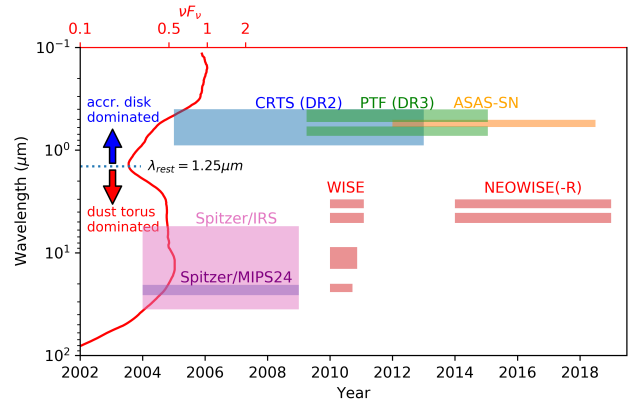}
		\caption{
		   Time coverage and wavelength sampling of the different
		   time-series datasets used in this work. We also show the
		   intrinsic SED shape of the normal AGN template at $z=0.15$ 
		   (mean redshift of the PG sample) along the $Y$-axis and 
		   highlight the separation between the UV--optical bump and 
		   the IR bump with dotted blue line.
		}
	\label{fig:data-summary}
    \end{center}
\end{figure}

The primary sample used for our mid-IR dust reverberation study is all 87 PG
quasars at z $<$ 0.5 \citep{Schmidt1983, Boroson1992}. For the 24~$\mum$
variability study, we also included 33 members of this PG sample plus another
106 quasars within the same redshift range to increase the statistical
significance.  Table~\ref{tab:data-summary} and Figure~\ref{fig:data-summary}
provide an overview of the time-series datasets that have been used in this
work. We present the details below.

\capstartfalse
\begin{deluxetable*}{ccclcccc}
    \tabletypesize{\footnotesize}
    \tablewidth{1.0\hsize}
    \tablecolumns{8}
    \tablecaption{Time-series datasets used in this work\label{tab:data-summary}
    }
    \tablehead{
    \colhead{name} &
    \colhead{sky coverage} &
    \colhead{$\lambda$} &
    \colhead{depth} &
	\colhead{time coverage} &
	\colhead{time gap} &
	\colhead{N$_{\rm epoch}$} &
	\colhead{N$_{\rm object}$}
}
\startdata
\multicolumn{8}{c}{Ground-based Optical Data}\\
\hline
CRTS (DR2)    & northern sky (survey) & unfiltered ($\sim$0.4--0.9~$\mum$) &   M$_V\sim$13--19 mag & 2005--2013 & 1--30 days & $\sim$100--150  & 83(PG)\\
ASAS-SN       & all sky (survey)        & V ($\sim0.55~\mum$) & M$_V\sim$10--17 mag  & 2012--2018 &  1-10 days & $\sim$200--300  & 87(PG)\\
PTF (DR3)     & northern sky (survey)   &  g$'$ ($\sim0.48~\mum$) & M$_{g'}\sim$14--21.3 mag  & 2009--2014$^a$  & & & 76(PG)\\
              &                         &  R ($\sim0.66~\mum$) & M$_{R}\sim$14--20.6 mag    & 2009--2014$^a$ &   \\
\hline
\multicolumn{8}{c}{Space-based Mid-IR Data}\\
\hline
WISE/NEOWISE  & all sky (survey) & W1 ($\sim3.4~\mum$) & W1$\sim$8--16.6 mag  & 2010--2018 &  6 months & 12--13 & 87(PG)\\
              &          & W2 ($\sim4.6~\mum$) & W2$\sim$7--16.0 mag  & 2010--2018 &  6 months & 12--13 & 87(PG)\\
              &          & W3 ($\sim12~\mum$)  & W3$\sim$3.8--10.8 mag & 2010        &  --       &  1--2 & 87(PG)\\
              &          & W4 ($\sim22~\mum$)  & W4$\sim-$0.4--6.7 mag & 2010        &  --       & 1--2 & 87(PG)+106\\
    {\it Spitzer} & all sky (targeted) & MIPS 24~$\mum$      & -- & 2004--2009  & 3--4 years   & 2--3  & 33(PG)+106 \\
              &                   & IRS spec. (5--35~$\mum$)  & -- &  2003--2009          & -- & 1 & 87(PG)
\enddata
\tablenotetext{a}{only selected g- and R-band data are available after Jan 1, 2013 (through Jan 28, 2015) in this data release}
\end{deluxetable*}
\capstarttrue

\subsubsection{Optical Data from Ground-based Transient Surveys}

We collected optical photometry of the PG quasars from the Catalina Real-Time
Transient Survey (CRTS; \citealt{CRTS2009}). This program utilizes two Steward
Observatory telescopes, a 1.5~m telescope on Mt. Lemmon, Arizona, and a 0.68~m
telescope on Mt. Bigelow, Arizona. Both survey the northern sky, covering
timescales from minutes to years. All images are unfiltered and processed using
the SExtractor package with the standard aperture photometry.  With a sequence
of 30~s exposures, the 1.5~m can reach as faint as $m_{V}\sim21.5$
while the 0.68-m limit is $m_{V}\sim19$.  Photometry for sources brighter
than $V=13$ can be problematic due to source saturation and nonlinearity of the
response.  We gathered the 2005-2013 photometry values from the public archive
of the second data release. According to \cite{Graham2017}, the published error
model of the CRTS archival photometry is problematic, and we follow their
methodology to make the necessary corrections to these data.

We also utilized the $V$-band optical data from the All-Sky Automated Survey for
Supernovae (ASAS-SN, \citealt{asassn}), which are publicly available at the
ASAS-SN light-curve servers \citep{Kochanek2017}. The data were obtained
through a global network of 20 14~cm diameter telescopes (camera lenses) with
commercial-level CCD cameras that have observed nearly the entire sky
continuously since 2012. The image FWHM is $\sim$16\arcsec with an
$\sim$8\arcsec pixel size.  The photometry is measured with a 2 pixel radius
aperture, and the background is estimated with a 7~pixel radius annulus. Typical
photometric uncertainties are less than 0.075 mag. The calibration is done by
observing nearby APASS stars. Typically, ASAS-SN photometry saturates at 10-11
mag, and the depth is roughly $m_{V}\sim17$ under good weather conditions. 

Finally, we gathered the optical time-sequence photometry for the PG quasars
from the Palomar Transient Factory (PTF; \citealt{PTF}) Data Release 3 (DR3),
which includes all data collected during the survey from 2009 March to 2012 December 
and some selected data obtained from 2013 to 2015 January. The PTF data nicely
fill the time gap between the CRTS and ASAS-SN light curves. This survey was
done with the Palomar 48 inch (1.2m) Samuel Oschin Telescope with a typical
2.\arcsec0 ~FWHM imaging resolution. The photometry was taken mostly through
two broadband filters, SDSS-$g'$ and Mould-$R$, reaching 4$\sigma$ limiting AB
magnitudes of $m_{g'}\sim21.3$ and $m_R\sim20.6$ in 60 second exposures. Both
bands saturate at $\sim14$~mag.

Since PG 1226+023 (also known as 3C 273) is saturated in CRTS and PTF, we have
looked for alternative datasets to build its optical light curve. In fact, this
object had been systematically monitored from late 2008 to the middle of 2018
as part of the Steward Observatory (SO) spectropolarmetric monitoring project
\citep{Smith2016}.\footnote{\url{http://james.as.arizona.edu/~psmith/Fermi/}}
These observations were carried out with the SPOL CCD spectropolarimeter
\citep{Schmidt1992} at either the SO 2.3~m Bok Telescope on Kitt Peak or the SO
1.54~m Kuiper Telescope on Mt. Bigelow. By convolving the spectra with a
standard Johnson $V$-band filter, $\sim300$ photometric measurements with high
signal-to-noise (S/N) covering 10 yr are available. We also collected the
$V$-band photometry of 3C 273 with the Light Curve Generator (LCG) on the AAVSO
website\footnote{\url{https://www.aavso.org/lcg}}, which archives photometry
from various facilities of bright variable objects.  We omitted measurements
flagged as being discrepant or having measurement uncertainties $>$0.1 mag. The
AAVSO data provide an optical light curve for 3C 273 from 2005 to 2019.

\subsubsection{Multi-epoch Mid-IR Photometry from \textit{WISE} and \textit{NEOWISE}}

The {\it WISE} mission performed an all-sky survey in four bands at 3.4, 4.6,
12, and 22~$\mum$, from 2010 January through 2010 October. Then the telescope
carried out a  4month \textit{NEOWISE} program using the first two bands
without cryogen before going into hibernation in 2011 February.  In 2013
December, {\it WISE} was reactivated and began the post-hibernation survey (the
\textit{NEOWISE} reactivation mission, or \textit{NEOWISE-R}).
\textit{WISE}/\textit{NEOWISE-R} observes the entire sky roughly every 6
months. All of the single-epoch images have been processed, and profile-fit
photometry of each detection has been carried out by a dedicated
pipeline.\footnote{See details in
\url{http://wise2.ipac.caltech.edu/docs/release/allwise/expsup/} and
\url{http://wise2.ipac.caltech.edu/docs/release/neowise/expsup/}}

We gathered all of the single-epoch profile-fit photometry measurements in the
$W1$ ($\sim3.4~\mum$) and $W2$ ($\sim4.6~\mum$) bands from the {\it WISE} and
\textit{NEOWISE} missions, up to the \textit{NEOWISE} 2019 Data Release, from
the NASA/IPAC Infrared Science Archive.  Typically, these data cover a time
period from 2010 January to 2018 December with a 3~yr gap between 2011 and
2014. Each object was observed for 12--13 epochs, with 10--20 exposures
acquired within each epoch. To construct the mid-IR light curves, we only used
detections from good-quality frame sets, following the suggestions given in the
online documentation.\footnote{That is, {frame quality score
\texttt{qual\_frame}}$>$0, frame image quality score {\texttt{qi\_fact}}$>$0,
South Atlantic Anomaly separation {\texttt{saa\_sep}}$>0$, and Moon masking
flag {\texttt{moon\_masked}}=0, see
\url{http://wise2.ipac.caltech.edu/docs/release/neowise/expsup/sec2_3.html}}

\subsubsection{10--25~$\mum$ measurements from {\it Spitzer} and {\it WISE} }\label{sec:data-24}

To test for variability at $\lambda>20~\mum$, we obtained MIPS 24~$\mu$m
measurements (PID 40053 and PID 50099; PI: George Rieke) of a heterogeneous
sample of quasars that had been measured at the same wavelength by {\it
Spitzer} for various previous observing programs. All of the observations were
conducted during the {\it Spitzer} cryogenic mission from 2004 to 2009 and
sample a time interval of 3--4 yr. The entire sample was observed twice,
while a subset of nearly 60 objects was observed during a third epoch roughly
1 yr after the second-epoch observation. To best utilize the instrumental
stability and minimize subtle possible systematic effects, these AGNs were
observed in exactly the same manner for all epochs. In this way, systematic
effects associated with different durations for data collection events (DCEs)
or a different total number of exposures are eliminated.  In all cases, flux
density measurements of the AGNs at 24 $\mu$m were obtained using the standard
MIPS photometry-mode astronomical observation template (AOT) with either 3 or
10~s DCEs.

All observations were processed using the standard reduction algorithms of the
MIPS Data Analysis Tool (DAT) with the same processing steps and calibrations
on all targets for both observations \citep{Gordon2005}. This included
subsampling the detector array pixels by a factor of 2 to produce final
mosaicked images of 363 $\times$ 401 pixels$^2$ ($\sim$ 7\farcm5 $\times$
8\farcm3). Aperture photometry using the DAOPHOT package \citep{stetson1987}
within IRAF was performed on the calibrated mosaicked 24 $\mu$m images. The
default parameters of the aperture photometry include an aperture of 24 pixels
($\sim$ 30$''$) and a sky annulus for background subtraction of 60--90$''$ from
the center of the photometric aperture. In a few cases, smaller apertures were
used to avoid contamination by nearby field objects. Flux densities were
calculated using an aperture correction of 1.105 and a conversion factor of
0.0454 MJy sr$^{-1}$ (DN s$^{-1}$)$^{-1}$ \citep{Engelbracht2007}. No color
corrections have been applied to the photometric results.

Because the reductions are identical for all the measurements of each object,
the relative brightness is determined very accurately; i.e., for repeated
measurements of standard stars, MIPS achieved repeatability of $\sim$ 0.4\%
\citep{Engelbracht2007}. Because the quasars are much fainter than the stars
and hence less immune to faint background structures, we ascribe a systematic
error of 0.7\% (see Section~\ref{sec:radio-quiet-24}) and combine it with the
statistical errors by root sum square.  We rejected a small number of sources
detected at S/N $<$ 5, where the systematic errors may be larger.
The measured fluxes and resulting errors for the remaining sample of 139
quasars are provided in Table~\ref{tab:agn-24-var}. 

\startlongtable
\begin{deluxetable*}{lclcccc}
\tablecaption{Quasar flux measurements at 24 $\mu$m  and the variability results\label{tab:agn-24-var}
}
        \tablewidth{1.0\hsize}
    \tablecolumns{7}
    \tablehead{
	\colhead{Name} & 
	\colhead{z}   &
	\colhead{JD}  & 
	\colhead{f24}  & 
	\colhead{error}  & 
	\colhead{stdev}   &
        \colhead{comments}\\
	\colhead{}      &
	\colhead{}      &
	\colhead{-2,450,000} &
	\colhead{mJy} &
	\colhead{mJy} &
	\colhead{of change}      &
	}
    \startdata
2MASX J00070361+1554240	& 0.114	& 3194.8	&	63.5	&	0.7	&		&	RQ	\\
	&  &	4676.2	&	62.9	&	0.7	&	0.6	&		\\
	&  &	5376$^a$	&	63.3	&	2.6	&	0.2	&		\\
PG0026+129	&  0.142 &	3747.4	&	43.2	&	0.7	&		&	RQ	\\
	&  &	4677.8	&	43.2	&	0.7	&	0.0	&		\\
	&  &	5382	&	44.1	&	3.8	&	0.2	&		\\
2MASS J00300421-2842259	& 0.278  &	3550.2	&	125.1	&	1.0	&		&	RQ, IRASF00275-2859	\\
	&  &	4675.4	&	124.7	&	1.0	&	0.3	&		\\
	&  &	5362	&	122.7	&	3.9	&	0.5	&		\\
2MASS J00411870+2816408	& 0.194 &	3195.3	&	74.5	&	0.8	&		&	RQ	\\
	&  &	4514.7	&	73.6	&	0.8	&	0.9	&		\\
	&  &	5392	&	76.6	&	2.5	&	1.2	&		\\
2MASX J00505570+2933281	& 0.136  &	3218.3	&	58.0	&	0.6	&		&		\\
	&  &	4514.7	&	57.3	&	0.7	&	0.8	&		\\
	&  &	5394	&	57.3	&	2.3	&	0.0	&		\\
PG0052+251	& 0.154  &	3217.8	&	68.7	&	0.7	&		&	RQ, W4 high due to strong 18 $\mu$m feature$^b$	\\
	&  &	4509.2	&	68.9	&	0.7	&	0.2	&		\\
	&  &	5393	&	76.0	&	2.4	&	2.9	&		\\
$\cdots$        & $\cdots$ &       $\cdots$ &      $\cdots$ &          $\cdots$ &            $\cdots$ &  $\cdots$
    \enddata
    \tablenotetext{a}{For midpoint of the {\it WISE} observations}
    \tablenotetext{b}{The strong 18$~\mu$m feature in the spectrum of this source \citep{Shi2014} combined with the $\sim$1~$\mu$m bluer bandpass of the W4 filter compared with the MIPS one can account for the higher signal seen by {\it WISE}.}
    \tablenotetext{c}{Steep-spectrum radio quasar (SSRQ) indicates that the spectral index between 1.4 and 5 GHz is $\le -0.7$. Radio data obtained from summary in NASA Extragalactic Database (NED) and other sources. (see Section~\ref{sec:larger-24mum} for details.) }
\tablenotetext{d}{Blazars are identified from \citet{Mao2016} and additional sources, see text.}
\tablenotetext{e}{Flat-spectrum radio quasar (FSRQ) indicates that the spectral index between 1.4 and 5 GHz is $> -0.7$.}
\tablenotetext{f}{This source is radio-intermediate by our criteria, but radio-loud using others \citep{laor2019}; given its flat spectrum also, we classify it as radio-loud.}
\tablecomments{
    (This table is available in its entirety in machine-readable form. A
	portion is shown here for guidance regarding its form and content.)
}
\end{deluxetable*}

To extend the time sampling to 2010, we collected {\it WISE} $W4$ measurements
nominally at 22 $\mu$m for 136 of these objects from the ALLWISE Source
Catalog.  One {\it WISE} measurement was rejected because of confusion with a
nearby source.  The $W4$ and MIPS [24] photometric bands are similar, with a
cut-on wavelength at 19.87~$\mu$m (half power) for the former and 20.80~$\mu$m
for the latter, with the long wavelength response determined by the Si:As IBC
detectors in both cases.

Low-resolution ($\lambda/\Delta\lambda\sim$60--130) mid-IR spectroscopic
observations of all 87 PG quasars were obtained using the {\it Spitzer} IRS.
By comparing them with the \textit{WISE} $W3$ and $W4$ photometric
measurements, we can study quasar variability at 10--22~$\mu$m. For the
\textit{Spitzer} mid-IR spectra, we adopted the optimal extraction products
from the Combined Atlas of Sources with \textit{Spitzer} IRS Spectra (CASSIS;
\citealt{cassis}).  Besides PG 0003+199 (no {Long-Low} module observation at
14--38~$\mum$) and PG 1352+183 (very poor {Short-Low} spectra at
5.2--14~$\mum$), the mid-IR spectra cover $\sim$5.2--24~$\mum$ for all objects
and with good S/Ns.  Compared with the {\it WISE} measurements in 2010, these
{\it Spitzer}/IRS spectra were typically obtained 5 yr earlier.  For PG
1226+023 (3C 273), multiple {\it Spitzer}/IRS observations were conducted
between 2004 and 2009, making possible the construction of mid-IR light curves
over a wide wavelength range. For the \textit{WISE} $W3$ and $W4$ measurements
of these PG quasars, since no convincing variability has been reported in these
bands over the 9-month cryogenic mission by the \textit{WISE} pipeline, we
adopted the profile-fit photometry values from the ALLWISE source catalog.

\subsection{Construction and Evaluation of the Light Curves}\label{sec:lc-build}

In the optical and mid-IR $W1$, $W2$ bands, we have sufficient data to sample 
light curves for many years. In this section, we discuss the procedures 
developed for this purpose.

\subsubsection{Optical}

Typically, the CRTS data cover MJD$\sim$53500--56500, and the ASAS-SN data cover
MJD$\sim$56000--58300, with an overlap of about 500 days. We rely on these
datasets to construct the optical light curves for most objects. The CRTS data
generally have smaller relative uncertainties. However, the absolute values of
the unfiltered photometry are hard to interpret, so we scaled the CRTS data to
match the ASAS-SN data in the region of overlap; the latter have photometry
measurements obtained through a standard V-band filter and calibrated by nearby
standard stars. Since CRTS and ASAS-SN observations have different
time samplings, we interpolated the observed values with the same time grid by
fitting a Damped Random Walk (DRW) model to the CRTS and ASAS-SN light curves
separately and computed an average scaling factor for the best-fit model light
curves with the shared time periods.  

\begin{figure}[htp]
    \begin{center}
	\includegraphics[width=1.0\hsize]{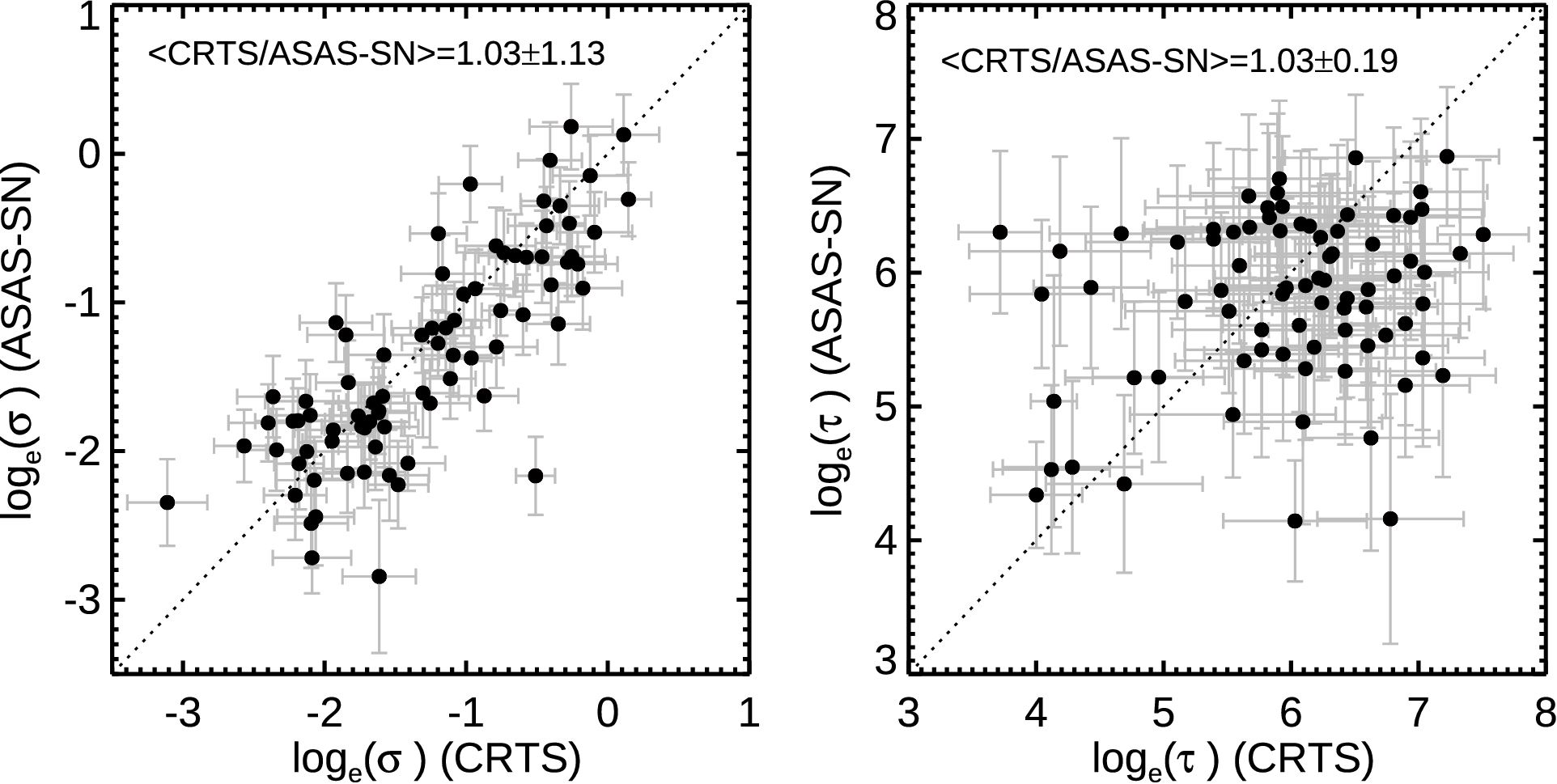}
		\caption{
	      Comparisons of the best-fit DRW model parameters from fitting the
	      ASAS-SN and CRTS light curves separately with the {\it JAVELIN}
	      code. The left panel shows the amplitude $\sigma$ of the fitted
	      covariance function and the left panel is the corresponding
	      damping timescale. We denote the average parameter ratio with the
	      corresponding standard deviation between CRTS and ASAS-SN light
	      curves on the top of each panel.
		}
	\label{fig:2lc_drw_par}
    \end{center}
\end{figure}

\begin{figure*}[htp]
    \begin{center}
	\includegraphics[width=1.0\hsize]{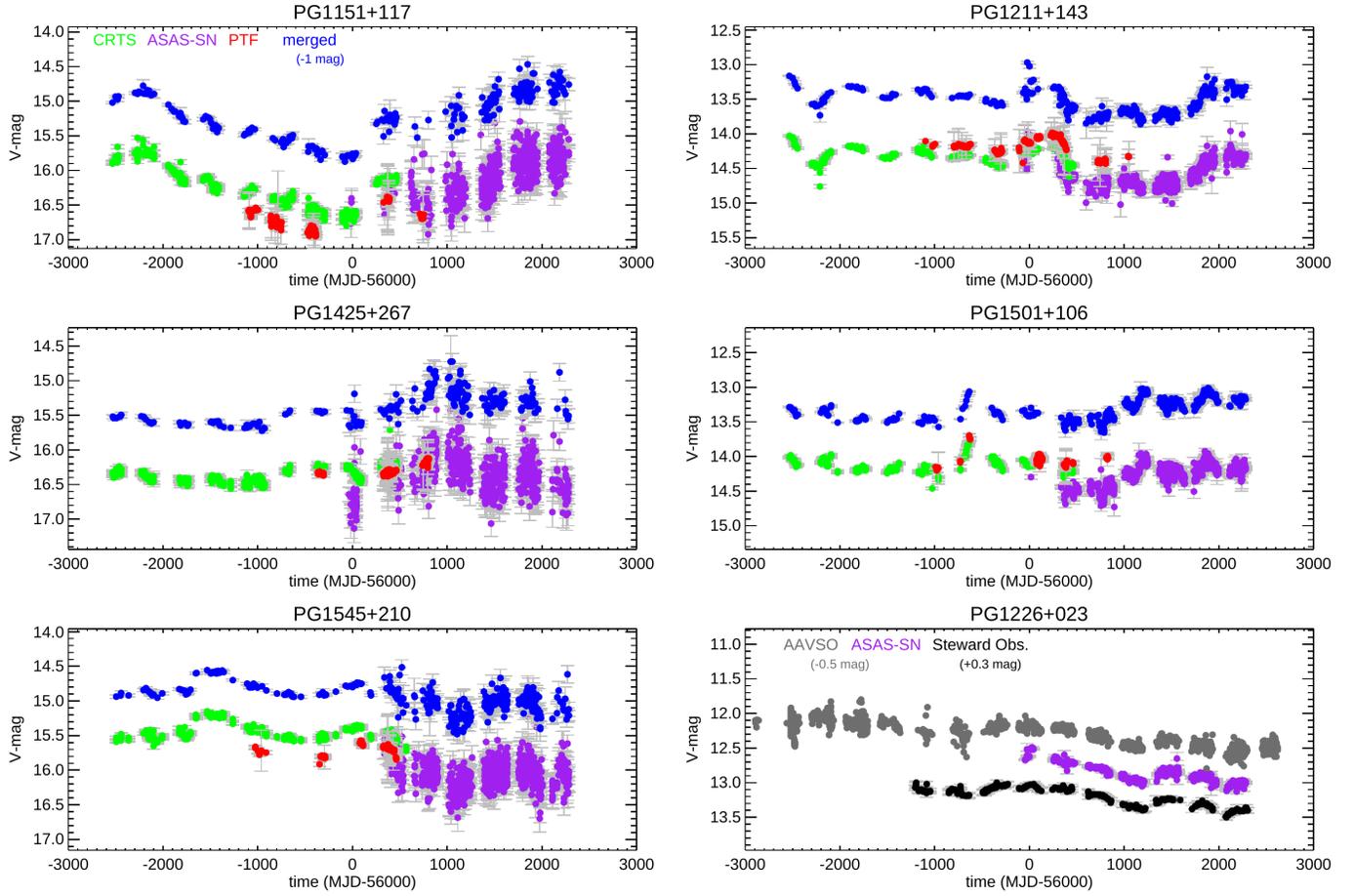}
		\caption{
			Example optical light curves of PG quasars.  The
			magnitude measurements from ASAS-SN (purple), CRTS
			(green) and PTF (red) are shown with the original
			values in their catalog without any shifts.  The final
			combined light curves (blue data points) are shifted by
			1 mag for clarity.  For PG 1226+023 (bottom-right
			panel), we compare its light curves from AAVSO (grey,
			shifted by $-$0.5 mag), ASAS-SN (purple, no shift) and
			observations obtained with two Steward Observatory
			telescopes (black, shifted by +0.3 mag).
		}
	\label{fig:opt_lc_example}
    \end{center}
\end{figure*}

To demonstrate the validity of using the combined unfiltered CRTS and V-band
ASAS-SN light curves to trace the accretion disk variability, we compare the
derived best-fit DRW parameters for these two datasets in
Figure~\ref{fig:2lc_drw_par}. We have used {\it JAVELIN} \citep{Zu2013} to fit the
slightly smoothed optical light curves\footnote{A one-day smoothing window is
introduced for both CRTS and ASAS-SN light curves to reduce the photometry
uncertainties before the fittings.} and the derived best-fit parameters using a
MCMC analysis. They are in excellent agreement despite the large uncertainties
of individual values, suggesting the CRTS and ASAS-SN light curves trace
identical variability properties. In terms of the wavelength dependence of the
AGN optical continuum variability, \cite{Jiang2017} found the average time
delays between $g$ ($\sim0.48~\mum$) and $z$ ($\sim0.91~\mum$)  are less than
six days for quasars with $L_\text{AGN, bol}\sim 10^{11}$--$10^{13}~L_\odot$.
This is at least one order of magnitude smaller than the dust-reverberation
lags our data can probe.  Consequently, we do not need to make any corrections
for the filter differences between the CRTS and ASAS-SN datasets. In fact, we
have also visually checked the CRTS and ASAS-SN combined light curves of
individual objects, finding that they have consistent features as revealed by
the PTF light curves.  Figure~\ref{fig:opt_lc_example} gives some examples of
how this model works on real data.

Among 87 PG quasars, four objects (PG 0804+761, PG 0838+761, PG 1100+772 and PG
1427+480) do not have CRTS observations. In these cases, we join the PTF data
and ASAS-SN data together to extend the time coverage of their optical light
curves. PG 1354+213 and PG 1416-129 are too faint for the ASAS-SN photometry to
be useful. We have also dropped the CRTS light curve of PG 1226+023 since the
photometry measurements are saturated.

For some objects, there are spike features in the optical light curves, which
commonly turned out to be one single photometry measurement larger (or smaller)
than the nearby average values by 1--2 mag. This is quite likely unrelated to
the AGN itself, but caused by e.g., poor seeing conditions, calibration errors,
or contamination such as cosmic rays. We rejected such data points by
introducing 3-sigma clipping and smoothed the final optical light curve by
averaging the remaining measurements taken within a single day to reduce the
noise.

In the bottom-right panel of Figure~\ref{fig:opt_lc_example}, we compare the
optical V-band light curves of PG 1226+023 from different sources. Although
AAVSO and ASAS-SN measurements have larger dispersions than the data obtained
from dedicated monitoring with the Steward Observatory (SO) telescopes, these
light curves reveal the same variability patterns for this object. We decide to
adopt the SO light curve for the correlation analysis between the optical and
mid-IR variability and combine it with the AAVSO data to increase the time sampling
range if required.

\subsubsection{{\it WISE} W1 and W2}\label{sec:wise-lc-construction}

A single {\it WISE} epoch normally contains 10--20 7.7-second individual
exposures with a total time coverage of about one day. For the mid-IR
variability traced by the {\it WISE} data, we have first visually investigated
the consistency of {\it WISE} W1 and W2 single-exposure photometry: they should
have similar variability features given their small wavelength difference. This
is confirmed for the available epochs of most objects. For PG 1259+593, PG
1534+580 and PG 1700+518, there are only 1--2 photometry spikes in the
single-epoch light curves and they can be easily rejected. There are
uncorrelated W1 and W2 variability signals in 1/3 of the epochs for PG
0923+129. Finally, PG 1617+175 has many abnormal W1-band photometry values in
the two epochs observed in the cryogenic survey, whose variability features are
not found in the W2 band. We have rejected these problematic measurements.

We are interested in the response signal of the dust emission; the signal in
the WISE bands will therefore arise near the inner radius of the torus around
the central engine and any variations should be much slower than one light day.
Therefore, we averaged the photometry values during each 1-day epoch to compile
the mid-IR light curve, i.e.,
\begin{equation}
    f_{epoch} = \frac{1}{N}\sum_{i=1}^Nf_i~.
\end{equation}
\noindent Standard deviations of the single-exposure photometry were also
derived to estimate the measurement uncertainties for each epoch. In addition,
uncertainties in photometric measurements ($\sigma_{i, pho}$) as well as the
system stability ($\sigma_{s.s.}$) need to be considered, so the total flux
uncertainty for a given epoch is
\begin{equation}
    \sigma_{epoch}^2 =  \frac{1}{N-1}\sum_{i=1}^N(f_i - \bar f)^2 + \frac{1}{N^2}\sum_{i=1}^N\sigma_{i, pho}^2 + \frac{1}{N}\sigma_{s.s.}^2~.
\end{equation}
In Appendix~\ref{app:pho_stable}, we study the photometric stability of WISE
and NEOWISE using mid-IR calibration stars, finding
$\sigma_{s.s.}\sim0.029$~mag for {\it WISE} measurements and
$\sigma_{s.s.}\sim0.016$~mag for NEOWISE measurements.

\vspace{10pt}

For both optical and mid-IR light curves, we converted magnitudes into flux to
better reflect the variability signals from the AGN, since various
contaminations (e.g., AGN host galaxy, nearby objects) and photometry zero
point uncertainties would only cause a constant flux shift.

In Table~\ref{tab:PG-data}, we provide a summary of available light curve data
and their basic properties for the PG sample. Interested readers may contact
the authors directly to get a copy of these light curves.

\startlongtable
\begin{deluxetable*}{lcccccccccccccc}
    \tabletypesize{\scriptsize}
    \tablewidth{1.0\textwidth}
    \tablecolumns{15}
    \tablecaption{Summary of Available Data and Basic Variability Properties\label{tab:PG-data}
    }
    \tablehead{
        \colhead{Name} &
        \multicolumn{3}{c}{Optical Sources} &
        \colhead{$\delta M_V$} &
        \colhead{MJD(V)} &
        \colhead{$N_\text{epoch}$} &
        \colhead{V}  &
        \colhead{$\Delta V$} &
        \colhead{MJD(IR)} &
        \colhead{$N_\text{epoch}$} &
        \colhead{W1}    &
        \colhead{$\Delta {W1}$} &
        \colhead{W2}    &
        \colhead{$\Delta {W2}$}\\
        &
        \colhead{\tiny CRTS}  &
        \colhead{\tiny ASAS-SN} &
        \colhead{\tiny PTF} & \\
	\colhead{(1)} &
	\colhead{(2)} &
	\colhead{(3)} &
	\colhead{(4)} &
	\colhead{(5)} &
	\colhead{(6)} &
	\colhead{(7)} &
	\colhead{(8)} &
	\colhead{(9)} &
	\colhead{(10)} &
	\colhead{(11)} &
	\colhead{(12)} &
	\colhead{(13)} &
	\colhead{(14)} &
	\colhead{(15)} 
        }
        \startdata
	  PG0003+158 &   I  &  I  & II    & 0.19  & 53648.3--58281.1 & 325 & 15.63 & 1.03 &  55376.1--58459.7 & 13 & 12.27 & 0.06 & 11.22 & 0.11\\
  PG0003+199 &   I  &  I  & II    & 0.33  & 53648.3--58281.1 & 382 & 14.16 & 0.46 &  55377.8--58461.4 & 14 &  8.75 & 0.54 &  7.84 & 0.43\\
  PG0007+106 &   I  &  I  & III   & 0.42  & 53520.5--58282.1 & 344 & 15.43 & 0.85 &  55375.0--58458.8 & 13 & 10.55 & 0.38 &  9.58 & 0.27\\
  PG0026+129 &   I  &  I  & III   & 0.32  & 53520.5--58282.1 & 381 & 15.33 & 1.82 &  55379.5--58463.5 & 13 & 11.10 & 0.24 & 10.10 & 0.19\\
  PG0043+039 &   I  &  I  & III   & 0.14  & 53563.4--58284.1 & 260 & 15.91 & 0.78 &  55559.1--58463.5 & 11 & 12.02 & 0.16 & 10.95 & 0.12\\
  PG0049+171 &   I  &  I  & N.A.  & 0.28  & 53636.4--58281.1 & 369 & 15.57 & 1.20 &  55203.5--58311.3 & 13 & 11.79 & 0.24 & 10.84 & 0.19\\
  PG0050+124 &   I  &  I  & III   & 0.41  & 53636.4--58273.1 & 380 & 14.12 & 0.54 &  55387.5--58309.8 & 12 &  8.77 & 0.30 &  7.77 & 0.21\\
  PG0052+251 &   I  &  I  & III   & 0.26  & 53561.4--58138.7 & 378 & 15.40 & 1.00 &  55206.4--58314.9 & 16 & 11.08 & 0.20 & 10.09 & 0.18\\
  PG0157+001 &   I  &  I  & II    & 0.22  & 53627.4--58158.5 & 315 & 15.46 & 0.82 &  55210.8--58319.8 & 13 & 10.80 & 0.43 &  9.81 & 0.31\\
  PG0804+761 & N.A. &  I  & I     & 0.22  & 55312.3--58271.7 & 388 & 14.25 & 0.75 &  55284.5--58398.4 & 14 &  9.37 & 0.33 &  8.34 & 0.22\\
  PG0838+770 & N.A. &  I  & I     & 0.22  & 55312.3--58257.8 & 180 & 15.95 & 1.00 &  55286.8--58400.7 & 15 & 11.58 & 0.31 & 10.66 & 0.27\\
  PG0844+349 &   I  &  I  & II    & 0.24  & 53495.2--58254.8 & 385 & 14.36 & 0.82 &  55305.2--58418.5 & 11 & 10.49 & 0.29 &  9.63 & 0.22\\
  PG0921+525 &   I  &  I  & II    & 0.27  & 53714.4--58260.8 & 383 & 14.36 & 0.93 &  55305.9--58422.9 & 18 & 10.24 & 0.55 &  9.30 & 0.41\\
  PG0923+201 &   I  &  I  & III   & 0.29  & 53469.2--58228.6 & 352 & 15.24 & 0.95 &  55319.5--58432.8 & 12 & 10.48 & 0.16 &  9.55 & 0.11\\
  PG0923+129 &   I  &  I  & II    & 0.45  & 53500.2--58264.5 & 421 & 14.26 & 0.56 &  55321.7--58434.8 & 12 & 10.16 & 0.36 &  9.27 & 0.31\\
  PG0934+013 &   I  &  I  & III   & 0.30  & 53464.2--58223.7 & 347 & 15.56 & 0.76 &  55327.8--58440.6 & 13 & 11.34 & 0.82 & 10.41 & 0.77\\
  PG0947+396 &   I  &  I  & III   & 0.24  & 53537.2--58252.8 & 168 & 16.41 & 1.02 &  55318.0--58431.2 & 15 & 11.75 & 0.44 & 10.59 & 0.43\\
  PG0953+414 &   I  &  I  & II    & 0.09  & 53711.4--58261.4 & 337 & 14.88 & 0.84 &  55318.3--58431.6 & 14 & 10.74 & 0.27 &  9.76 & 0.17\\
  PG1001+054 &   I  &  I  & II    &-0.08  & 53464.2--58246.6 & 298 & 16.17 & 1.82 &  55332.6--58445.1 & 12 & 11.42 & 0.24 & 10.41 & 0.25\\
  PG1004+130 &   I  &  I  & II    & 0.19  & 53527.2--58264.5 & 256 & 15.33 & 0.60 &  55330.7--58443.3 & 13 & 11.58 & 0.18 & 10.48 & 0.13\\
  PG1011-040 &   I  &  I  & III   & 0.31  & 53496.1--58230.6 & 364 & 14.93 & 0.78 &  55340.5--58453.2 & 13 & 11.08 & 0.25 & 10.21 & 0.24\\
  PG1012+008 &   I  &  I  & III   & 0.27  & 53464.2--58228.6 & 335 & 15.62 & 1.02 &  55339.3--58448.9 & 12 & 11.36 & 0.35 & 10.37 & 0.23\\
  PG1022+519 &   I  &  I  & II    & 0.21  & 53714.4--58271.8 & 263 & 15.41 & 0.81 &  55318.4--58431.8 & 15 & 11.35 & 0.42 & 10.53 & 0.42\\
  PG1048+342 &   I  &  I  & III   & 0.22  & 53496.0--58275.8 & 227 & 16.52 & 0.71 &  55332.0--58444.6 & 15 & 12.36 & 0.24 & 11.34 & 0.27\\
  PG1048-090 &   I  &  I  & III   &-0.51  & 53707.7--58233.5 & 345 & 15.44 & 0.62 &  55350.8--58462.9 & 14 & 11.98 & 0.06 & 10.97 & 0.06\\
  PG1049-005 &   I  &  I  & N.A.  & 0.23  & 53464.3--58230.7 & 303 & 15.99 & 0.99 &  55347.7--58459.8 & 12 & 11.22 & 0.10 & 10.10 & 0.07\\
  PG1100+772 & N.A. &  I  & I     & 0.23  & 55294.4--58281.8 & 338 & 15.26 & 0.64 &  55297.3--58410.9 & 14 & 11.29 & 0.17 & 10.23 & 0.17\\
  PG1103-006 &   I  &  I  & III   &-0.07  & 53464.3--58231.9 & 310 & 16.05 & 1.11 &  55351.0--58462.9 & 13 & 12.21 & 0.08 & 11.09 & 0.10\\
  PG1114+445 &   I  &  I  & II    & 0.18  & 53531.2--58280.8 & 261 & 15.83 & 0.86 &  55331.7--58451.9 & 18 & 10.88 & 0.10 &  9.73 & 0.11\\
  PG1115+407 &   I  &  I  & N.A.  & 0.16  & 53527.2--58280.8 & 307 & 15.76 & 0.65 &  55334.1--58446.5 & 15 & 11.33 & 0.22 & 10.42 & 0.13\\
  PG1116+215 &   I  &  I  & III   & 0.17  & 53469.3--58283.8 & 373 & 14.71 & 1.83 &  55345.2--58457.6 & 12 & 10.02 & 0.26 &  9.01 & 0.23\\
  PG1119+120 &   I  &  I  & III   & 0.26  & 53498.2--58283.8 & 368 & 14.62 & 0.96 &  55349.6--58461.6 & 12 & 10.55 & 0.27 &  9.55 & 0.14\\
  PG1121+422 &   I  &  I  & N.A.  & 0.14  & 53527.2--58280.8 & 261 & 16.25 & 1.17 &  55334.3--58452.1 & 17 & 12.13 & 0.38 & 11.15 & 0.38\\
  PG1126-041 &   I  &  I  & II    & 0.19  & 53496.2--58230.5 & 375 & 14.48 & 0.63 &  55357.4--58263.6 & 13 &  9.70 & 0.30 &  8.78 & 0.27\\
  PG1149-110 &   I  &  I  & III   & 0.54  & 53479.2--58275.8 & 360 & 15.11 & 3.18 &  55364.7--58271.2 & 12 & 11.06 & 0.55 & 10.34 & 0.57\\
  PG1151+117 &   I  &  I  & II    & 0.13  & 53466.2--58283.8 & 230 & 16.15 & 1.40 &  55356.8--58259.8 & 12 & 11.87 & 0.51 & 10.87 & 0.30\\
  PG1202+281 &   I  &  I  & II    & 0.14  & 53470.3--58254.9 & 238 & 16.35 & 1.16 &  55352.1--58463.8 & 14 & 11.42 & 0.44 & 10.48 & 0.32\\
  PG1211+143 &   I  &  I  & II    & 0.13  & 53466.3--58284.0 & 364 & 14.55 & 0.89 &  55360.1--58266.3 & 11 & 10.06 & 0.26 &  8.96 & 0.27\\
  PG1216+069 &   I  &  I  & I     & 0.13  & 55270.4--58282.8 & 272 & 13.42 & 0.57 &  55364.1--58270.3 & 12 & 11.63 & 0.14 & 10.68 & 0.10\\
  PG1226+023 & III  &  I  & N.A.  & 0.00  & 55956.1--58283.8 & 299 & 12.87 & 0.61 &  55370.4--58274.3 & 12 &  8.41 & 0.16 &  7.45 & 0.11\\
  PG1229+204 &   I  &  I  & II    & 0.29  & 53469.3--58282.8 & 424 & 14.89 & 0.66 &  55361.5--58267.6 & 15 & 10.64 & 0.48 &  9.73 & 0.42\\
  PG1244+026 &   I  &  I  & II    & 0.31  & 53767.3--58283.8 & 276 & 15.82 & 0.74 &  55374.2--58277.9 & 12 & 11.63 & 0.08 & 10.58 & 0.06\\
  PG1259+593 &   I  &  I  & II    & 0.06  & 53767.3--58279.8 & 309 & 15.44 & 0.62 &  55336.9--58452.8 & 16 & 11.25 & 0.18 & 10.22 & 0.13\\
  PG1302-102 &   I  &  I  & N.A.  & 0.16  & 53496.2--58274.7 & 347 & 15.21 & 0.68 &  55208.7--58286.7 & 13 & 11.38 & 0.24 & 10.28 & 0.17\\
  PG1307+085 &   I  &  I  & N.A.  & 0.14  & 53466.3--58283.8 & 315 & 15.62 & 0.74 &  55377.2--58280.7 & 12 & 11.41 & 0.28 & 10.36 & 0.23\\
  PG1309+355 &   I  &  I  & II    & 0.15  & 53526.2--58284.9 & 360 & 15.36 & 0.64 &  55362.7--58268.9 & 17 & 11.14 & 0.13 & 10.07 & 0.08\\
  PG1310-108 &   I  &  I  & N.A.  & 0.26  & 53496.2--58274.7 & 349 & 15.24 & 0.62 &  55210.2--58288.5 & 13 & 11.21 & 0.16 & 10.08 & 0.20\\
  PG1322+659 &   I  &  I  & II    &-0.04  & 53860.4--58279.8 & 281 & 15.56 & 0.87 &  55330.8--58451.9 & 18 & 11.48 & 0.33 & 10.48 & 0.23\\
  PG1341+258 &   I  &  I  & II    & 0.06  & 53470.3--58283.9 & 375 & 15.63 & 1.16 &  55204.0--58281.2 & 16 & 11.49 & 0.37 & 10.66 & 0.30\\
  PG1351+236 &   I  &  I  & III   & 0.27  & 53470.3--58284.9 & 356 & 15.27 & 0.80 &  55207.3--58291.2 & 17 & 11.37 & 0.31 & 10.74 & 0.34\\
  PG1351+640 &   I  &  I  & II    & 0.00  & 53767.5--58284.9 & 390 & 14.77 & 0.76 &  55337.1--58453.2 & 17 & 10.28 & 0.08 &  9.19 & 0.11\\
  PG1352+183 &   I  &  I  & II    & 0.18  & 53469.4--58284.9 & 272 & 16.13 & 0.80 &  55208.9--58291.4 & 16 & 11.71 & 0.31 & 10.68 & 0.28\\
  PG1354+213 &   I  & III & II    & 0.00  & 53469.4--56447.2 &  80 & 16.64 & 0.78 &  55208.4--58291.3 & 17 & 12.69 & 0.50 & 11.56 & 0.48\\
  PG1402+261 &   I  &  I  & II    & 0.08  & 53470.3--58283.9 & 363 & 15.61 & 0.65 &  55208.4--58285.9 & 16 & 10.60 & 0.25 &  9.54 & 0.22\\
  PG1404+226 &   I  &  I  & N.A.  & 0.13  & 53469.4--58284.9 & 298 & 15.80 & 0.77 &  55209.7--58287.8 & 15 & 11.81 & 0.24 & 10.89 & 0.19\\
  PG1411+442 &   I  &  I  & II    & 0.22  & 53509.4--58284.8 & 355 & 14.73 & 0.67 &  55372.8--58276.5 & 15 & 10.11 & 0.16 &  9.07 & 0.16\\
  PG1415+451 &   I  &  I  & II    & 0.17  & 53509.4--58284.8 & 317 & 15.79 & 0.74 &  55372.7--58276.5 & 15 & 11.33 & 0.44 & 10.44 & 0.33\\
  PG1416-129 &   I  & III & N.A.  & 0.00  & 53498.3--56478.5 & 100 & 16.92 & 0.74 &  55222.9--58306.4 & 13 & 12.30 & 0.43 & 11.32 & 0.37\\
  PG1425+267 &   I  &  I  & II    & 0.18  & 53470.3--58281.9 & 224 & 16.36 & 1.00 &  55212.5--58293.7 & 17 & 11.98 & 0.09 & 10.88 & 0.07\\
  PG1426+015 &   I  &  I  & II    & 0.25  & 53464.4--58283.9 & 378 & 14.32 & 0.66 &  55221.0--58304.2 & 13 &  9.95 & 0.26 &  9.06 & 0.17\\
  PG1427+480 & N.A. &  I  & I     & 0.25  & 54962.3--58284.8 & 402 & 16.39 & 2.28 &  55373.0--58276.8 & 15 & 12.25 & 0.15 & 11.18 & 0.11\\
  PG1435-067 &   I  &  I  & N.A.  & 0.22  & 53497.3--58283.9 & 330 & 15.80 & 0.84 &  55225.1--58308.8 & 13 & 11.40 & 0.29 & 10.34 & 0.20\\
  PG1440+356 &   I  &  I  & II    & 0.17  & 53480.3--58281.9 & 390 & 14.61 & 0.72 &  55211.4--58292.7 & 17 &  9.93 & 0.23 &  9.07 & 0.23\\
  PG1444+407 &   I  &  I  & III   & 0.16  & 53509.4--58281.9 & 313 & 15.71 & 0.75 &  55209.3--58291.4 & 18 & 11.36 & 0.15 & 10.33 & 0.14\\
  PG1448+273 &   I  &  I  & II    & 0.15  & 53470.4--58283.9 & 382 & 14.72 & 0.64 &  55217.2--58299.9 & 18 & 10.94 & 0.41 & 10.05 & 0.33\\
  PG1501+106 &   I  &  I  & II    & 0.28  & 53466.4--58282.9 & 403 & 14.29 & 0.64 &  55225.8--58309.5 & 14 & 10.19 & 0.30 &  9.20 & 0.21\\
  PG1512+370 &   I  &  I  & II    & 0.23  & 53480.3--58281.9 & 277 & 16.19 & 1.04 &  55217.9--58300.5 & 18 & 12.19 & 0.09 & 11.07 & 0.10\\
  PG1519+226 &   I  &  I  & II    & 0.20  & 53506.3--58284.9 & 355 & 15.86 & 1.08 &  55225.9--58309.6 & 16 & 10.82 & 0.34 &  9.88 & 0.21\\
  PG1534+580 &   I  &  I  & II    & 0.31  & 53856.4--58281.8 & 344 & 14.55 & 0.44 &  55203.3--58278.5 & 16 & 10.41 & 0.51 &  9.44 & 0.47\\
  PG1535+547 &   I  &  I  & II    & 0.17  & 53880.5--58284.8 & 404 & 14.67 & 0.71 &  55203.1--58291.3 & 18 &  9.91 & 0.55 &  9.07 & 0.42\\
  PG1543+489 &   I  &  I  & II    & 0.06  & 53531.3--58277.9 & 202 & 16.24 & 0.82 &  55216.8--58299.2 & 19 & 11.52 & 0.13 & 10.40 & 0.13\\
  PG1545+210 &   I  &  I  & II    & 0.37  & 53506.4--58284.9 & 324 & 15.98 & 0.96 &  55232.7--58320.9 & 17 & 11.62 & 0.24 & 10.56 & 0.23\\
  PG1552+085 &   I  &  I  & III   & 0.12  & 53466.4--58284.9 & 379 & 15.59 & 0.81 &  55241.0--58325.8 & 15 & 11.58 & 0.37 & 10.66 & 0.31\\
  PG1612+261 &   I  &  I  & II    & 0.06  & 53470.4--58281.9 & 413 & 15.42 & 0.99 &  55241.2--58325.9 & 16 & 11.01 & 0.15 & 10.10 & 0.15\\
  PG1613+658 &   I  &  I  & II    & 0.15  & 53856.4--58281.9 & 330 & 14.65 & 0.57 &  55355.0--58464.9 & 18 & 10.16 & 0.14 &  9.15 & 0.09\\
  PG1617+175 &   I  &  I  & II    & 0.12  & 53469.4--58283.8 & 420 & 15.16 & 0.88 &  55245.4--58330.4 & 16 & 10.40 & 0.30 &  9.48 & 0.27\\
  PG1626+554 &   I  &  I  & II    & 0.19  & 53505.4--58281.8 & 341 & 15.65 & 1.03 &  55221.1--58304.2 & 17 & 11.39 & 0.24 & 10.47 & 0.18\\
  PG1700+518 &   I  &  I  & II    & 0.18  & 53505.4--58281.2 & 374 & 14.89 & 0.49 &  55242.8--58327.9 & 17 & 10.26 & 0.04 &  9.20 & 0.04\\
  PG1704+608 &   I  &  I  & II    & 0.18  & 53856.4--58284.9 & 347 & 15.34 & 0.91 &  55223.3--58306.5 & 16 & 10.79 & 0.06 &  9.73 & 0.06\\
  PG2112+059 &   I  &  I  & III   & 0.19  & 53466.5--58284.0 & 415 & 15.42 & 0.77 &  55330.9--58416.8 & 13 & 10.99 & 0.12 &  9.83 & 0.07\\
  PG2130+099 &   I  &  I  & II    & 0.31  & 53466.5--58284.0 & 419 & 14.54 & 0.76 &  55336.4--58422.1 & 14 &  9.56 & 0.34 &  8.59 & 0.25\\
  PG2209+184 &   I  &  I  & II    & 0.41  & 53480.5--58284.0 & 385 & 15.54 & 0.87 &  55348.9--58439.7 & 15 & 11.40 & 0.50 & 10.61 & 0.62\\
  PG2214+139 &   I  &  I  & II    & 0.31  & 53554.4--58282.1 & 396 & 14.42 & 0.49 &  55348.5--58439.6 & 13 &  9.77 & 0.12 &  8.88 & 0.07\\
  PG2233+134 &   I  &  I  & II    & 0.19  & 53531.4--58284.0 & 276 & 16.34 & 0.96 &  55355.1--58440.4 & 15 & 12.04 & 0.19 & 10.88 & 0.16\\
  PG2251+113 &   I  &  I  & II    & 0.33  & 53531.4--58282.0 & 340 & 15.67 & 0.63 &  55358.4--58443.3 & 12 & 10.98 & 0.03 & 10.07 & 0.04\\
  PG2304+042 &   I  &  I  & II    & 0.51  & 53553.4--58281.6 & 362 & 15.14 & 2.28 &  55358.5--58443.5 & 12 & 11.04 & 1.01 & 10.42 & 1.05\\
  PG2308+098 &   I  &  I  & II    & 0.17  & 53506.5--58282.0 & 283 & 16.06 & 1.11 &  55361.6--58446.5 & 11 & 12.02 & 0.24 & 10.93 & 0.23\\

        \enddata
         \tablecomments{
	     Col. (2)-(5): the optical data status: I -- used to build the
	     light curve, II -- used to check the data consistency, III -- not
	     useful either due to limited time samplings or poor data quality,
	     N.A. -- data is not available;
	     col. (6), (10): the range of Modified Julian Date that the final optical and IR light curve covers;
	     col. (7), (11): number of data points in the light curves;
	     col. (8), (12), (14): the average magnitude of the light curves;
	     col. (9), (13), (15): the variation amplitude of the light curves.
    }
\end{deluxetable*}

\subsection{Converting the Mid-IR Measurements at $\lambda>10~\mum$ for Comparison}

We do not have sufficient data to construct light curves in the 10--25~$\mu$m
range, but can assess variability by comparing the individual measurements. We
describe here how we extended this comparison beyond just the MIPS 24 $\mu$m
data.

\subsubsection{WISE W4 to MIPS [24]}

To convert the {\it WISE} W4 measurements to the MIPS [24] system, we first
eliminated all the quasars in our sample that we found to vary, plus those with
W4 measurement errors $\ge$ 0.08 magnitudes. These cuts left 95 quasars with
high-weight measurements in both systems. We took the zero point for [24] to be
7.17 Jy and based the [24] magnitude on the weighted average of the two
measurements of each quasar. A linear fit, which is specific to quasar SEDs,
results in a transformation equation of 
\begin{equation} 
    W4 - [24] = 0.014 \pm 0.036 + (0.024 \pm 0.015) \times (W3-W4)~.
\label{trans} 
\end{equation}

We applied Equation~\ref{trans} to the {\it WISE} data and assigned the date
corresponding to the averaged modified Julian Date of the {\it WISE}
single-exposures from the {\it WISE} All-Sky Data Release Source Catalog.  To
allow for uncertainties in the transformation as well as potential systematic
uncertainties in the WISE measurements, we augmented the quoted errors by 0.02
magnitudes (see Appendix~\ref{app:pho_stable}), as the root sum square with the
nominal error. We then calculated the change between the weighted average of
the first {\it Spitzer} measurements and the {\it WISE} one.
Table~\ref{tab:agn-24-var} shows the transformed {\it WISE} measurement, its
estimated uncertainty, and the number of standard deviations difference between
it and the {\it Spitzer} measurements.  Sources with changes $> 3 \sigma$
between the {\it Spitzer} and {\it WISE} measurements and among {\it Spitzer}
measurements themselves are indicated as variable in the comments column.

The slope coefficient in Equation~\ref{trans} is consistent with zero, so we
have tested the reliability of the results by simply averaging the ratios of
the two measurements to derive a color-independent transformation. Using these
results slightly increased the apparent significance level of the changes,
typically by about 0.1 sigma. In general, there would be no change in the
designation of variable sources with this alternative transformation with the
exception of Cygnus A,  which is changed from being marginally not variable
(2.5 $\sigma$) to marginally variable (3.4 $\sigma$). That is, the results are
relatively robust against modest errors in the transformation. This is
not surprising, given how closely the W4 and MIPS [24] bands resemble each
other.

The MIPS-measured sample includes 33 of the 87 z $<$ 0.5 PG quasars (plus 106
additional AGN). To investigate the longer wavelength AGN IR variability for
the complete PG sample, we compared their {\it WISE} W3 ($\sim12~\mum$) and W4
($\sim22~\mum$) measurements and {\it Spitzer}/IRS spectra \citep{spitzerirs}. 
We adopted the profile-fit photometry from the ALLWISE Source Catalog, whose 
measurements were carried out on the co-added image atlas since little 
variability was detected in PG quasars at these wavelengths. Due to the wide 
passbands of W3 and W4 as well as the calibration uncertainty of the W4 Relative 
Spectral Response \citep{WISE}, we introduced flux corrections by increasing 
the W3 band by 17\% and decreasing the W4 band by 10\%, based on the typical 
$f_\nu\sim \nu^{-1}$ SED shape of unobscured quasars at these wavelengths, 
as suggested by the Explanatory Supplement to the WISE All-Sky Data Release 
Products.\footnote{See details in
    \url{http://wise2.ipac.caltech.edu/docs/release/allsky/expsup/sec4_4h.html}.}

\subsubsection{Synthesis Photometry from Spitzer/IRS Spectra}

To be compared with the measurements in {\it WISE} W3 and W4 bands, we stitched
the {\it Spitzer}/IRS spectra from different module observations together and
computed synthetic photometry by convolving the spectral flux with the
corresponding photometry Relative Spectral Response curves. For the synthesis
photometry uncertainty, we convolved the total error spectra, which include
statistical and systematic errors\footnote{See
\url{http://irs.sirtf.com/Smart/CassisProducts}}, with the relative photometry
curves.

\section{Reverberation Analysis Methods}\label{sec:method}

In contrast to the reverberation mapping analysis of the optical emission
lines, the transfer functions of the dust mid-IR response are greatly
complicated by the radiation transfer of an IR optically-thick torus, whose
structures, density profile, grain properties and the inclination angle to the
observer are poorly known \citep[e.g.,][]{Kawaguchi2011, Almeyda2017}. With the
sparse cadence of the infrared light curves, it is not very meaningful
to fit these data with the simulated reverberation response of any specific
torus model.  Instead, we develop a ``minimalist'' method to retrieve the most
important properties of the torus by comparing the IR and optical light curves.

\subsection{Constraining the Optical Continuum Variability}
 
The optical continuum variability of bright AGNs is well described by a damped
random walk (DRW) \citep[e.g.,][]{Kelly2009, Kozlowski2010, MacLeod2010,
Zu2011}. In this model, the quasar optical light curves can be reproduced as a
stochastic process and the covariance function follows $S(\Delta t) = \sigma^2
\exp(-|\Delta t/\tau_\text{DRW}|)$, where $\sigma$ is the amplitude and
$\tau_\text{DRW}$ is the damping timescale.  We use {\it JAVELIN} \citep{Zu2013} to
fit the optical light curves with the DRW model and fill the time gaps with the
model values.\footnote{The DRW model is also known as the simplest of
    the continuous-time autoregressive moving average process models
    (CARMA(1,0) or CAR(1)). Despite its wide use, evidence for deviations
    from the CAR(1) model for AGN optical light curves
    has been found \citep[e.g.,][]{Mushotzky2011, Graham2014} and some
    authors have suggested that higher order CARMA models should be adopted
    \citep[e.g.,][]{Kelly2014, Kasliwal2017}. However, as demonstrated in
    \cite{Kasliwal2017}, the behavior of the DRW model and more
    sophisticated CARMA models are identical for time-scales longer than
    $\sim$10~days. Our optical light curve data typically have significant flux
    uncertainties and their average time-samplings are $\sim$5--10 days.
    To be compared with the IR light curves (with time-sampling intervals
    $\sim200$~days), the fitted optical light curves need be further
smoothed (see later). As a result, it is not necessary to use
higher-order CARMA models to describe the optical light curves.}

\begin{figure}[htp]
    \begin{center}
	\includegraphics[width=1.0\hsize]{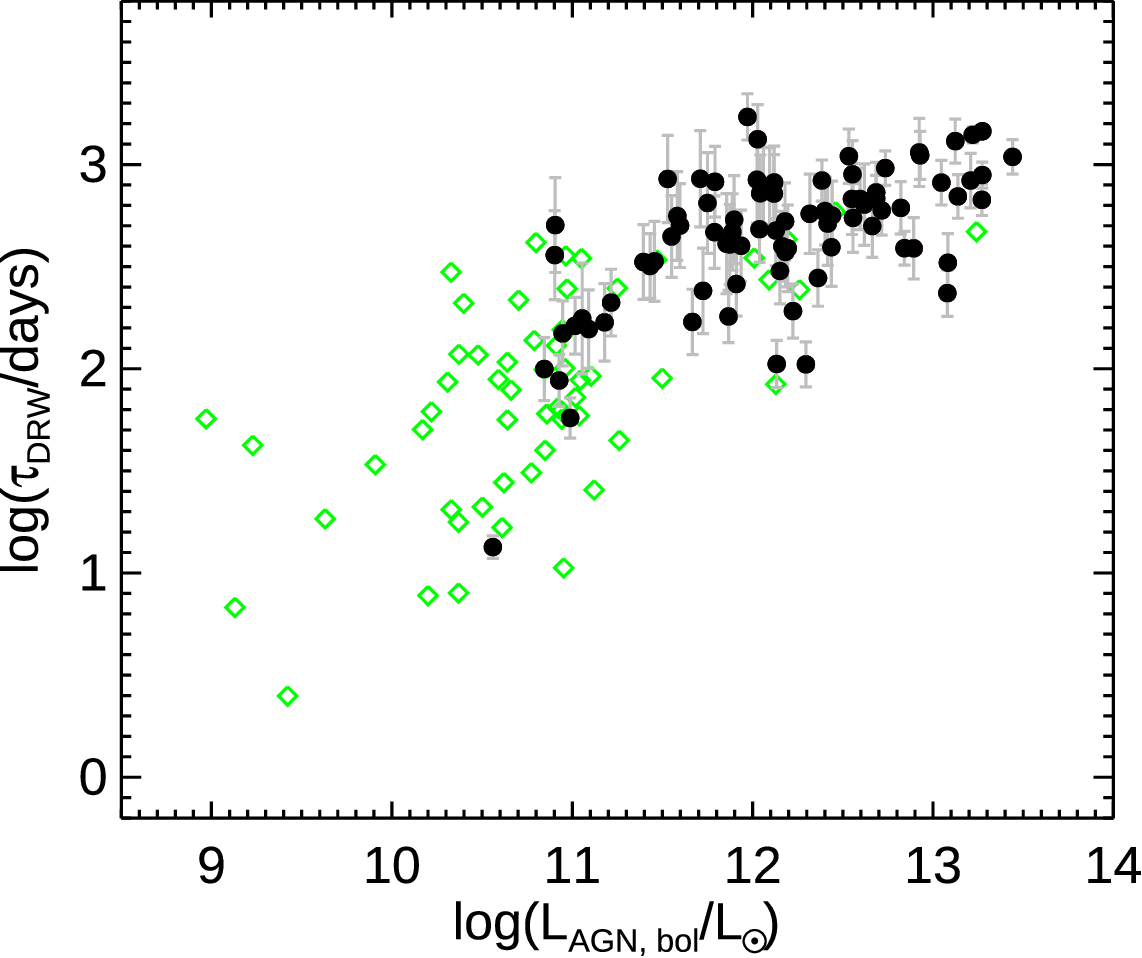}
		\caption{
		    Rest-frame DRW damping timescale, $\tau_\text{DRM}$, of the
		    optical light curves as a function of AGN bolometric
		    luminosity, $L_\text{AGN, bol}$, for PG quasars (filled
		    black dots) and the AGN sample studied by \cite{Zu2011}
		    (open green dots).
		}
	\label{fig:drw_tau}
    \end{center}
\end{figure}

The damping timescale $\tau_{DRW}$ is known to have correlations with AGN
luminosity \citep{Kelly2009, Zu2011}. In Figure~\ref{fig:drw_tau}, we confirm
such a correlation also exists by modeling our stitched optical light curves;
the trend extends the work by \cite{Zu2011} to higher luminosity.
The consistency demonstrated in Figure~\ref{fig:drw_tau} shows that,
despite their relatively low quality, the optical light curves
compiled from these ground transient surveys should be good enough to produce
meaningful results. In addition, the characteristic variability timescales, as
traced by $\tau_\text{DRW}$, are so long that averaging flux measurements and
smoothing light curves (introduced later) would not introduce much in the way
of systematic uncertainty. Finally, the optical variability timescales are
typically larger than the $\sim$200 days per epoch cadences of the mid-IR light
curves, thus the dust reverberation signals should be detected with the
sparsely sampled {\it WISE} data.

\subsection{Retrieving Dust Reverberation Signals}\label{sec:model}

Traditional analyses of the dust reverberation signals were focused on
calculating time lags between two light curves by cross-correlation. Although
we use this method later to confirm the results, the low cadence of our data
led us to develop an alternative method. In addition to the dust time lag, the
scaling factor between the IR and optical variability amplitudes gives another
important piece of information regarding the AGN torus. We therefore decided to
introduce a simple model based only on single-value time lags and amplitudes to
retrieve useful reverberation signals and mitigate the influences of many data
quality issues.

As discussed in Appendix~\ref{app:model}, if the viewing angle is not perfectly
face-on, the dust response lags on the same radius will have a range, resulting
in a smoothing response to the optical variation features in the IR light curve
(see also \citealt{Kawaguchi2011}). Therefore, for simplicity, we apply a
top-hat function, $b(\tau)$, to smooth the optical DRW model fitting curve,
$F(t)_{\rm OPT}$, so 
\begin{equation}\label{eqn:opt_ir_cor}
    <F(t)_{\rm OPT}>_{\tau_W} =  \int^{+\tau_{\rm W}/2}_{-\tau_{\rm W}/2}F(t-\tau)_{\rm OPT}b(\tau) d\tau ~~,
\end{equation}
where $\tau_{\rm W}$ is the width of the boxcar. On average, the smoothing
window size should be correlated with the size of the dust emission structure,
$R$, which is traced by the average time lag, $\Delta t$, between the IR and
optical light curves. By default, the fitted optical DRW model is smoothed
    on-the-fly with $\tau_{\rm W}=\Delta t/2$ and a maximum value of $\tau_{\rm
    W}=200$~days. This upper limit is set to be similar to the time gaps among
the NEOWISE epochs as well as in the optical light curve to avoid over-smoothing.

We assume that, to first order, the IR dust emission light curve $F(t)_{\rm
IR}$ can be described as a scaled version of the smoothed optical light curve
$<F(t')_{\rm OPT}>$ with a constant time lag $\Delta t= t'-t$: 
\begin{equation}
    F(t)_\text{IR, dust} = AMP \times <F(t-\Delta t)_{\rm OPT}>_{\tau_{\rm W}} + F_\text{const.} ~~,
\end{equation}
where $AMP$ is the ratio between the optical and IR flux variation amplitudes,
and $F_{\rm const}$ is the systematic, time-insensitive flux shift between the
optical and IR bands. Physically, $AMP$ reflects the efficiency of the dust IR
energy transfer from the optical variation signal, which is related to the
amount of dust; $\Delta t$ is the light travel time from the accretion disk to
the dust torus at the studied wavelength. The constant, $F_{\rm const}$, is
determined by (1) the AGN SED averaged over a long period of time, (2) systemic
uncertainties between the optical and mid-IR flux zero-points, and (3)
contamination from the host galaxy emission or nearby sources.

The AGN accretion disk could also produce some variability in the mid-IR light
curves. With spectropolarimetry observations of a small sample of bright
quasars, \cite{Kishimoto2008} showed that the quasar accretion disk power-law
continuum extends into the near-IR ($\lambda\sim2~\mum$) and is consistent with
a $F_\nu\propto \nu^{1/3}$ shape.  It is not clear at what wavelength the
accretion disk emission would transfer to a Rayleigh-Jeans slope; typically a
value of 3--5~$\mum$ is suggested in the literature \citep[e.g.,][]{Honig2010,
Stalevski2016}. Considering the lack of constraints, we assume the same
$F_\nu\propto \nu^{1/3}$ spectral shape is valid to describe the accretion disk
emission from the optical to the mid-IR bands. Due to the weak wavelength
dependence of the accretion disk variability \citep[e.g.,][]{Jiang2017}, the
accretion disk emission in the mid-IR can be assumed to change simultaneously
with the optical. As a result, we have
\begin{equation}\label{eqn:ad-spectrum}
    F(t)_\text{IR, accr. disk} = F(t)_{\rm OPT}\left(\frac{\nu_{\rm IR}}{\nu_{\rm OPT}}\right)^{1/3}~~,
\end{equation}
and $\nu_{\rm IR}/\nu_{\rm OPT}\sim$0.16 and $\sim$0.12  for the {\it WISE} W1 
and W2 bands, respectively. To recover the dust reverberation signals, the
contribution of IR variability by the accretion disk itself needs to be removed
according to 
\begin{equation}\label{eqn:ir-dust-signal}
    F(t)_\text{IR, dust} = F(t)_\text{IR} - F(t)_\text{IR, accr. disk} ~~,
\end{equation}
where $F(t)_\text{IR}$ is the observed {\it WISE} light curve. As pointed out at the
very end of Section~\ref{sec:lc-build}, various time-insensitive contaminations
and uncertainties in the optical light curves would only cause a flux offset.
With our linear models to relate the optical and IR variability, these factors
are included in $F_{\rm const}$ and should not influence the measurements of
$\Delta t$ and $AMP$.

\begin{figure*}
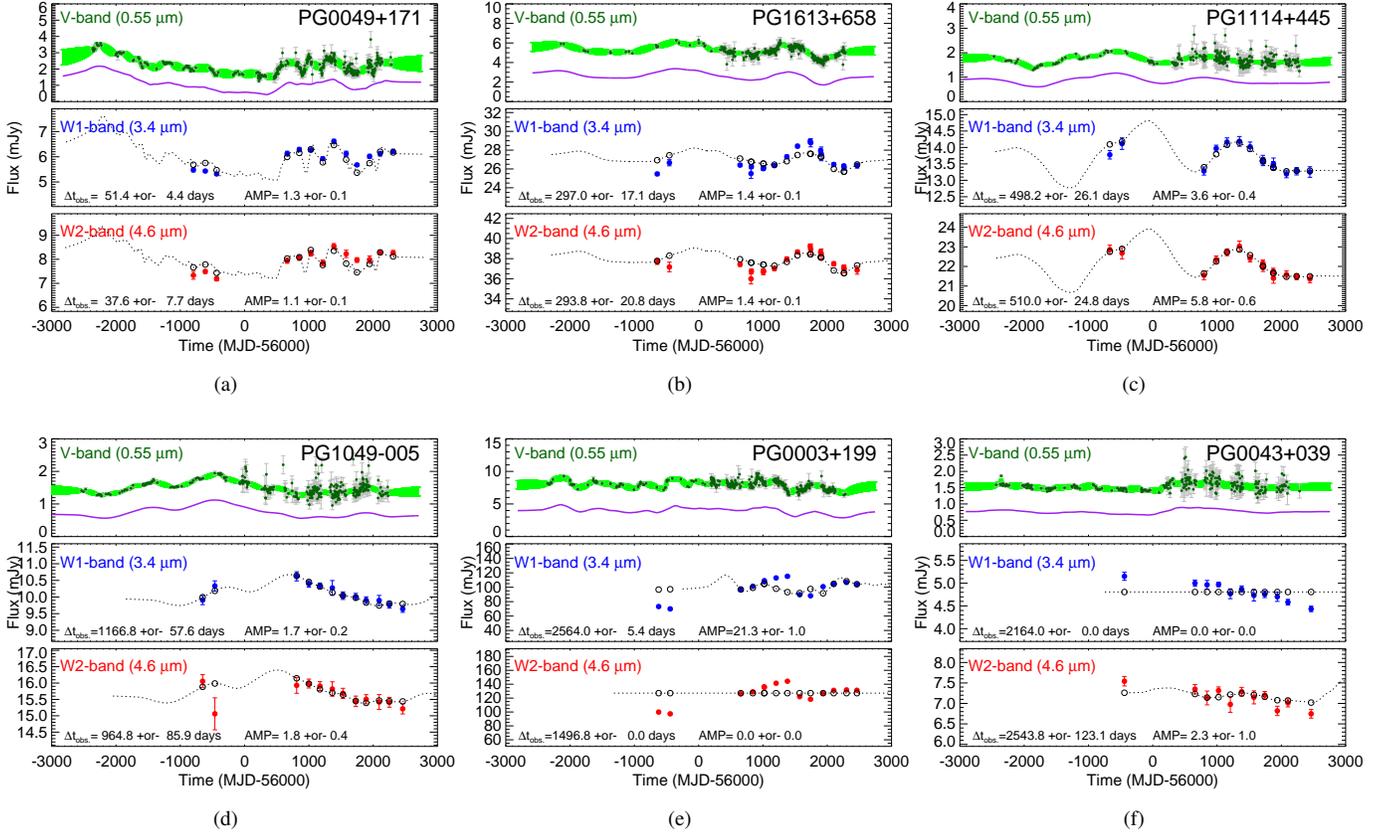

    \gridline{\fig{f5set_5.ps}{0.33\textwidth}{(a)}
	  \fig{f5set_74.ps}{0.33\textwidth}{(b)}
	  \fig{f5set_28.ps}{0.33\textwidth}{(c)}}
\gridline{\fig{f5set_25.ps}{0.33\textwidth}{(d)}
          \fig{f5set_1.ps}{0.33\textwidth}{(e)}
          \fig{f5set_4.ps}{0.33\textwidth}{(f)}}
\caption{
		    Representative optical and mid-IR light curves and
		    corresponding DRW and time-lag fits of PG quasars. Objects
		    in Panels (a)-(d) have convincing dust reveration signals
		    with different mid-IR time lags from $\sim60$ to
		    $\sim1400$ days; Panel (e) gives an example whose IR light
		    curves are uncorrelated with the optical ones; Panel (f) is
		    a case where the mid-IR and/or the optical light curves do not
		    contain enough features to get any convincing time lag
		    measurement.
		    All measurements are presented in the observed frame. The
		    optical (green), 3.4~$\mum$ (blue) and 4.6~$\mum$ (red)
		    data are presented in the top, middle and bottom
		    sub-panels, respectively. The thick green line in the top
		    sub-panel is a DRW model constrained by the optical light
		    curve, with the solid purple line representing a smoothed
		    version of the model scaled down for clarity. The dashed
		    lines in the middle and bottom sub-panels are the model
		    based on the optical light curve to fit the IR, as
		    described in Section~\ref{sec:model}. The model mid-IR
		    fluxes at the observed epochs are shown as open circles. We
		    denote the best-fit parameters with their errors for the
		    mid-IR reverberation signals on the bottom of the
		    corresponding sub-panels.  \\ (The complete figure set (87
		    images) for all PG quasars is available in the online
		    journal)
    }
    \label{fig:lc_example}
\end{figure*}

Based on the model described above, we fitted the smoothed and delayed optical
light curves (interpolated by the DRW model) to the {\it WISE} W1 and W2 ones
separately with the Levenberg-Marquardt least-squares fitting procedure as
implemented in the {\sc IDL} MPFIT \citep{MPFIT} package.
Figure~\ref{fig:lc_example} provides some examples of how this approach applies
to real data.


\figsetstart
\figsetnum{5}
\figsettitle{Optical and Mid-IR Light Curves and RM Model Fittings of PG Quasars}

\figsetgrpstart
\figsetgrpnum{5.1}
\figsetgrptitle{PG0003+158}
\figsetplot{f5set_0.ps}
\figsetgrpnote{Optical and mid-IR light curves and corresponding DRW and time-lag fits of one PG quasars.  All measurements are presented in the observed frame. The optical (green), 3.4 μm (blue) and
4.6 μm (red) data are presented in the top, middle and bottom panels, respectively. The thick green line in the top panel is a DRW
model constrained by the optical light curve, with the solid purple line representing a smoothed version of the model scaled down for clarity.
The dashed lines in the middle and bottom panels are the model based on the optical light curve to fit the IR, as described in Section 3.2.
The model mid-IR fluxes at the observed epochs are shown as open circles. We denote the best-fit parameters with their errors for the mid-IR
reverberation signals on the bottom of the corresponding panels.

}
\figsetgrpend

\figsetgrpstart
\figsetgrpnum{5.2}
\figsetgrptitle{PG0003+199}
\figsetplot{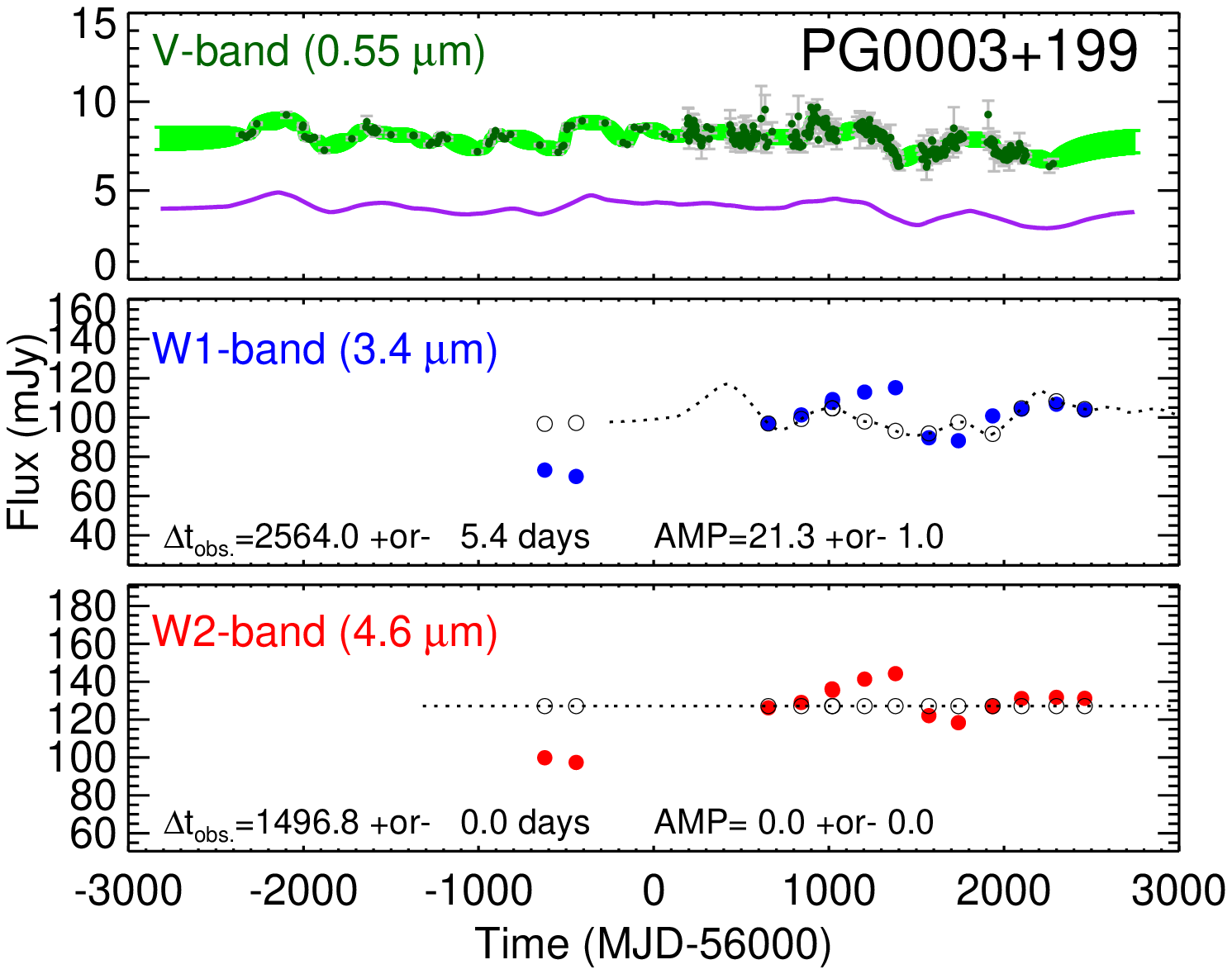}
\figsetgrpnote{Optical and mid-IR light curves and corresponding DRW and time-lag fits of one PG quasars.  All measurements are presented in the observed frame. The optical (green), 3.4 μm (blue) and
4.6 μm (red) data are presented in the top, middle and bottom panels, respectively. The thick green line in the top panel is a DRW
model constrained by the optical light curve, with the solid purple line representing a smoothed version of the model scaled down for clarity.
The dashed lines in the middle and bottom panels are the model based on the optical light curve to fit the IR, as described in Section 3.2.
The model mid-IR fluxes at the observed epochs are shown as open circles. We denote the best-fit parameters with their errors for the mid-IR
reverberation signals on the bottom of the corresponding panels.

}
\figsetgrpend

\figsetgrpstart
\figsetgrpnum{5.5}
\figsetgrptitle{PG0043+039}
\figsetplot{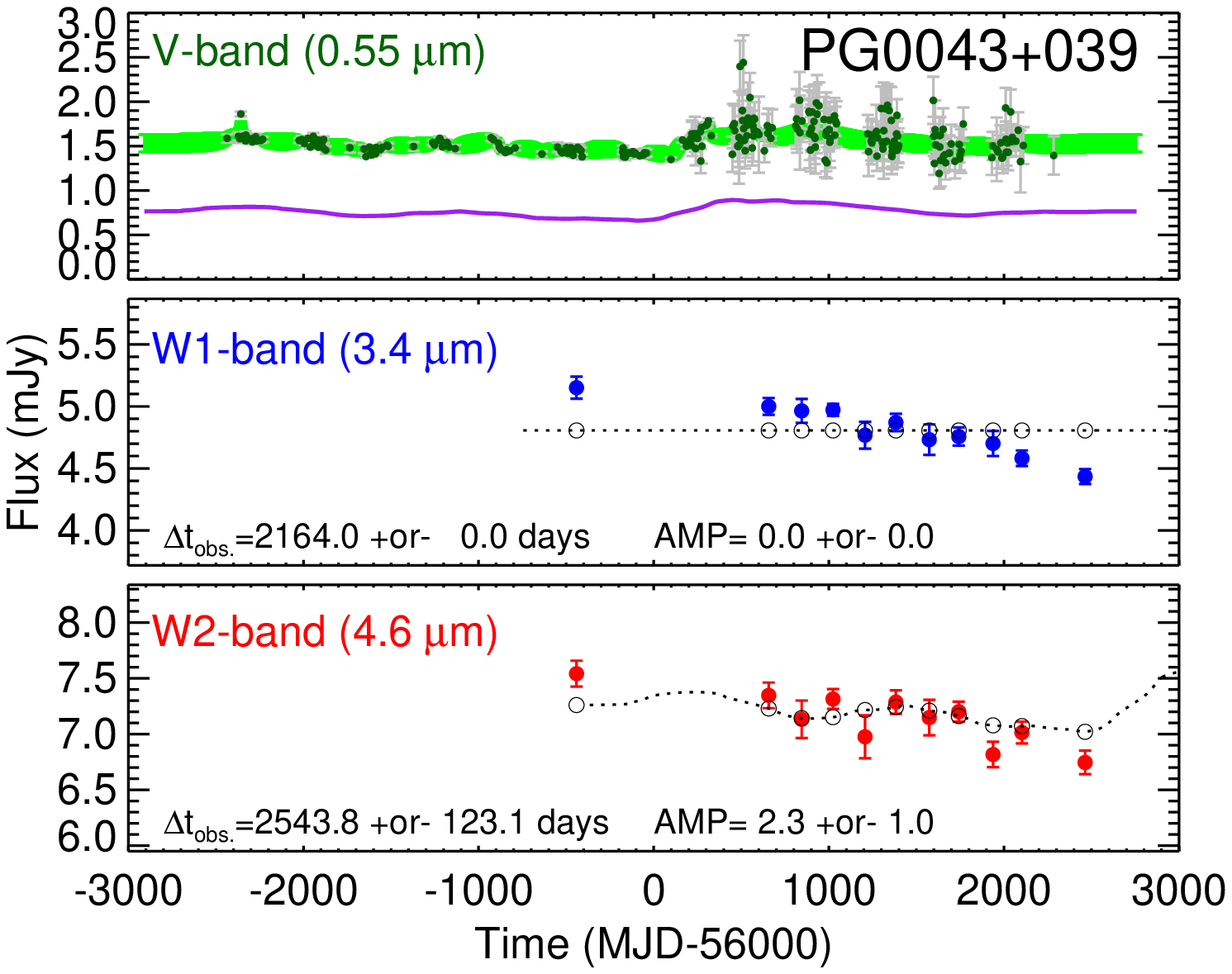}
\figsetgrpnote{Optical and mid-IR light curves and corresponding DRW and time-lag fits of one PG quasars.  All measurements are presented in the observed frame. The optical (green), 3.4 μm (blue) and
4.6 μm (red) data are presented in the top, middle and bottom panels, respectively. The thick green line in the top panel is a DRW
model constrained by the optical light curve, with the solid purple line representing a smoothed version of the model scaled down for clarity.
The dashed lines in the middle and bottom panels are the model based on the optical light curve to fit the IR, as described in Section 3.2.
The model mid-IR fluxes at the observed epochs are shown as open circles. We denote the best-fit parameters with their errors for the mid-IR
reverberation signals on the bottom of the corresponding panels.

}
\figsetgrpend

\figsetgrpstart
\figsetgrpnum{5.6}
\figsetgrptitle{PG0049+171}
\figsetplot{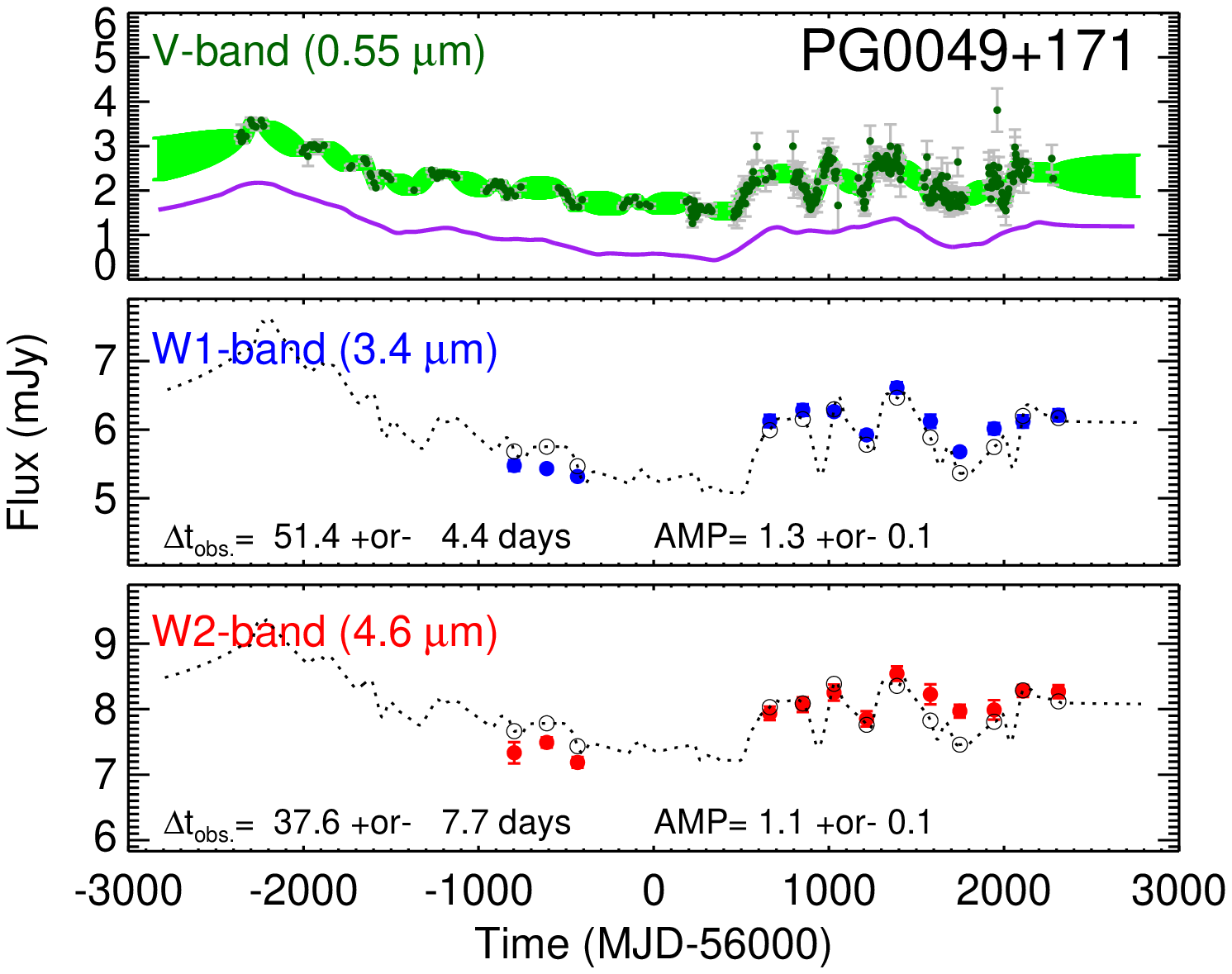}
\figsetgrpnote{Optical and mid-IR light curves and corresponding DRW and time-lag fits of one PG quasars.  All measurements are presented in the observed frame. The optical (green), 3.4 μm (blue) and
4.6 μm (red) data are presented in the top, middle and bottom panels, respectively. The thick green line in the top panel is a DRW
model constrained by the optical light curve, with the solid purple line representing a smoothed version of the model scaled down for clarity.
The dashed lines in the middle and bottom panels are the model based on the optical light curve to fit the IR, as described in Section 3.2.
The model mid-IR fluxes at the observed epochs are shown as open circles. We denote the best-fit parameters with their errors for the mid-IR
reverberation signals on the bottom of the corresponding panels.

}
\figsetgrpend

\figsetgrpstart
\figsetgrpnum{5.26}
\figsetgrptitle{PG1049-005}
\figsetplot{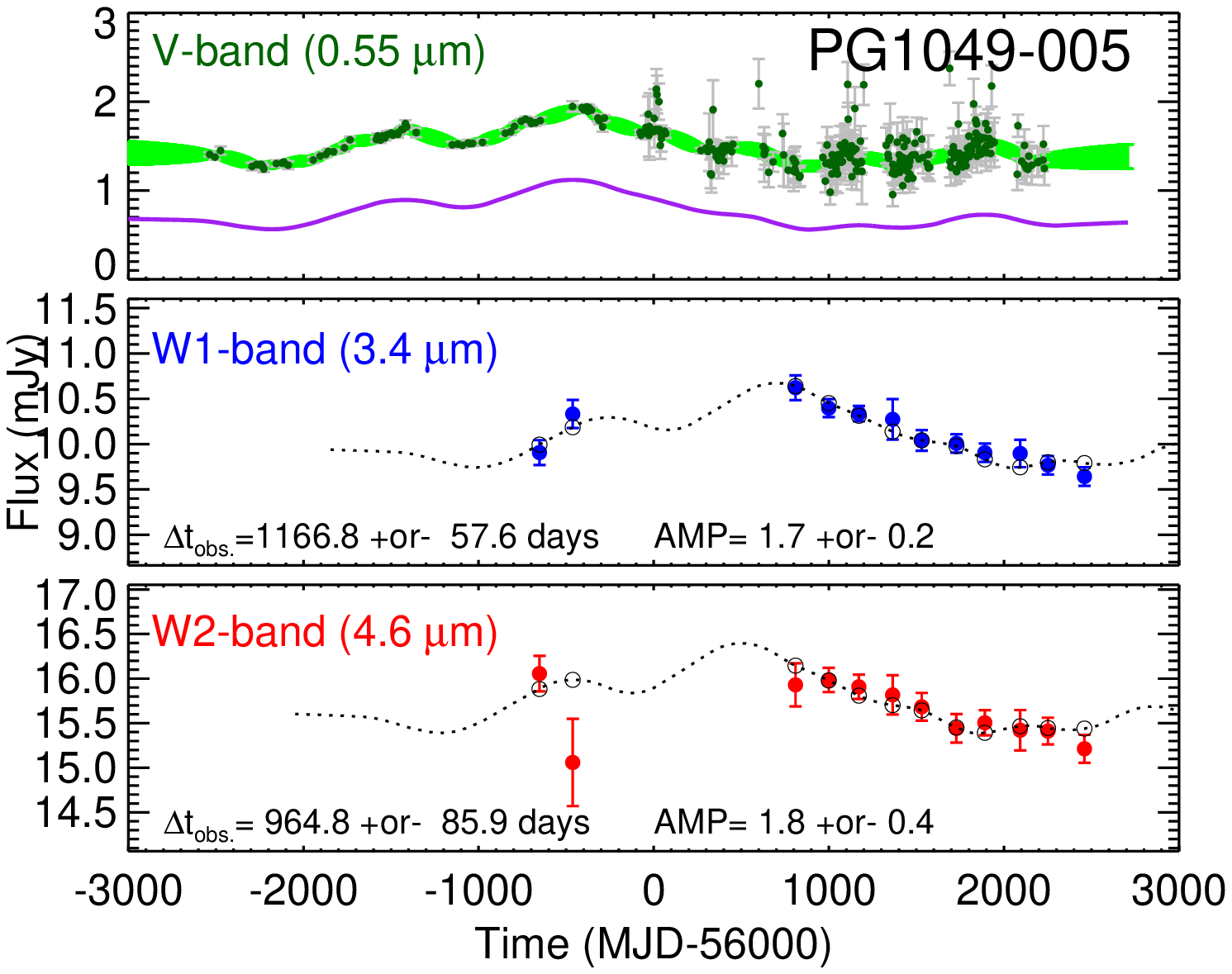}
\figsetgrpnote{Optical and mid-IR light curves and corresponding DRW and time-lag fits of one PG quasars.  All measurements are presented in the observed frame. The optical (green), 3.4 μm (blue) and
4.6 μm (red) data are presented in the top, middle and bottom panels, respectively. The thick green line in the top panel is a DRW
model constrained by the optical light curve, with the solid purple line representing a smoothed version of the model scaled down for clarity.
The dashed lines in the middle and bottom panels are the model based on the optical light curve to fit the IR, as described in Section 3.2.
The model mid-IR fluxes at the observed epochs are shown as open circles. We denote the best-fit parameters with their errors for the mid-IR
reverberation signals on the bottom of the corresponding panels.

}
\figsetgrpend

\figsetgrpstart
\figsetgrpnum{5.29}
\figsetgrptitle{PG1114+445}
\figsetplot{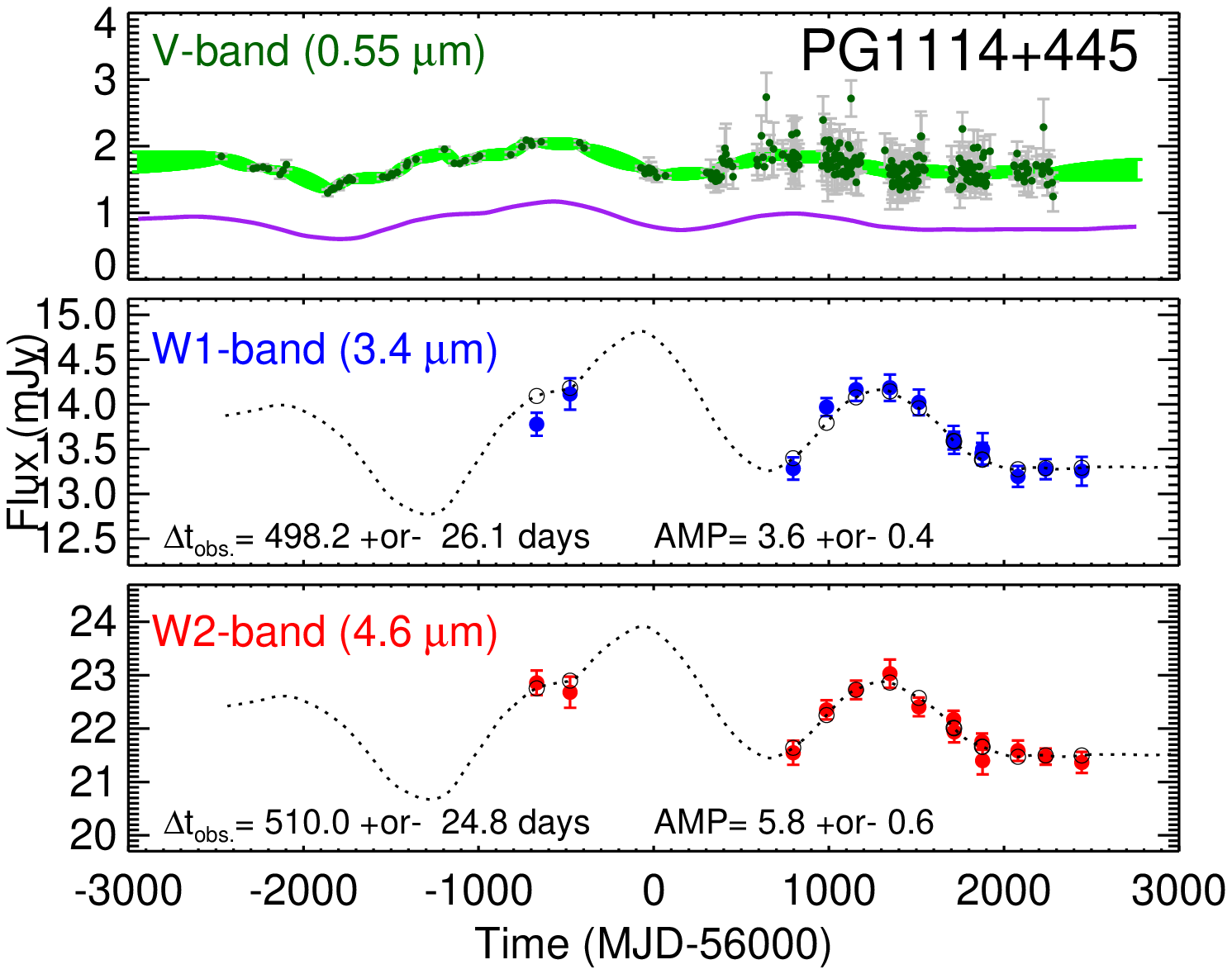}
\figsetgrpnote{Optical and mid-IR light curves and corresponding DRW and time-lag fits of one PG quasars.  All measurements are presented in the observed frame. The optical (green), 3.4 μm (blue) and
4.6 μm (red) data are presented in the top, middle and bottom panels, respectively. The thick green line in the top panel is a DRW
model constrained by the optical light curve, with the solid purple line representing a smoothed version of the model scaled down for clarity.
The dashed lines in the middle and bottom panels are the model based on the optical light curve to fit the IR, as described in Section 3.2.
The model mid-IR fluxes at the observed epochs are shown as open circles. We denote the best-fit parameters with their errors for the mid-IR
reverberation signals on the bottom of the corresponding panels.

}
\figsetgrpend

\figsetgrpstart
\figsetgrpnum{5.75}
\figsetgrptitle{PG1613+658}
\figsetplot{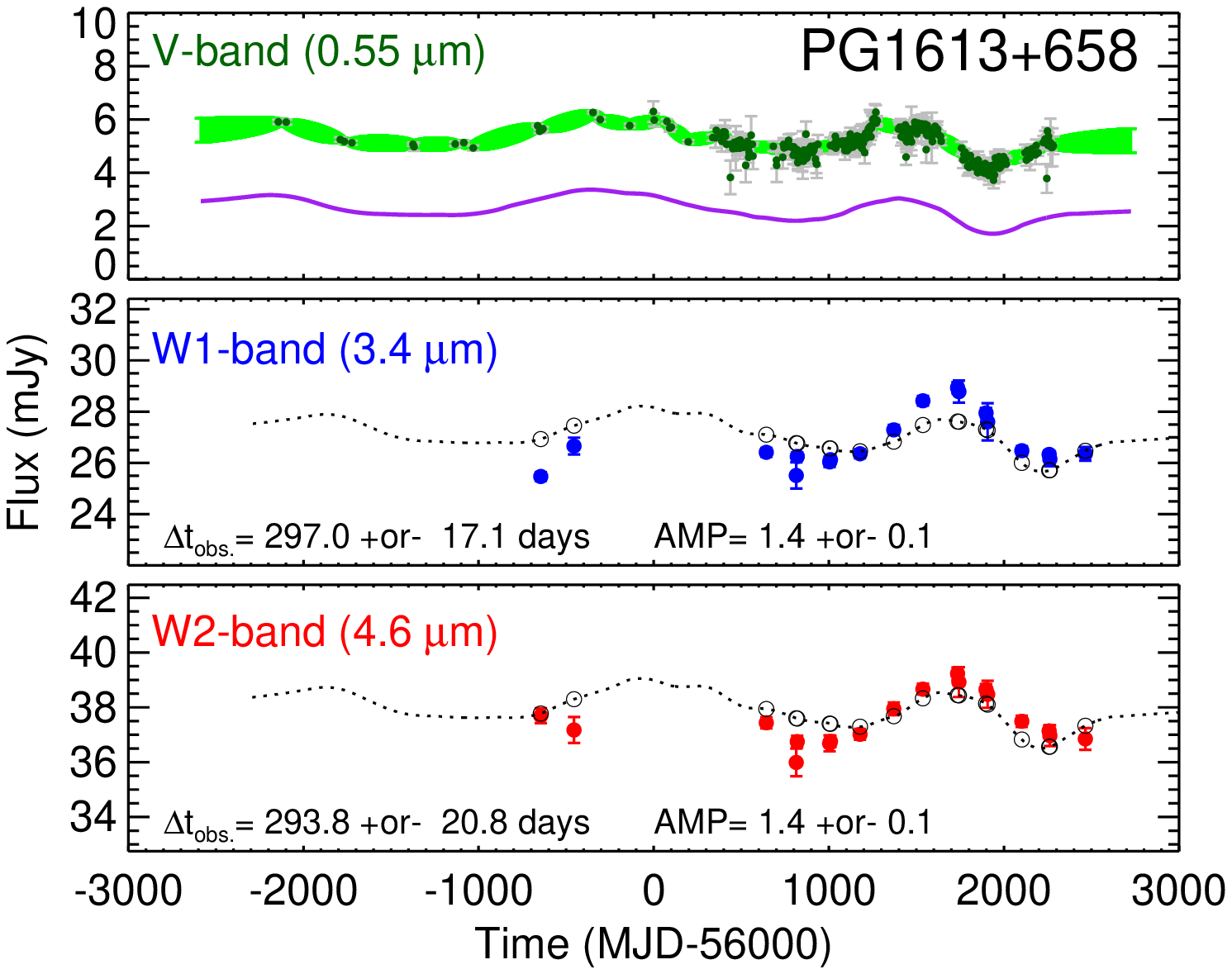}
\figsetgrpnote{Optical and mid-IR light curves and corresponding DRW and time-lag fits of one PG quasars.  All measurements are presented in the observed frame. The optical (green), 3.4 μm (blue) and
4.6 μm (red) data are presented in the top, middle and bottom panels, respectively. The thick green line in the top panel is a DRW
model constrained by the optical light curve, with the solid purple line representing a smoothed version of the model scaled down for clarity.
The dashed lines in the middle and bottom panels are the model based on the optical light curve to fit the IR, as described in Section 3.2.
The model mid-IR fluxes at the observed epochs are shown as open circles. We denote the best-fit parameters with their errors for the mid-IR
reverberation signals on the bottom of the corresponding panels.

}
\figsetgrpend

\figsetend

\subsection{Detecting the Time Lags}\label{sec:lag-detection}

Two methods have been adopted to find the most likely time lags. We first use
the classical $\chi^2$ minimization technique to get best-fit values of ${\rm
AMP}$, $\Delta t$ and $F_{\rm const}$ by fitting the rescaled and shifted
optical DRW light curve model to the {\it WISE} W1 and W2 light curves
separately. To explore the influences of different initial values on the final
results, we change initial guesses of $\Delta t/{\rm day}$ from 0 to 3000 with
an increment of 10. For objects with multiple local minimized $\chi^2$ values, the
parameter values with the smallest $\chi^2$ were usually adopted in the end.
For the time lag uncertainty, the error provided by {\sc MPFIT} only accounts
for the measurement uncertainties of the mid-IR fluxes. To estimate the
uncertainty introduced by the DRW interpolation of the optical light curve and
the optical data itself, we compute 1000 (optical) DRW models with the Monte
Carlo (MC) method around the best-fit parameters of the optical light curves
within the observing constraints from {\it JAVELIN}. Adopting the same initial
values of the model parameters for the MPFIT code, the same observed mid-IR
light curve is then repeatedly fitted by all the 1000 mock optical DRW light
curves with the same RM model using {\it MPFIT}.  For the majority of objects,
the distribution of the best-fit time lags with these optical mock light curves
is symmetric and can be approximated by a Gaussian. We trim away outliers with
3-$\sigma$ clipping and compute the standard deviation to represent the
1-$\sigma$ uncertainty caused by the optical data and DRW model. The final
time-lag uncertainty is a combination of this optical uncertainty based on MC
simulation and the IR uncertainty reported by {\it MPFIT}.


\figsetstart
\figsetnum{6}
\figsettitle{Cross-Correlation Functions between the Optical and IR Light Curves of PG Quasars}

\figsetgrpstart
\figsetgrpnum{6.1}
\figsetgrptitle{PG0003+158}
\figsetplot{f6set_0.ps}
\figsetgrpnote{We calculate the CCF(t) with different smoothing windows: τ_W (black curves),
τ W /2 (red curves), τW /10(green curves) where τ_W is assumed to be 200 days. The histograms show the best-fit time lag distribution by MC
modeling of the optical light curve (blue regions).
}
\figsetgrpend

\figsetgrpstart
\figsetgrpnum{6.2}
\figsetgrptitle{PG0003+199}
\figsetplot{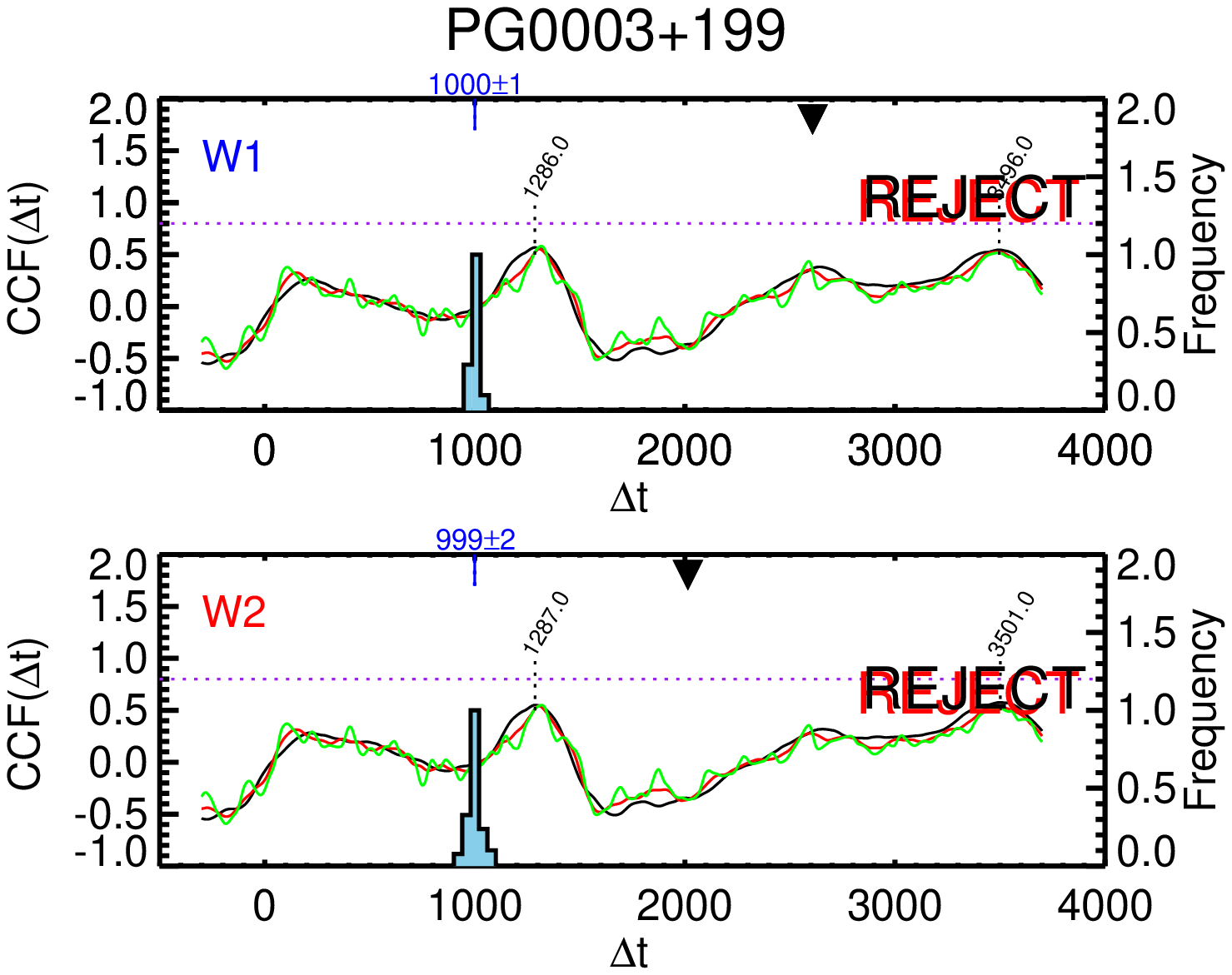}
\figsetgrpnote{We calculate the CCF(t) with different smoothing windows: τ_W (black curves),
τ W /2 (red curves), τW /10(green curves) where τ_W is assumed to be 200 days. The histograms show the best-fit time lag distribution by MC
modeling of the optical light curve (blue regions).
}
\figsetgrpend

\figsetgrpstart
\figsetgrpnum{6.3}
\figsetgrptitle{PG0007+106}
\figsetplot{f6set_2.ps}
\figsetgrpnote{We calculate the CCF(t) with different smoothing windows: τ_W (black curves),
τ W /2 (red curves), τW /10(green curves) where τ_W is assumed to be 200 days. The histograms show the best-fit time lag distribution by MC
modeling of the optical light curve (blue regions).
}
\figsetgrpend

\figsetgrpstart
\figsetgrpnum{6.4}
\figsetgrptitle{PG0026+129}
\figsetplot{f6set_3.ps}
\figsetgrpnote{We calculate the CCF(t) with different smoothing windows: τ_W (black curves),
τ W /2 (red curves), τW /10(green curves) where τ_W is assumed to be 200 days. The histograms show the best-fit time lag distribution by MC
modeling of the optical light curve (blue regions).
}
\figsetgrpend

\figsetgrpstart
\figsetgrpnum{6.5}
\figsetgrptitle{PG0043+039}
\figsetplot{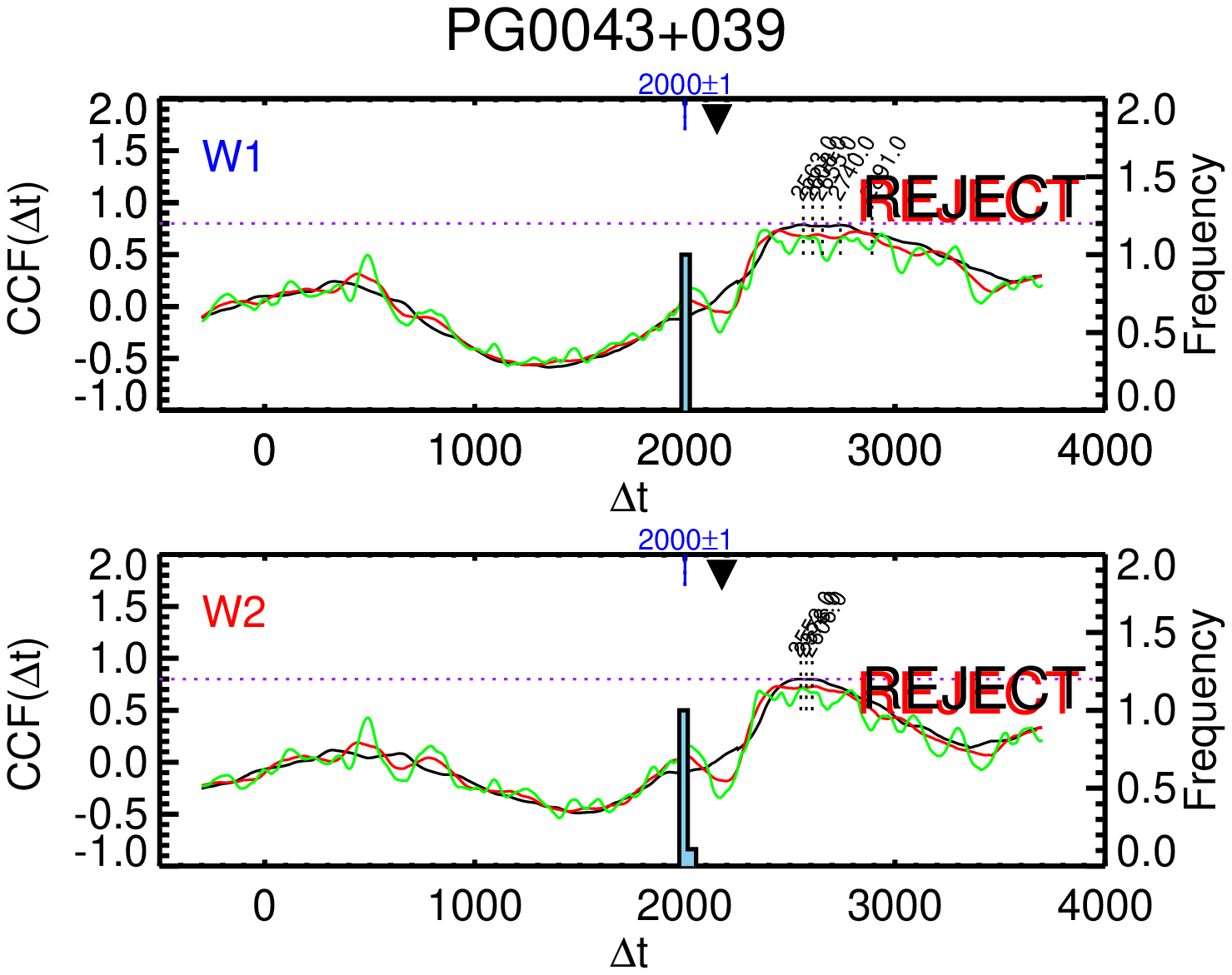}
\figsetgrpnote{We calculate the CCF(t) with different smoothing windows: τ_W (black curves),
τ W /2 (red curves), τW /10(green curves) where τ_W is assumed to be 200 days. The histograms show the best-fit time lag distribution by MC
modeling of the optical light curve (blue regions).
}
\figsetgrpend

\figsetgrpstart
\figsetgrpnum{6.6}
\figsetgrptitle{PG0049+171}
\figsetplot{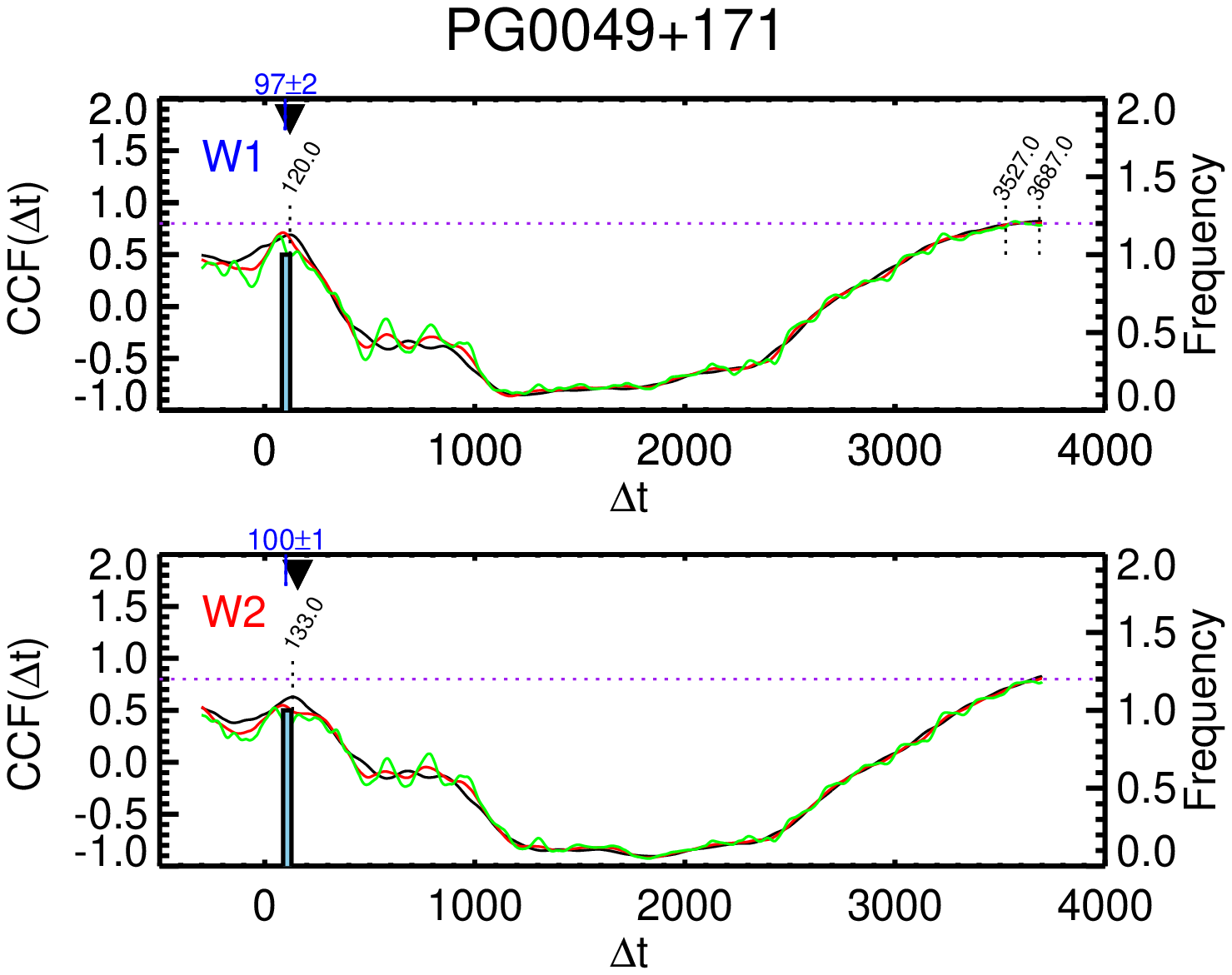}
\figsetgrpnote{We calculate the CCF(t) with different smoothing windows: τ_W (black curves),
τ W /2 (red curves), τW /10(green curves) where τ_W is assumed to be 200 days. The histograms show the best-fit time lag distribution by MC
modeling of the optical light curve (blue regions).
}
\figsetgrpend

\figsetgrpstart
\figsetgrpnum{6.26}
\figsetgrptitle{PG1049-005}
\figsetplot{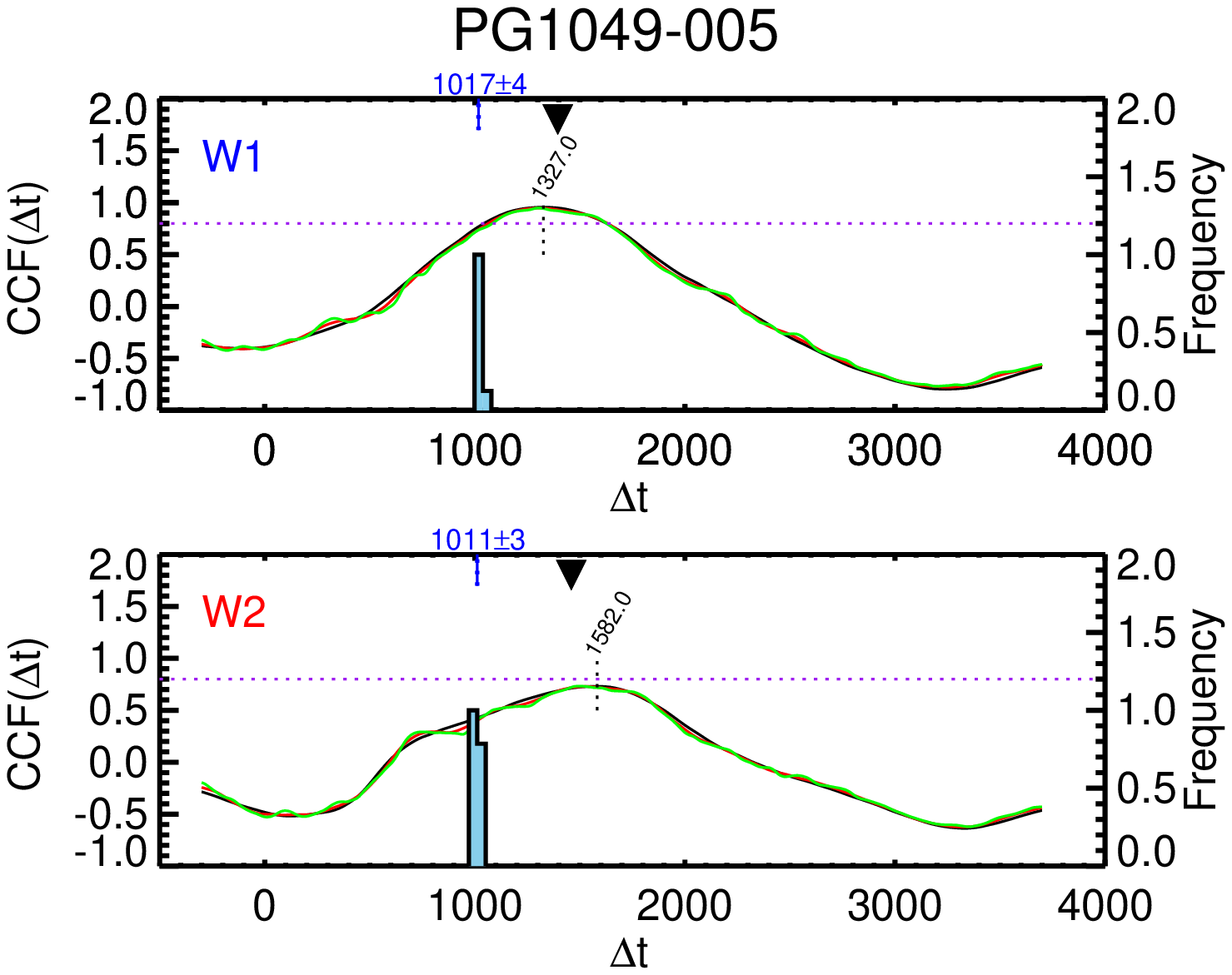}
\figsetgrpnote{We calculate the CCF(t) with different smoothing windows: τ_W (black curves),
τ W /2 (red curves), τW /10(green curves) where τ_W is assumed to be 200 days. The histograms show the best-fit time lag distribution by MC
modeling of the optical light curve (blue regions).
}
\figsetgrpend

\figsetgrpstart
\figsetgrpnum{6.27}
\figsetgrptitle{PG1100+772}
\figsetplot{f6set_26.ps}
\figsetgrpnote{We calculate the CCF(t) with different smoothing windows: τ_W (black curves),
τ W /2 (red curves), τW /10(green curves) where τ_W is assumed to be 200 days. The histograms show the best-fit time lag distribution by MC
modeling of the optical light curve (blue regions).
}
\figsetgrpend

\figsetgrpstart
\figsetgrpnum{6.28}
\figsetgrptitle{PG1103-006}
\figsetplot{f6set_27.ps}
\figsetgrpnote{We calculate the CCF(t) with different smoothing windows: τ_W (black curves),
τ W /2 (red curves), τW /10(green curves) where τ_W is assumed to be 200 days. The histograms show the best-fit time lag distribution by MC
modeling of the optical light curve (blue regions).
}
\figsetgrpend

\figsetgrpstart
\figsetgrpnum{6.29}
\figsetgrptitle{PG1114+445}
\figsetplot{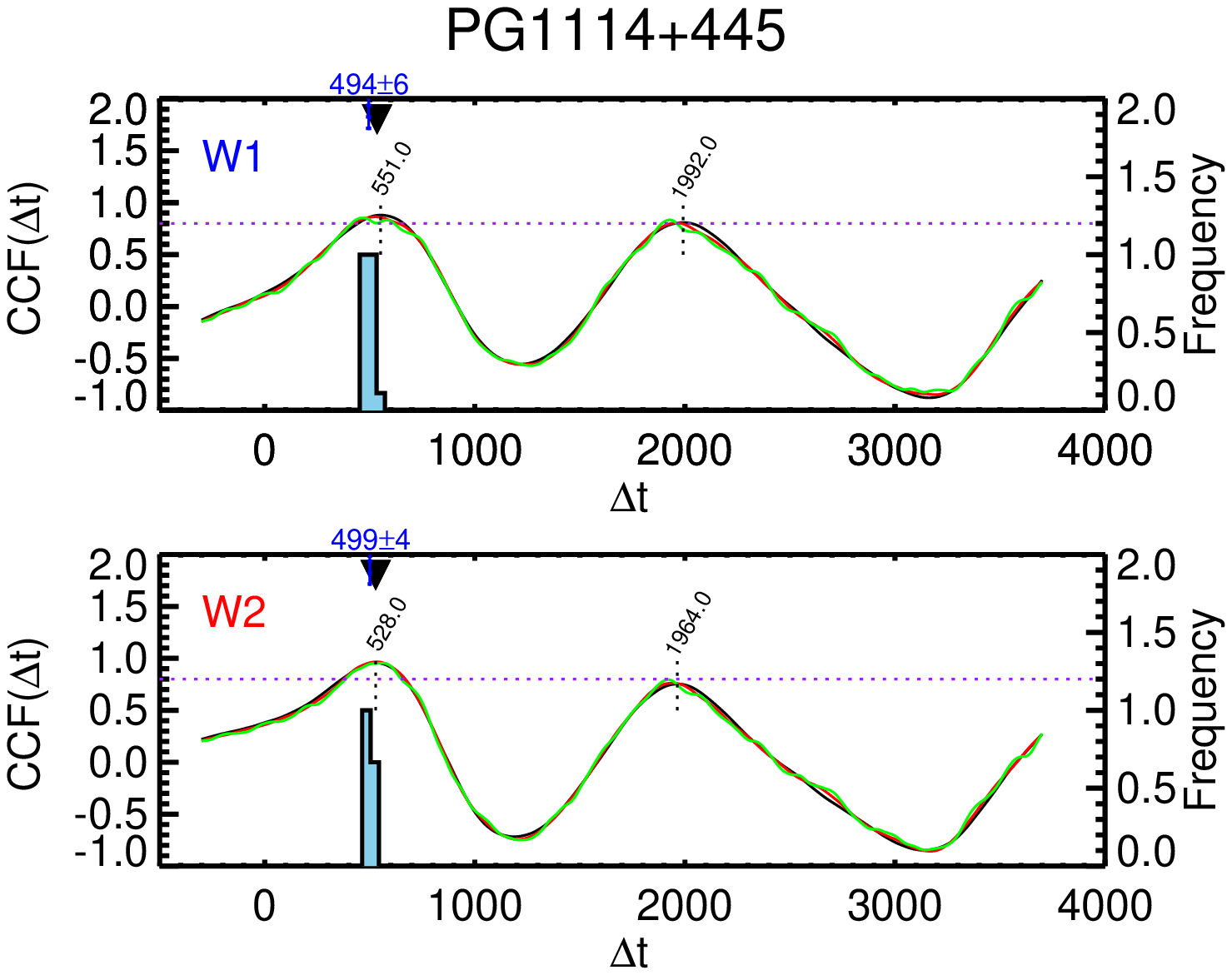}
\figsetgrpnote{We calculate the CCF(t) with different smoothing windows: τ_W (black curves),
τ W /2 (red curves), τW /10(green curves) where τ_W is assumed to be 200 days. The histograms show the best-fit time lag distribution by MC
modeling of the optical light curve (blue regions).
}
\figsetgrpend

\figsetgrpstart
\figsetgrpnum{6.75}
\figsetgrptitle{PG1613+658}
\figsetplot{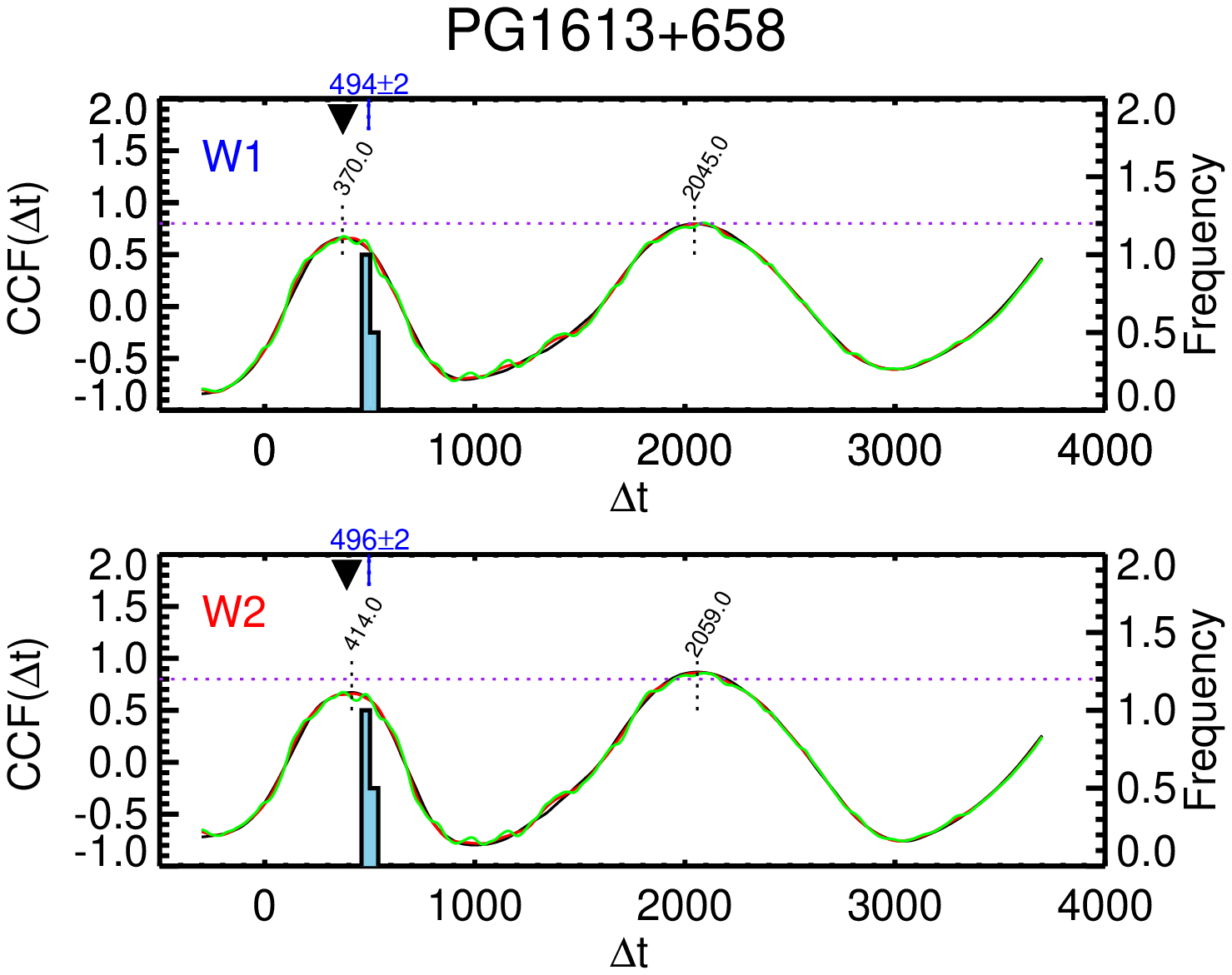}
\figsetgrpnote{We calculate the CCF(t) with different smoothing windows: τ_W (black curves),
τ W /2 (red curves), τW /10(green curves) where τ_W is assumed to be 200 days. The histograms show the best-fit time lag distribution by MC
modeling of the optical light curve (blue regions).
}
\figsetgrpend

\figsetend

\begin{figure*}[htbp]
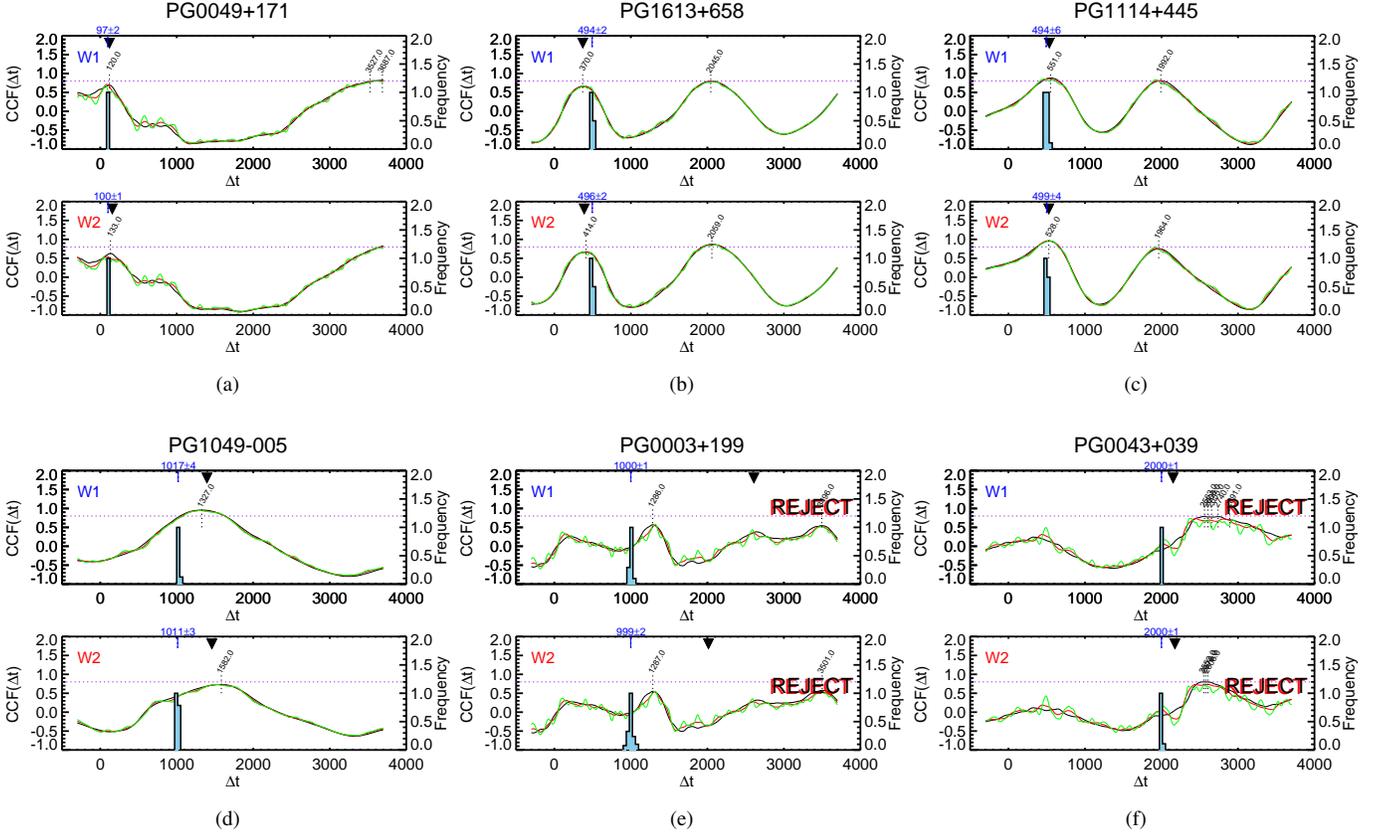

    \gridline{\fig{f6set_5.ps}{0.33\textwidth}{(a)}
	  \fig{f6set_74.ps}{0.33\textwidth}{(b)}
	  \fig{f6set_28.ps}{0.33\textwidth}{(c)}}
    \gridline{\fig{f6set_25.ps}{0.33\textwidth}{(d)}
	  \fig{f6set_1.ps}{0.33\textwidth}{(e)}
	  \fig{f6set_4.ps}{0.33\textwidth}{(f)}}
\caption{
		    Cross-correlation functions between the optical and IR
		    light curves of example PG quasars. The plots here are
		    matched to the objects in Figure~\ref{fig:lc_example} with
		    the same panel IDs. Panels (a)-(d) are the cases where we
		    think the time lag measurements are reasonable and Panels
		    (e)-(f) are objects whose fittings should be rejected. In
		    each sub-panel, we calculate the $CCF(t)$ with different
		    smoothing windows: $\tau_W$ (black curves), $\tau_W$/2 (red
		    curves), $\tau_W$/10(green curves) where $\tau_W$ is
		    assumed to be 200 days. The histograms show the best-fit
		    time lag distribution by MC modeling of the optical light
		    curve (blue regions).  \\ (The complete figure set (87
		    images) for all PG quasars is available in the online
		    journal)
    }
    \label{fig:ccf_example}
\end{figure*}

\begin{figure}[htp]
    \begin{center}
	\includegraphics[width=1.0\hsize]{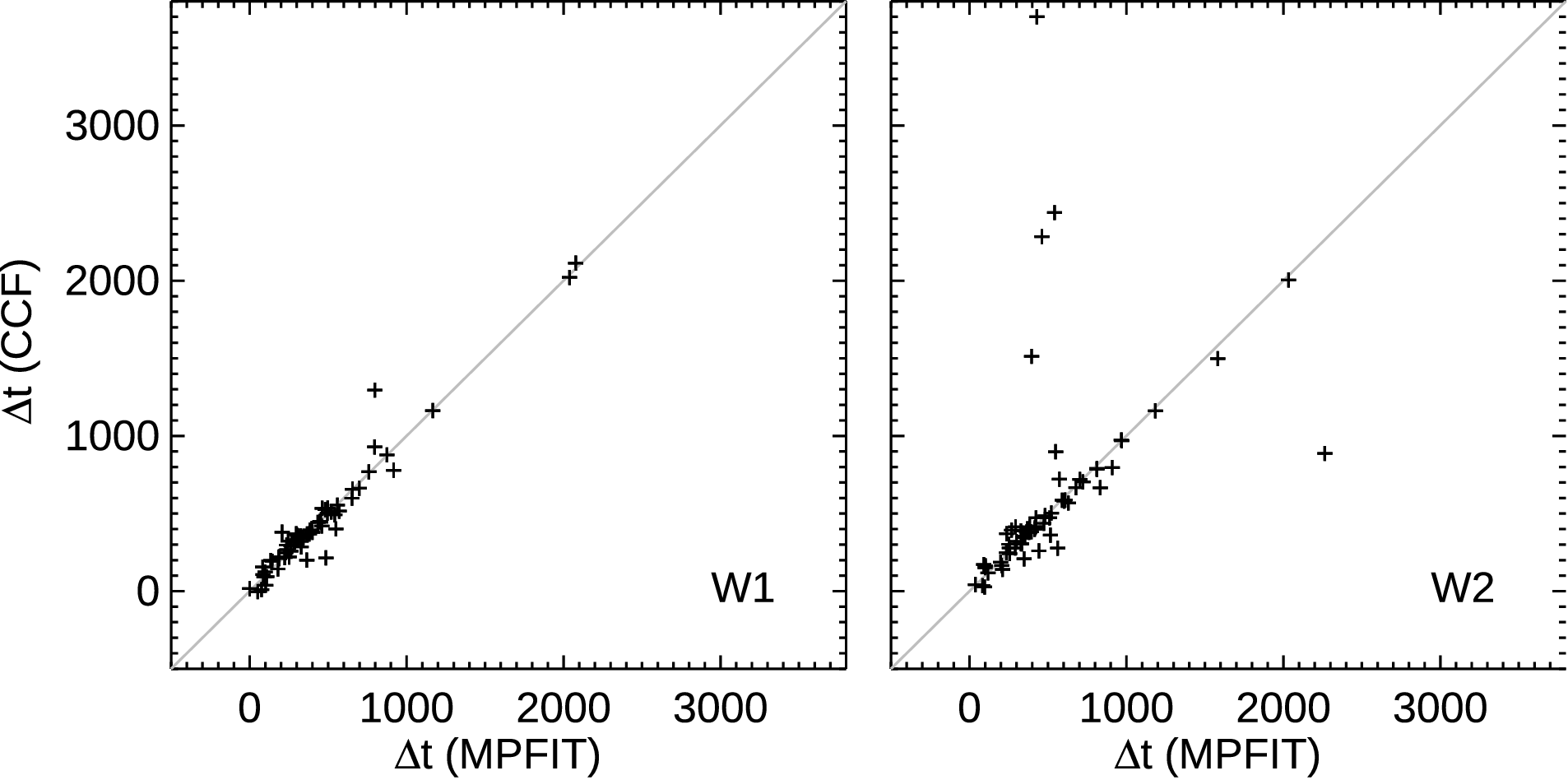}
		\caption{
		    Comparison of mid-IR time lags between the results from
		    $\chi^2$ fittings and cross-correlation analysis. The grey
		    diagonal lines are the 1:1 relation.
		    }
	\label{fig:lag_check}
    \end{center}
\end{figure}

After removing the accretion disk variability in the AGN IR emission following
Equations~\ref{eqn:ad-spectrum} and \ref{eqn:ir-dust-signal}, we also performed
a cross-correlation analysis to constrain the time lags between the optical and
the mid-IR light curves. For each object, the interpolated optical light curve
based on the DRW model was shifted with $\Delta t$= $-300$--3500 days and the
cross-correlation functions $CCF(\Delta t)$ to the {\it WISE} mid-IR light curves
were calculated. We provide several examples in Figure~\ref{fig:ccf_example}.
In general, we adopt $CCF(\Delta t)>0.8$ as the criterion for a strong
correlation signal. Due to the limited time sampling, multiple $CCF(\Delta)$
peaks for the optical-MIR light curve pair can be found in many objects. In
such cases, we inspect the peak locations sorted by the correlation values with
the following steps: (1) we first reject any peaks close to the maximum of the
explored $\Delta t$ range, which are nonphysical given the possible torus sizes
for the AGN luminosity range; (2) we compare the other peak locations to the
time lag derived from the $\chi^2$ fitting and adopt the one with the smallest
discrepancy. For the vast majority of the sample, such a peak location has the
maximum $CCF$ value. In Figure~\ref{fig:lag_check}, we compare the time lags
from light-curve fitting with the maximum $CCF(\Delta t)$ from
cross-correlation analysis. In general, these two methods yield consistent
$\Delta t$, particularly for the W1 band.

By default, the final time lags reported in Table~\ref{tab:PG-rever-result}
are adopted from the light-curve fitting method. However, in cases like
PG~0049+171 where the MPFIT code returns suspicious results, e.g.,
$\Delta_{\rm W1}>\Delta_{\rm W2}$, we choose to adopt the time lags inferred
from the cross-correlation analysis.  The reasons for doing this is that model
fittings with MPFIT are sensitive to the initial values and the fittings can
be stopped at some local minimum of the parameter space. In addition, given
the sparse sampling of the light curves, the solution is also not unique. We
have inspected each object carefully to check if the fitting results make
sense.

We can have a more intuitive idea on how the dust reverberation signals are
retrieved by looking at representative objects individually in
Figures~\ref{fig:lc_example} and~\ref{fig:ccf_example}. In both figures, Panels
(a)-(d) present four examples with convincing measurements of time lags over a
broad range and Panels (e)-(f) are two cases whose dust RM fittings are
rejected. For objects with convincing time lags, both optical and mid-IR light
curves have clear features like small dents (Panel (a)), bumps (Panel (b), (c))
or sometimes just a simple slope (Panel (d)). For these objects, typically
there are clear peaks in their cross-correlation curve with CCF$\gtrsim0.8$ and
the corresponding time lag is identical to that obtained by RM model fitting.
Objects like the one in Panel (e) are rejected because their optical and mid-IR
light curves seem uncorrelated. As shown in Panel (e) of
Figure~\ref{fig:lc_example}, both optical and mid-IR light curves have strong
features but the RM model fitting poorly reproduces the mid-IR data. For such
objects, the peak of the CCF values are typically smaller than 0.6, suggesting
a weak, if non-zero, correlation. As will be argued in
Section~\ref{sec:disc-origin}, there could be other explanations for the IR
variability of such objects. Objects like Panel (f) do not have enough features
in the mid-IR light curves and their optical light curves are also typically of
very low variability amplitude. Thus, the data itself cannot allow the model to
determine the fitting parameters. In Figure~\ref{fig:ccf_example}, it can be
seen that the cross-correlation analysis does not show any peaks for this source with
CCF$\gtrsim0.5$ for any reasonable time lags (the W1-band dust time lags for
quasars are very unlikely to have $\Delta t>2000$~days, see
Section~\ref{sec:lag-lum}). We will reject such objects for the reverberation
analysis.

\section{Results}\label{sec:result}

\subsection{Mid-IR Variability and Dust Reverberation Signals}\label{sec:result-var-summary}

Table~\ref{tab:PG-rever-result} summarizes the derived parameters from fitting
the optical and {\it WISE} W1 and W2 mid-IR light curves with the reverberation
mapping model introduced above.  Among the 87 PG quasars, we detect convincing
dust time lag signals in 67 objects. For those objects without time-lag
measurements, eight have data completeness or quality issues, eight lack clear
variability features (e.g., very smooth light curves) to pin down a best-fit
model (e.g., PG 0043+039, as shown in Figure~\ref{fig:lc_example}), and
the remaining four objects have uncorrelated optical and mid-IR temporal
variations (e.g., PG 0003+199, as shown in Figure~\ref{fig:lc_example}).
In addition, there are nine quasars whose {\it WISE} flux during the first two
or three epochs falls well above or below the time-shifted DRW model prediction
in both bands and we rely on the later NEOWISE epochs to extract the dust RM
signals.

For the study of the time lag -- AGN luminosity relation, we will focus on the
67 PG quasars ($\sim$77\% of the sample) with convincing time lag measurements
and assume the torus properties are stable over timescales of several years or
more.

\startlongtable
\begin{deluxetable*}{lclccccccccccc}
    \tabletypesize{\scriptsize}
    \tablewidth{1.0\textwidth}
    \tablecolumns{14}
    \tablecaption{Summary of the Mid-IR Reverberation Properties
    \label{tab:PG-rever-result}
    }
    \tablehead{
        \colhead{name} &
        \colhead{$z$}  &
        \colhead{type} &
        \colhead{$L_\text{AGN, bol}$} &
        \colhead{$\log_e(\sigma)$} &
        \colhead{$\log_e(\tau)$} &
        \colhead{$\Delta t_\text{W1, $\chi^2$}$}  &
        \colhead{$AMP_\text{W1}$} &
        \colhead{$\Delta t_\text{W1, CCF}$} &
        \colhead{$\Delta t_\text{W2, $\chi^2$}$}  &
        \colhead{$AMP_\text{W2}$} &
        \colhead{$\Delta t_\text{W2, CCF}$} &
        \colhead{tag$_{W1}$}    &
	\colhead{tag$_{W2}$}   \\
	\colhead{(1)}  &
	\colhead{(2)}  &
	\colhead{(3)}  &
	\colhead{(4)}  &
	\colhead{(5)}  &
	\colhead{(6)}  &
	\colhead{(7)}  &
	\colhead{(8)}  &
	\colhead{(9)}  &
	\colhead{(10)}  &
	\colhead{(11)}  &
	\colhead{(12)}  &
	\colhead{(13)}  &
	\colhead{(14)} 
        }
        \startdata
 PG 0003+158 &	  0.45 &  HDD   &  13.3      & $-$1.64 & 6.95  & 1084.1$\pm$402.0&     0.2$\pm$0.1&  3700 (0.55) & 1057.5$\pm$462.0&   0.1$\pm$0.2&   3700 (0.34)   &0  &0   \\
 PG 0003+199 &    0.03 &  Norm  &  11.2      & $-$0.42 & 5.43  & 2607.5$\pm$288.3&    20.0$\pm$1.0&  1287 (0.57) & 2014.3$\pm$332.7&   0.0$\pm$0.0&   3501 (0.58)   &0  &0   \\
 PG 0007+106 &    0.09 &  Norm  &  11.9      & $-$1.02 & 5.70  &  280.0$\pm$ 41.0&     4.7$\pm$0.2&  316  (0.96) &  313.6$\pm$ 34.5&   4.5$\pm$0.2&   332  (0.92)   &1  &1   \\
 PG 0026+129 &    0.14 &  HDD   &  12.4      & $-$1.04 & 6.53  &  557.2$\pm$ 41.4&     1.7$\pm$0.2&  503  (0.91) &  624.9$\pm$ 33.7&   2.1$\pm$0.2&   1488 (0.90)   &1  &1   \\
 PG 0043+039 &    0.38 &  HDD?  &  13.0      & $-$2.30 & 4.60  & 2152.8$\pm$187.5&     0.0$\pm$0.0&  2564 (0.79) & 2176.2$\pm$264.7&   0.0$\pm$0.0&   2553 (0.80)   &0  &0   \\
 PG 0049+171 &    0.06 &  HDD   &  11.6      & $-$0.60 & 6.45  &  120.0$\pm$ 11.4&     0.8$\pm$0.1&  3688 (0.82) &  157.0$\pm$ 28.7&   0.6$\pm$0.1&   3700 (0.83)   &1  &1   \\
 PG 0050+124 &    0.06 &  Norm  &  12.3      & $-$0.32 & 4.76  &  290.4$\pm$ 42.8&    14.9$\pm$0.4&  325  (0.95) &  348.0$\pm$ 43.8&  15.2$\pm$0.5&   354  (0.95)   &1  &1   \\
 PG 0052+251 &    0.16 &  HDD   &  12.6      & $-$0.94 & 6.52  &  400.5$\pm$ 42.9&     1.1$\pm$0.1&  2569 (0.77) &  463.4$\pm$ 60.7&   0.9$\pm$0.1&   2555 (0.81)   &1  &1   \\
 PG 0157+001 &    0.16 &  Norm  &  12.6      & $-$1.51 & 6.67  &  916.5$\pm$  0.4&    17.8$\pm$0.5&  244  (0.96) &  910.8$\pm$  0.4&  17.3$\pm$0.7&   252  (0.95)   &1  &1   \\
 PG 0804+761 &    0.10 &  WDD   &  12.6      &  0.11 & 6.68  &  659.9$\pm$ 22.7&     5.2$\pm$0.1&  665  (0.92) &  661.2$\pm$ 23.8&   5.2$\pm$0.2&   676  (0.92)   &1  &1   \\
 PG 0838+770 &    0.13 &  Norm  &  11.9      & $-$1.71 & 6.20  &  347.7$\pm$  5.1&     5.0$\pm$0.1&  339  (0.97) &  362.1$\pm$ 52.8&   5.8$\pm$0.2&   364  (0.98)   &1  &1   \\
 PG 0844+349 &    0.06 &  HDD?  &  11.9      & $-$0.32 & 5.31  &  230.8$\pm$ 31.2&     1.5$\pm$0.1&  255  (0.93) &  288.0$\pm$ 31.5&   1.3$\pm$0.1&   287  (0.92)   &1  &1   \\
 PG 0921+525 &    0.04 &  Norm  &  11.0      & $-$0.00 & 5.18  &  101.8$\pm$  3.0&     3.4$\pm$0.1&  159  (0.85) &  104.0$\pm$  3.3&   2.7$\pm$0.1&   187  (0.82)   &1  &1   \\
 PG 0923+201 &    0.19 &  WDD   &  12.6      & $-$1.23 & 7.04  & 1200.8$\pm$ 92.4&     4.9$\pm$0.4&  1155 (0.62) & 1327.8$\pm$151.5&   4.9$\pm$0.5&   1305 (0.87)   &0  &0   \\
 PG 0923+129 &    0.03 &  Norm  &  11.0      & $-$1.06 & 4.12  &  306.1$\pm$ 11.9&     8.0$\pm$0.4&  86   (0.81) &  298.8$\pm$ 16.5&   7.5$\pm$0.6&   1272 (0.76)   &0  &0   \\
 PG 0934+013 &    0.05 &  Norm  &  10.9      & $-$1.48 & 6.34  &  330.4$\pm$ 10.3&     9.3$\pm$0.2&  291  (0.97) &  367.9$\pm$ 68.8&  11.9$\pm$0.3&   482  (0.96)   &1  &1   \\
 PG 0947+396 &    0.21 &  Norm  &  12.4      & $-$1.68 & 6.64  & 1963.5$\pm$ 28.2&     7.3$\pm$0.3&  1915 (0.89) & 2021.3$\pm$100.4&  11.9$\pm$0.5&   1915 (0.84)   &0  &0   \\
 PG 0953+414 &    0.24 &  WDD   &  12.9      & $-$0.42 & 7.30  &  913.0$\pm$ 35.6&     2.8$\pm$0.1&  911  (0.97) & 1153.6$\pm$ 49.9&   2.1$\pm$0.2&   1134 (0.98)   &1  &1   \\
 PG 1001+054 &    0.16 &  Norm  &  12.0      & $-$1.60 & 6.39  &  256.0$\pm$ 47.9&     6.3$\pm$0.4&  715  (0.86) &  255.0$\pm$ 91.4&   9.4$\pm$0.8&   370  (0.92)   &1  &1   \\
 PG 1004+130 &    0.24 &  Norm  &  12.6      & $-$1.30 & 6.80  &  796.0$\pm$ 93.7&     1.6$\pm$0.2&  707  (0.83) &  532.0$\pm$107.8&   0.9$\pm$0.1&   692  (0.68)   &1  &1   \\
 PG 1011$-$040 &    0.06 &  HDD   &  11.7      & $-$1.26 & 6.87  &  424.8$\pm$ 35.9&     3.0$\pm$0.2&  415  (0.96) &  510.8$\pm$ 47.7&   3.0$\pm$0.2&   480  (0.95)   &1  &1   \\
 PG 1012+008 &    0.19 &  Norm  &  12.2      & $-$1.67 & 6.51  &  519.8$\pm$ 20.1&     5.8$\pm$0.2&  536  (0.98) &  570.5$\pm$ 94.3&   5.4$\pm$0.3&   2463 (0.97)   &1  &1   \\
 PG 1022+519 &    0.05 &  HDD?  &  11.2      & $-$1.78 & 5.23  &   94.0$\pm$ 47.5&     6.6$\pm$0.2&  94   (0.93) &  118.0$\pm$ 43.5&   7.2$\pm$0.3&   125  (0.94)   &1  &1   \\
 PG 1048+342 &    0.17 &  Norm  &  11.8      & $-$2.14 & 6.94  &  276.0$\pm$ 61.5&     2.1$\pm$0.1&  241  (0.87) &  565.8$\pm$115.5&   2.9$\pm$0.2&   572  (0.85)   &1  &1   \\
 PG 1048$-$090 &    0.34 &  WDD   &  12.8      & $-$1.29 & 6.78  &  589.0$\pm$155.2&     0.0$\pm$0.0&  2632 (0.54) &  549.7$\pm$ 55.3&   0.0$\pm$0.0&   2488 (0.71)   &0  &0   \\
 PG 1049$-$005 &    0.36 &  Norm  &  13.0      & $-$1.71 & 7.08  & 1284.6$\pm$108.3&     1.6$\pm$0.2&  1327 (0.96) & 1059.3$\pm$130.8&   1.7$\pm$0.4&   1583 (0.73)   &1  &0   \\
 PG 1100+772 &    0.31 &  HDD   &  13.1      & $-$1.38 & 6.13  &  720.8$\pm$ 88.1&     1.3$\pm$0.1&  722  (0.59) &  724.4$\pm$144.8&   1.6$\pm$0.2&   725  (0.58)   &0  &0   \\
 PG 1103$-$006 &    0.43 &  Norm  &  12.8      & $-$1.80 & 6.38  &  593.9$\pm$ 51.1&     0.6$\pm$0.2&  2223 (0.46) &  823.0$\pm$273.4&   0.8$\pm$0.3&   2371 (0.31)   &0  &0   \\
 PG 1114+445 &    0.14 &  Norm  &  12.3      & $-$1.73 & 6.55  &  534.8$\pm$ 21.7&     3.3$\pm$0.4&  551  (0.88) &  527.9$\pm$ 37.0&   5.4$\pm$0.6&   528  (0.96)   &1  &1   \\
 PG 1115+407 &    0.15 &  HDD   &  12.4      & $-$1.93 & 5.83  &  465.9$\pm$ 18.3&     5.0$\pm$0.3&  422  (0.83) &  455.4$\pm$ 30.7&   4.3$\pm$0.5&   273  (0.74)   &1  &1   \\
 PG 1116+215 &    0.18 &  WDD   &  12.9      & $-$0.40 & 7.28  &  819.6$\pm$ 98.6&     4.6$\pm$0.2&  1895 (0.85) &  967.4$\pm$ 90.0&   4.9$\pm$0.3&   1139 (0.83)   &1  &1   \\
 PG 1119+120 &    0.05 &  Norm  &  11.5      & $-$0.69 & 5.92  &  108.0$\pm$  6.7&     3.1$\pm$0.2&  42   (0.94) &  101.0$\pm$  6.7&   2.7$\pm$0.2&   29   (0.91)   &1  &1   \\
 PG 1121+422 &    0.23 &  HDD   &  12.5      & $-$1.48 & 7.28  &  912.1$\pm$116.1&     2.8$\pm$0.1&  497  (0.94) &  515.2$\pm$ 82.4&   3.2$\pm$0.2&   2277 (0.88)   &1  &1   \\
 PG 1126$-$041 &    0.06 &  Norm  &  11.8      & $-$0.41 & 6.27  &  554.1$\pm$ 28.3&     7.0$\pm$0.3&  566  (0.87) &  661.1$\pm$ 25.4&   7.3$\pm$0.4&   731  (0.88)   &1  &1   \\
 PG 1149$-$110 &    0.05 &  Norm  &  11.1      & $-$1.38 & 5.15  &   86.9$\pm$ 23.1&     7.6$\pm$0.2&  110  (0.85) &  211.6$\pm$ 23.2&   7.6$\pm$0.2&   143  (0.79)   &1  &1   \\
 PG 1151+117 &    0.18 &  Norm  &  12.0      & $-$1.25 & 7.69  &  295.9$\pm$ 27.3&     2.4$\pm$0.1&  341  (0.96) &  414.0$\pm$ 63.5&   2.3$\pm$0.2&   414  (0.87)   &1  &1   \\
 PG 1202+281 &    0.17 &  Norm  &  12.1      & $-$1.69 & 6.93  &  361.0$\pm$ 39.3&     7.8$\pm$0.4&  358  (0.96) &  439.1$\pm$ 44.5&   7.9$\pm$0.5&   429  (0.95)   &1  &1   \\
 PG 1211+143 &    0.09 &  Norm  &  12.1      & $-$0.07 & 4.79  &  365.6$\pm$ 90.2&     2.7$\pm$0.2&  358  (0.88) &  557.5$\pm$ 47.6&   3.3$\pm$0.3&   2015 (0.77)   &1  &1   \\
 PG 1216+069 &    0.33 &  HDD   &  13.0      & $-$0.10 & 3.23  & 1016.9$\pm$ 23.9&     0.3$\pm$0.1&  985  (0.30) & 1012.9$\pm$ 76.3&   0.2$\pm$0.1&   703  (0.34)   &0  &0   \\
 PG 1226+023 &    0.16 &  WDD   &  13.4      &  1.09 & 7.21  & 1918.4$\pm$ 67.9&     0.0$\pm$0.0&  933  (0.75) & 1684.4$\pm$280.6&   0.0$\pm$0.0&   951  (0.59)   &0  &0   \\
 PG 1229+204 &    0.06 &  Norm  &  11.5      & $-$0.70 & 6.87  &  266.3$\pm$ 71.2&     5.3$\pm$0.1&  261  (0.98) &  328.0$\pm$  8.5&   5.2$\pm$0.1&   305  (0.97)   &1  &1   \\
 PG 1244+026 &    0.05 &  Norm  &  11.1      & $-$2.36 & 5.27  &  118.9$\pm$ 70.3&     0.8$\pm$0.3&  1101 (0.60) &  180.0$\pm$177.3&   0.7$\pm$0.5&   1205 (0.46)   &0  &0   \\
 PG 1259+593 &    0.47 &  WDD   &  13.3      & $-$1.54 & 7.25  &  746.2$\pm$ 55.9&     2.3$\pm$0.1&  732  (0.98) &  784.4$\pm$ 70.3&   2.2$\pm$0.2&   794  (0.93)   &1  &1   \\
 PG 1302$-$102 &    0.29 &  HDD   &  13.2      & $-$1.18 & 7.05  &  643.8$\pm$ 33.2&     1.6$\pm$0.1&  619  (0.90) &  681.0$\pm$ 82.6&   1.8$\pm$0.2&   662  (0.91)   &1  &1   \\
 PG 1307+085 &    0.16 &  Norm  &  12.2      & $-$1.51 & 6.20  &  357.6$\pm$ 46.1&     2.8$\pm$0.2&  334  (0.91) &  505.9$\pm$ 57.7&   3.1$\pm$0.2&   368  (0.87)   &1  &1   \\
 PG 1309+355 &    0.18 &  Norm  &  12.4      & $-$1.74 & 6.59  &  404.6$\pm$ 62.0&     2.5$\pm$0.3&  419  (0.78) &  531.8$\pm$ 92.0&   1.6$\pm$0.5&   1453 (0.58)   &1  &1   \\
 PG 1310$-$108 &    0.04 &  Norm  &  10.9      & $-$1.68 & 4.56  &  152.0$\pm$ 30.2&     2.0$\pm$0.3&  208  (0.76) &  363.9$\pm$ 26.5&   4.4$\pm$0.4&   391  (0.83)   &1  &1   \\
 PG 1322+659 &    0.17 &  Norm  &  12.1      & $-$1.29 & 6.38  &  524.2$\pm$ 54.3&     2.8$\pm$0.1&  657  (0.89) &  745.0$\pm$115.0&   3.0$\pm$0.1&   748  (0.86)   &1  &1   \\
 PG 1341+258 &    0.09 &  HDD?  &  11.7      & $-$2.25 & 5.27  &  351.1$\pm$246.1&     5.8$\pm$0.3&  356  (0.71) &  383.9$\pm$ 18.5&   6.9$\pm$0.4&   383  (0.71)   &1  &1   \\
 PG 1351+236 &    0.05 &  Norm  &  10.8      & $-$2.08 & 4.70  &   79.6$\pm$ 48.9&     7.9$\pm$0.4&  11   (0.70) &   83.8$\pm$ 32.9&   5.1$\pm$0.4&   38   (0.60)   &1  &1   \\
 PG 1351+640 &    0.09 &  Norm  &  12.2      & $-$0.37 & 6.11  &  631.2$\pm$ 29.5&     1.1$\pm$0.1&  1509 (0.87) &  808.6$\pm$ 51.8&   1.5$\pm$0.2&   953  (0.88)   &1  &1   \\
 PG 1352+183 &    0.16 &  Norm  &  11.9      & $-$1.70 & 6.20  &  272.0$\pm$ 10.1&     3.1$\pm$0.2&  247  (0.91) &  316.0$\pm$ 78.4&   3.5$\pm$0.3&   329  (0.94)   &1  &1   \\
 PG 1354+213 &    0.30 &  Norm  &  12.4      & $-$2.04 & 7.06  &  911.5$\pm$124.1&     4.4$\pm$0.2&  897  (0.91) & 1061.4$\pm$169.3&   5.2$\pm$0.3&   1066 (0.88)   &0  &0   \\
 PG 1402+261 &    0.16 &  Norm  &  12.4      & $-$1.44 & 6.18  &  368.1$\pm$ 54.6&     5.6$\pm$0.4&  1972 (0.91) &  414.6$\pm$ 82.4&   8.8$\pm$0.6&   2011 (0.88)   &1  &1   \\
 PG 1404+226 &    0.10 &  HDD?  &  11.7      & $-$2.16 & 5.63  &  214.7$\pm$ 13.2&     2.4$\pm$0.2&  1893 (0.86) &  272.0$\pm$ 30.7&   2.9$\pm$0.4&   1977 (0.89)   &1  &1   \\
 PG 1411+442 &    0.09 &  Norm  &  12.0      & $-$0.55 & 6.89  &  441.7$\pm$ 46.2&     2.3$\pm$0.1&  704  (0.88) &  445.2$\pm$ 77.6&   2.5$\pm$0.2&   1000 (0.84)   &1  &1   \\
 PG 1415+451 &    0.11 &  Norm  &  11.7      & $-$1.79 & 6.64  &  300.0$\pm$232.3&     4.3$\pm$0.1&  350  (0.95) &  677.4$\pm$245.2&   0.0$\pm$0.0&   433  (0.91)   &1  &1   \\
 PG 1416$-$129 &    0.13 &  Norm  &  11.6      & $-$2.32 & 6.41  &  424.8$\pm$426.7&     2.8$\pm$0.3&  $-$162 (0.53) &  511.0$\pm$113.4&   1.9$\pm$0.3&   530  (0.44)   &0  &0   \\
 PG 1425+267 &    0.37 &  Norm  &  12.7      & $-$2.00 & 7.25  &  744.1$\pm$ 99.5&     0.8$\pm$0.1&  749  (0.88) &  760.5$\pm$321.7&   0.6$\pm$0.2&   738  (0.70)   &1  &1   \\
 PG 1426+015 &    0.09 &  Norm  &  12.0      & $-$0.11 & 6.77  &  264.0$\pm$ 24.7&     3.9$\pm$0.2&  314  (0.95) &  352.0$\pm$ 64.5&   3.3$\pm$0.2&   359  (0.92)   &1  &1   \\
 PG 1427+480 &    0.22 &  Norm  &  12.2      & $-$2.19 & 5.97  &  342.6$\pm$ 46.7&     1.4$\pm$0.1&  332  (0.86) &  356.0$\pm$ 52.0&   1.1$\pm$0.2&   352  (0.78)   &1  &1   \\
 PG 1435$-$067 &    0.13 &  HDD   &  12.2      & $-$1.88 & 5.43  &  244.0$\pm$ 42.1&     4.8$\pm$0.2&  314  (0.86) &  356.1$\pm$ 63.6&   4.5$\pm$0.4&   362  (0.82)   &1  &1   \\
 PG 1440+356 &    0.08 &  Norm  &  11.9      & $-$0.35 & 6.29  &  255.3$\pm$ 40.0&     4.3$\pm$0.2&  290  (0.95) &  264.9$\pm$ 30.4&   4.7$\pm$0.2&   296  (0.92)   &1  &1   \\
 PG 1444+407 &    0.27 &  Norm  &  12.7      & $-$1.95 & 6.52  &  399.5$\pm$ 16.4&     2.6$\pm$0.2&  405  (0.93) &  448.1$\pm$ 61.1&   2.6$\pm$0.3&   505  (0.74)   &1  &1   \\
 PG 1448+273 &    0.06 &  Norm  &  11.4      & $-$0.75 & 5.88  &  280.0$\pm$ 31.7&     2.8$\pm$0.1&  289  (0.98) &  371.3$\pm$ 49.0&   2.6$\pm$0.1&   378  (0.97)   &1  &1   \\
 PG 1501+106 &    0.04 &  Norm  &  11.4      & $-$0.05 & 5.91  &  113.9$\pm$ 16.4&     2.1$\pm$0.1&  151  (0.85) &  112.0$\pm$ 11.9&   2.0$\pm$0.1&   2703 (0.77)   &1  &1   \\
 PG 1512+370 &    0.37 &  Norm  &  12.7      & $-$1.95 & 6.91  & 1352.4$\pm$257.2&     0.5$\pm$0.1&  3152 (0.82) & 2122.8$\pm$478.4&   1.2$\pm$0.3&   3383 (0.68)   &0  &0   \\
 PG 1519+226 &    0.14 &  Norm  &  12.0      & $-$1.68 & 7.40  &  193.8$\pm$113.1&     7.4$\pm$0.2&  228  (0.99) &  356.5$\pm$105.0&   6.2$\pm$0.3&   365  (0.98)   &1  &1   \\
 PG 1534+580 &    0.03 &  Norm  &  10.9      & $-$0.67 & 5.08  &  128.0$\pm$  2.9&     3.2$\pm$0.2&  240  (0.67) &  139.3$\pm$ 12.7&   4.1$\pm$0.1&   376  (0.75)   &0  &0   \\
 PG 1535+547 &    0.04 &  Norm  &  10.9      & $-$0.54 & 5.99  &  180.0$\pm$  1.4&     8.5$\pm$0.1&  157  (0.92) &  207.7$\pm$  0.2&   7.7$\pm$0.2&   219  (0.92)   &1  &1   \\
 PG 1543+489 &    0.40 &  Norm  &  13.1      & $-$2.88 & 5.85  & 2052.2$\pm$606.7&     8.9$\pm$0.9&  2002 (0.74) & 2066.0$\pm$276.3&  13.4$\pm$1.3&   1974 (0.72)   &1  &1   \\
 PG 1545+210 &    0.27 &  WDD   &  12.7      & $-$1.37 & 6.90  &  774.8$\pm$ 51.3&     1.8$\pm$0.1&  794  (0.94) &  835.3$\pm$ 74.3&   2.1$\pm$0.1&   804  (0.94)   &1  &1   \\
 PG 1552+085 &    0.12 &  HDD?  &  12.1      & $-$1.37 & 6.85  &  212.0$\pm$ 30.2&     3.3$\pm$0.1&  243  (0.98) &  264.0$\pm$ 22.9&   3.5$\pm$0.1&   261  (0.98)   &1  &1   \\
 PG 1612+261 &    0.13 &  Norm  &  12.0      & $-$0.98 & 6.77  &  627.3$\pm$ 39.7&     1.2$\pm$0.2&  624  (0.41) &  664.2$\pm$ 62.0&   2.0$\pm$0.3&   1262 (0.71)   &0  &0   \\
 PG 1613+658 &    0.13 &  Norm  &  12.4      & $-$0.65 & 6.43  &  371.9$\pm$ 63.3&     1.6$\pm$0.2&  2046 (0.80) &  390.4$\pm$ 37.1&   1.8$\pm$0.2&   2059 (0.87)   &1  &1   \\
 PG 1617+175 &    0.11 &  WDD   &  12.1      & $-$0.71 & 6.75  &  472.4$\pm$ 34.7&     3.3$\pm$0.1&  1187 (0.97) &  575.8$\pm$ 26.7&   4.4$\pm$0.2&   1183 (0.97)   &1  &1   \\
 PG 1626+554 &    0.13 &  HDD   &  12.2      & $-$1.14 & 6.10  &  346.8$\pm$  7.4&     1.6$\pm$0.1&  381  (0.90) &  414.5$\pm$  9.2&   1.4$\pm$0.1&   416  (0.81)   &1  &1   \\
 PG 1700+518 &    0.28 &  Norm  &  13.1      & $-$1.69 & 7.49  &  292.4$\pm$669.7&     1.4$\pm$0.3&  $-$300 (0.80) &  967.9$\pm$216.2&   0.0$\pm$0.0&   3700 (0.68)   &0  &0   \\
 PG 1704+608 &    0.37 &  Norm  &  13.1      & $-$1.15 & 6.93  & 1650.3$\pm$744.1&     0.7$\pm$0.1&  3700 (0.60) & 1689.5$\pm$815.7&   0.8$\pm$0.1&   1660 (0.70)   &0  &0   \\
 PG 2112+059 &    0.47 &  Norm  &  13.3      & $-$1.44 & 7.74  & 2265.0$\pm$105.9&     0.9$\pm$0.2&  646  (0.95) & 2614.9$\pm$413.5&   1.1$\pm$1.0&   883  (0.67)   &1  &1   \\
 PG 2130+099 &    0.06 &  Norm  &  11.9      & $-$0.27 & 6.41  &  525.3$\pm$ 44.7&     5.8$\pm$0.2&  521  (0.89) &  801.3$\pm$ 46.6&   5.5$\pm$0.2&   1075 (0.89)   &1  &1   \\
 PG 2209+184 &    0.07 &  HDD?  &  11.6      & $-$1.02 & 6.23  &  108.0$\pm$ 10.4&     5.3$\pm$0.1&  19   (0.95) &  187.0$\pm$ 13.3&   7.7$\pm$0.2&   189  (0.97)   &1  &1   \\
 PG 2214+139 &    0.07 &  WDD   &  11.9      & $-$0.59 & 6.13  &  324.6$\pm$165.6&     4.2$\pm$0.4&  2545 (0.86) &  378.4$\pm$622.9&   3.0$\pm$0.4&   2622 (0.85)   &1  &0   \\
 PG 2233+134 &    0.32 &  Norm  &  12.7      & $-$2.30 & 6.73  &  455.1$\pm$ 39.0&     2.2$\pm$0.2&  449  (0.90) &  426.9$\pm$ 47.7&   2.7$\pm$0.4&   420  (0.83)   &1  &1   \\
 PG 2251+113 &    0.32 &  WDD   &  12.9      & $-$2.40 & 6.30  &  579.7$\pm$116.8&     2.7$\pm$0.8&  576  (0.85) &  598.0$\pm$101.8&   3.5$\pm$1.0&   591  (0.91)   &0  &0   \\
 PG 2304+042 &    0.04 &  Norm  &  10.6      &  0.11 & 2.66  &   86.0$\pm$  3.6&     3.7$\pm$0.0&  156  (0.97) &   98.2$\pm$  1.5&   3.8$\pm$0.1&   171  (0.97)   &1  &1   \\
 PG 2308+098 &    0.43 &  HDD   &  13.2      & $-$1.11 & 7.68  &  703.1$\pm$ 50.7&     1.3$\pm$0.1&  548  (0.88) & 1418.8$\pm$129.5&   1.2$\pm$0.1&   1493 (0.83)   &0  &0   \\

        \enddata
         \tablecomments{
	     Col. (1): object name; Col. (2): redshift. Col. (3): IR SED type
	     from \cite{Lyu2017}. Col. (4): AGN bolometric luminosity
	     estimated from IR SED decompositions \citep{Lyu2017} with
	     bolometric corrections described in
	     Section~\ref{sec:lum-estimate}. Col.  (5)-(6): DRW model
	     amplitude ($\sigma$) and timescale ($\tau$) from the fitting the
	     combined V-band light curve with the {\it Javelin} code, both in
	     natural logs. Col. (7)-(8): IR dust time lag (obsered frame) and
	     IR-optical variability amplitude factor from fitting the {\it
	     WISE} W1-band light curves with our model. We have made
	     necessary corrections as stated in
	     Section~\ref{sec:lag-detection}. This column presents the
	     finally adopted values of the time lags (in the observed frame)
	     then will be used in our further analysis in
	     Section~\ref{sec:result}. Col. (9): W1-band dust time lag
	     suggested by cross-correlation analysis with the peak CCF value
	     in brackts. Col. (10)--Col. (12): similar to Col.  (6)--(8) but
	     for {\it WISE} W2-band. Col. (13), (14): if the results are used
	     for reverberation analysis, 1--yes, 0--no.\\
	{\it Comments on individual objects}: the results of PG 0003+129, PG
	1048-090 and PG 1244+026 are dropped since their IR light curves
	contain large uncertainties. The mid-IR photometry of several epochs of
	PG 1226+069 are saturated, making the mid-IR light curve unuseful.
	The optical light curves of PG 0923+129, PG 1216+069, PG 1354+213 and
	PG 1416-129 have limited time (only good CRTS or ASAS-SN data are
	available). PG 0043+039, PG 0947+396, PG 1100+772, PG 1103+006, PG
	1512+370, PG 1700+518, PG 1704+608 and PG 2308+098 do not show enough
	variability features so that meaningful time lags cannot be
	constrained. The optical and IR light curves of PG 0003+199, PG
	0923+201, PG 1534+580 and PG 1612+261 seem uncorrelated with peak
	CCF$<$0.7, so their results have been also dropped.
	}
\end{deluxetable*}


\subsection{The IR Time Lag -- AGN Luminosity Correlation}\label{sec:lag-lum}

The inner size of the dust torus is physically determined by grain sublimation.
Since carbon and silicate dust are the dominant species of interstellar grains,
a similar grain mixture is also commonly assumed for the AGN torus. The
sublimation temperatures, $T_\text{sub}$, of grains are estimated to be
1500--1800~K for graphite and 800-1000 K for silicates.  Consequently, only
graphite grains would survive at the innermost regions and silicate grains
would be distributed at larger radii.  Assuming an optically-thin environment,
\cite{Barvainis1987} provided an estimate of the graphite sublimation radius
\begin{equation}\label{eqn:sub_c}
    \frac{R_{\rm sub, C}}{\rm pc} = 1.3 \left(\frac{L_{\rm UV}}{10^{46} {\rm erg~s}^{-1}} \right)^{0.5} \left( \frac{T_{\rm sub}}{1500 K} \right)^{-2.8} \left( \frac{a}{0.05~\mum}\right)^{-0.5}~,
\end{equation}
where $L_{\rm UV}$ is the AGN UV luminosity.  Following \cite{Kishimoto2007},
we also introduce a $a^{-1/2}$ term to approximate the $R_\text{sub}$
dependence on the grain size $a$.  For silicate dust grains, adopting the
absorption efficiency of astronomical silicate \citep{Draine1984, Laor1993} and
repeating the derivations in \cite{Barvainis1987}, we have
\begin{equation}\label{eqn:sub_s}
    \frac{R_{\rm sub, S}}{\rm pc} = 2.7 \left(\frac{L_{\rm UV}}{10^{46} {\rm erg~s}^{-1}} \right)^{0.5} \left( \frac{T_{\rm sub}}{1000 K} \right)^{-2.8} \left( \frac{a}{0.05~\mum}\right)^{-0.5}.
\end{equation}
If we take $L_{\rm UV}=0.165~L_{\rm AGN, bol}$ \citep{Risaliti2004} and assume
$T_{\rm sub}=$1500 K and $T_{\rm sub}=$1000 K for graphite and silicate dust
grains, for an AGN with $L_{\rm AGN, bol}=10^{12}~L_\odot$, $R_{\rm sub,
C}\sim0.33$~pc and $R_{\rm sub, S}\sim$0.69~pc.

For an infinitely thin ring viewed from a perfect face-on observing angle, the
time lag is directly related to the dust radius as $\Delta t = R_\text{d}/c$.
In real situations, the relation would be dependent on the observing angle as
well as the dust distribution (some additional discussion is provided in
Appendix~\ref{app:model}). However, for a large sample of type-1 AGNs, these
effects would be smeared out. To first order, we adopt the following linear
equation to fit the data:
\begin{equation}\label{eqn:fit-model}
    \log(\Delta t) = \alpha + \beta \log(L_{\rm AGN})~.
\end{equation}
In the following analysis and discussion, we have applied the dilation factor
$(1+z)$ for the time lag (i.e., $\Delta t=\Delta t_{\bf int.} = \Delta t_{\bf
obs.}/(1+z)$). Due to the lack of constraints on the wavelength-dependent torus
size, we will not make K corrections for individual quasars but treat the
sample as a whole and compute the correction with an averaged redshift.

\subsubsection{AGN Luminosity Estimation}\label{sec:lum-estimate}

\begin{figure}[htp]
    \begin{center}
	\includegraphics[width=1.0\hsize]{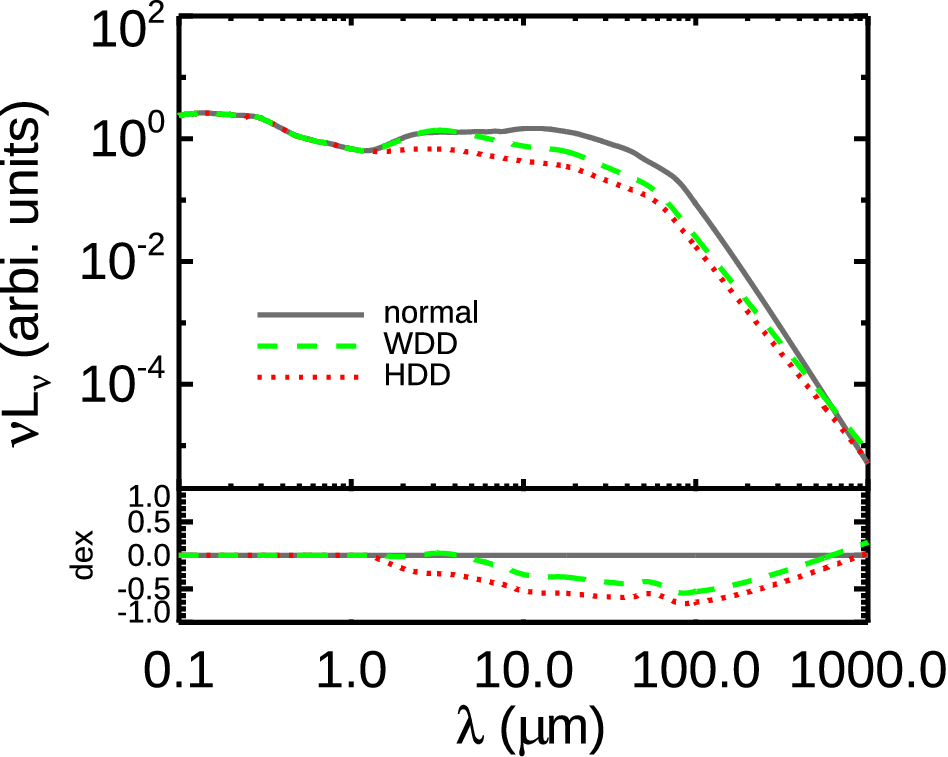}
		\caption{
		    Comparison of the normal, WDD and HDD AGN templates
		    developed in \cite{Lyu2017}. The bottom panel highlights
		    the infrared flux differences in dex of these templates by
		    normalizing their SEDs by the normal AGN template.
		    }
	\label{fig:agn_sed}
    \end{center}
\end{figure}

To make a comparison of the AGN bolometric luminosity $L_\text{AGN, bol}$ in
the literature, we use the integrated AGN IR emission of PG quasars as the
best estimator. As demonstrated in \cite{Lyu2017}, despite their identical
UV-optical SED shape, unobscured type-1 quasars present intrinsic SED
variations in the IR that can be grouped into normal, warm-dust-deficient (WDD)
and hot-dust-deficient (HDD) types (see the comparison in
Figure~\ref{fig:agn_sed}). After building the intrinsic AGN templates,
\cite{Lyu2017} fit the observed SEDs of PG quasars with an empirical SED model.
The AGN intrinsic SED types were determined by comparing the fitted $\chi^2$ of
the model. We convert the AGN-heated IR luminosity derived from the
optical-to-IR SED fittings in \cite{Lyu2017} to the monochromatic luminosity at
5100~\AA~with the following scaling factors:
\begin{align}
    \lambda L_\lambda(0.51~\mum)& = 0.47 L_\text{\rm NORM, 8-1000~\mum} \\
		               & = 1.02 L_\text{\rm WDD, 8-1000~\mum} \\
                               & = 1.75 L_\text{\rm HDD, 8-1000~\mum}
\end{align}
where NORM, WDD, HDD represent the normal \cite{Elvis1994}-like,
warm-dust-deficient, and hot-dust-deficient AGN templates as characterized in
\cite{Lyu2017, Lyu2017b}.  We then convert $L_\lambda(0.51~\mum)$ to
$L_\text{AGN, bol}$ with the updated quasar bolometric correction from
\cite{Runnoe2012}:
\begin{equation}\label{eqn:bol-corr}
    \log \left(\frac{L_{\rm AGN, bol}}{\text{erg~s}^{-1}}\right) =4.89 +0.91\log \left(\frac{\lambda L_\lambda (0.51\mum)}{\text{erg~s}^{-1}}\right)
\end{equation}

We realize that there are several alternative tracers for $L_{\rm AGN, bol}$
widely used in the literature, such as hard X-ray luminosity, mid-IR AGN
continuum luminosity, and the mid-IR \OIV$\lambda$25.89~$\mum$ emission line
luminosity \citep{Melendez2008, Diamond-Stanic2009, Rigby2009}.  Although these
tracers could be less affected by dust obscuration and might be isotropic,
whether they can be applied over a wide AGN luminosity range that covers the
quasar population is a question. For example, the application of \OIV~emission
line has been only discussed for the Seyfert population ($L_\text{AGN,
bol}\lesssim10^{11} L_\odot$) and there is evidence suggesting the structures
of AGN narrow-line regions could evolve at high luminosity
\citep[e.g.,][]{Netzer2004}. Using the single-band mid-IR continuum is also
complicated: (1) the fraction of the AGN luminosity that is reprocessed in the
mid-IR (7--15~$\mum$) can vary intrinsically by 0.3--0.9 dex \citep{Lyu2017};
(2) the possible addition of reprocessed emission from AGN-heated polar dust
can easily lead to scatter exceeding 1 dex for the same $L_\text{AGN, bol}$ \citep{Lyu2018}.
Given these uncertainties, we opt to use the V-band bolometric correlation
throughout this work.

The ambiguities in calculating AGN bolometric luminosity have contributions
from both measured qualities (e.g., the photometric errors) and bolometric
correction factors. In many cases, the latter can dominate. Given the likely
variation from source to source, we adopt 0.3 dex as the relative
$L_\text{AGN}$ uncertainty for all objects. 

\subsubsection{Correlation between Mid-IR Time Lag and AGN Luminosity} 

\begin{figure*}[htp!]
    \begin{center}
	\includegraphics[width=1.0\hsize]{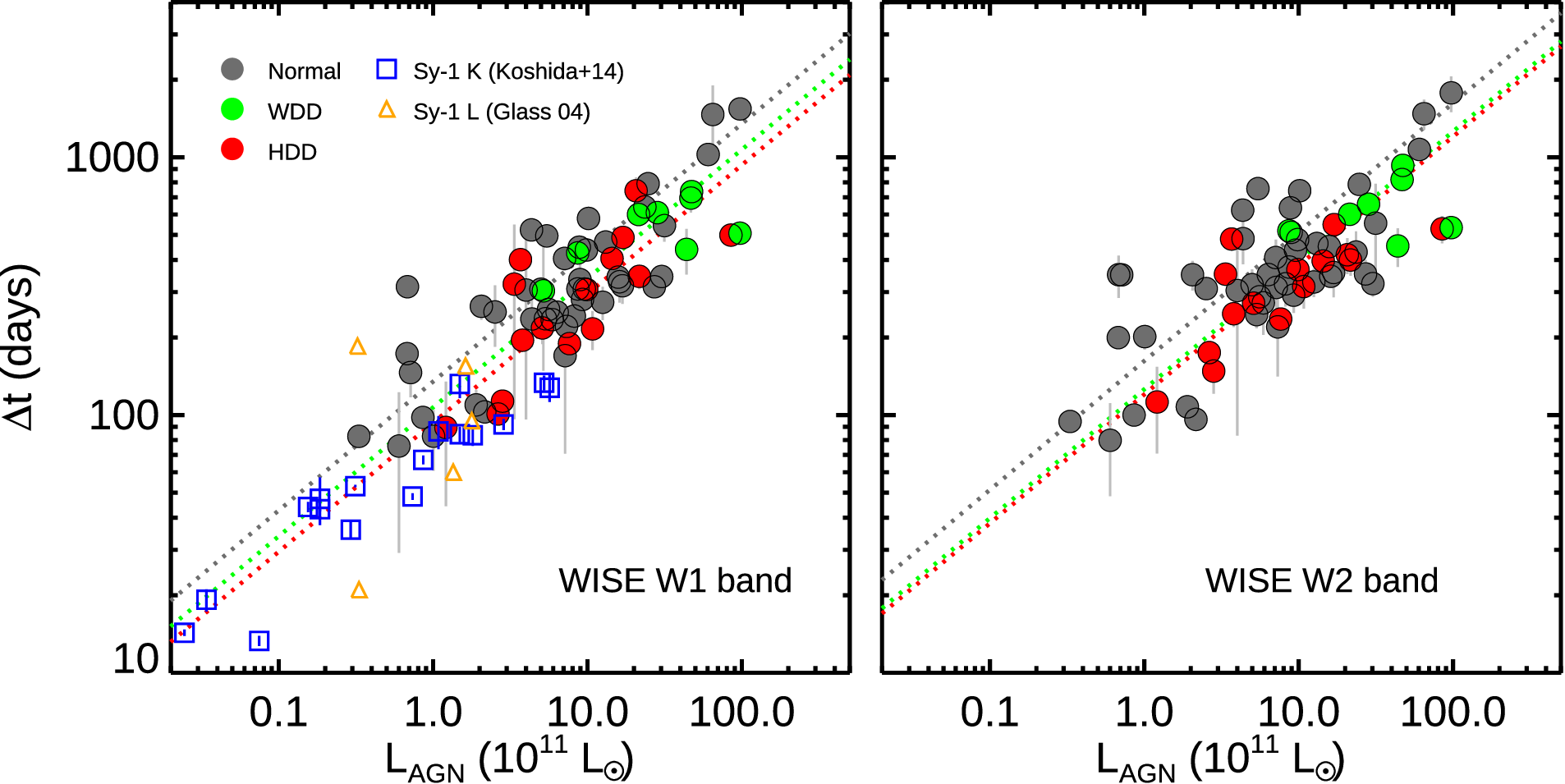}
		\caption{
		    Dust time lags between the {\it WISE} mid-IR band and
		    optical band light curves plotted against the AGN
		    luminosity for PG quasars. The dotted lines are the fitted
		    correlations for normal (grey), WDD (green) and HDD (red)
		    quasars assuming $\Delta t\propto L_{\rm AGN}^{0.5}$. We
		    also show the K-band time-lag measurements of the 17 Seyfert-1
		    nuclei studied in \cite{Koshida2014}  and the L-band time-lag 
		    measurements of the 5 Seyfert-1 nuclei in \cite{Glass2004} with 
		    our estimation of their AGN bolometric luminosities in the left panel.
		    }
	\label{fig:tl_sed}
    \end{center}
\end{figure*}

In Figure~\ref{fig:tl_sed}, we present the time lags of the mid-IR emission in
the {\it WISE} W1 and W2 bands of PG quasars as a function of their AGN bolometric
luminosities. It is clear that these two quantities are strongly correlated.
From fitting these measurements with equation~\ref{eqn:fit-model} with the IDL
program {\sc fitexy}, we have
\begin{equation}\label{eqn:wise_w1}
    \Delta t_{\rm torus, W1}/{\rm day} = 10^{2.10\pm0.06} (L_{\rm AGN, SED}/10^{11}L_\odot) ^{0.47\pm0.06}
\end{equation}
for the W1 band, and
\begin{equation}\label{eqn:wise_w2}
    \Delta t_{\rm torus, W2}/{\rm day} = 10^{2.20\pm0.06} (L_{\rm AGN, SED}/10^{11}L_\odot) ^{0.45\pm0.05}
\end{equation}
for the W2 band. These correlations closely follow the expected $\Delta t\propto
L_{\rm AGN}^{0.5}$ relation. At a given $L_{\rm AGN}$, the time lag
differences in the W1 and W2 bands are small with the mean value of $\Delta
t_{\rm torus, W2}/\Delta t_{\rm torus, W1}\sim1.21\pm0.36$, and a
median value at 1.15. A linear fit to both W1 and W2 time
lags yields
\begin{equation}
    \Delta t_{\rm torus, W2} = (1.17\pm0.11)\Delta t_{\rm torus, W1} + (0.21\pm54.3) ~.
\end{equation}
Despite the large uncertainty, the fitted intercept is very close to zero,
indicating that the time lag in W2 is always statistically larger than in W1
by a constant factor.

The average redshift of the 67 fitted PG quasars is 0.15$\pm$0.10.  Assuming a
normal AGN template, the corresponding intrinsic wavelengths of the {\it WISE}
W1 and W2 band filters are 2.94$^{+0.28}_{-0.24}~\mum$ and
4.03$^{+0.38}_{-0.33}~\mum$. For an AGN with $L_\text{AGN,
bol}=10^{12}~L_\odot$, these results correspond to a dust emission region size
at $\lambda_{\rm rest}\sim3~\mum$ of $\sim$0.31 pc for the W1 band and at
$\lambda_{\rm rest}\sim4~\mum$ of $\sim0.37$ pc for the W2 band. The similar
sizes at the W1 and W2 bands strongly support a compact torus, where the
emission at the rest frame $\sim$3--5~$\mum$ is dominated by dust grains with
similar temperatures.

\subsubsection{Comparison with Previous Dust Reverberation Studies}

Previous ground-based K-band reverberation mapping has been only focused on
relatively low-luminosity AGNs ($L_{\rm AGN}\lesssim10^{11}~L_\odot$) in
Seyfert galaxies. However, given the SED analysis in \cite{Lyu2018}, it is
likely that Seyfert-1 nuclei and quasars share similar torus properties. For
the 17 Seyfert-1 objects in \cite{Koshida2014}, we adopted the average K-band
time lags estimated with the CCF methods and the  $\alpha_\nu=1/3$ accretion-disk
component model, and converted the weighted averaged V-band absolute magnitude
to $L_\text{AGN, bol}$ following Equation~\ref{eqn:bol-corr}.  (See
Appendix~\ref{app:koshida} for further discussion of the consistency of their
and our measurements of time lags and AGN luminosities.) Applying the same
correlation analysis to Seyfert-1 K-band time lag and AGN bolometric luminosity
yields
\begin{equation}
    \Delta t_{\rm torus, K}/{\rm day} = 10^{1.86\pm0.06} (L_{\rm AGN, SED}/10^{11}L_\odot)^{0.45\pm0.07}
~.
\end{equation}
For $L_{\rm AGN, bol}=10^{12}~L_\odot$, the corresponding K-band torus size
($\lambda_{\rm rest}\sim2.1~\mum$) is 0.17 pc, which is about half of the WISE
W1 size. Such a difference could provide important insights to the properties
of the AGN innermost torus, which will be discussed in detail in another paper.
Both the K-band and W1 behavior are consistent with a general scaling as
$L^{0.5}$ over a range of 10, 000 in luminosity, demonstrating that similar
circumnuclear torus structures are common to AGNs in general.

\cite{Glass2004} provided preliminary time lag measurements between the U 
($\sim0.36~\mum$) and L ($\sim3.4~\mum$) bands of five Seyfert-1 nuclei:
Fairall~9, Akn~120, NGC 3783, ESO141-G55 and NGC7469. We have analyzed the IR
SEDs of all these AGNs with our empirical templates in \cite{Lyu2018}.
Following Section~\ref{sec:lum-estimate}, we estimate their $L_{\rm AGN}$ and
plot their locations in Figure~\ref{fig:tl_sed}. These objects have a very
limited luminosity range ($L_{\rm AGN}=10^{10.5}\sim10^{11.2}~L_\odot$), so it
is not possible to pin down a meaningful lag-luminosity relation given the
uncertain time lag measurements and small sample. Nevertheless, they distribute
around the prediction of our time-lag correlations based on the {\it WISE} W1
and W2 band data of PG quasars.

The optically reddened type-1 AGN NGC 6418 is the only object with published
robust mid-IR dust reverberation results. With high cadence {\it
Spitzer}/IRAC monitoring, \cite{Vazquez2015} found this object had time lags of
37.2 days at 3.6~$\mum$ and 47 days at 4.5~$\mum$.  Based on the $H\alpha$
luminosity of NGC 6418, they determined a lower limit to the AGN bolometric
luminosity, $L_\text{AGN, bol}\ge 5\times 10^9~L_\odot$.  With our previous
developed AGN templates \citep{Lyu2017, Lyu2017b}, we can estimate an upper
limit on its AGN bolometric luminosity to be $\lesssim$6.3--20 $\times 10^9
L_\odot$ (depending on if the AGN is normal, WDD or HDD) based on its {\it
WISE} W4 flux and SED shape.  If NGC 6418 shares similar torus properties as PG
quasars, given the similar bandpasses of {\it WISE} and {\it Spitzer/IRAC}, its
AGN bolometric luminosity should be about 8-9$\times10^9 L_\odot$ based on
equations \ref{eqn:wise_w1} and \ref{eqn:wise_w2} for the reported time lags.
This value is well above the lower limit constrained by \cite{Vazquez2015} and
below the upper limit given by its mid-IR emission. As a result, we conclude
that there is no evidence for strong differences in the torus structures
between this low-luminosity AGN and quasars in the mid-IR.

\subsection{PG quasar variability at 10--25~$\mum$}\label{sec:result-24mum}

Among the 87 PG quasars at $z\lesssim0.5$, 33 have repeated MIPS measurements
at 24~$\mum$. In Table~\ref{tab:agn-24-var}, we present the flux change
significance, $S_{i,j}$, between (1) the first two MIPS measurements, (2) the
third measurement (if acquired) and the previous measurement that shows the
greatest difference, and (3) the average of the MIPS measurements and the WISE
measurement if the latter is available. The significance of a change in flux
is given by 
\begin{equation}
    S_{i,j} = \frac{|f_i - f_j|}{\sqrt{\sigma_i^2+\sigma_j^2}} ~,
\end{equation}
where $f_i$, $f_j$ are the two flux measurements and $\sigma_i$, $\sigma_j$ are
the corresponding uncertainties. We identify variability when one of these
values satisfies $S_{i,j}\ge3$ (similar to the three-sigma criterion given its
definition). Among 26 radio-quiet PG objects, only one, PG 1535+547, was found
to vary at 24~$\mum$. For the remaining seven radio-loud quasars, two out of
three flat-spectrum ($\alpha>-0.7$)\footnote{We assume a radio spectrum with
$f_\nu\propto\nu^\alpha$ where $f_\nu$ is the observed flux density and $\nu$
is the observed frequency.} but none of the four steep-spectrum ($\alpha<-0.7$)
objects were found to vary.  It appears that AGN 24~$\mum$ variability is
related with the radio-band classification. In the upcoming section, we will
expand the statistics with a larger sample of 139 quasars with similar
observations.

We also explore the 12$~\mum$ and 22~$\mum$ variability of the 87 PG quasars by
comparing {\it Spitzer}/IRS synthetic photometry to the {\it WISE} measurements (see
Table~\ref{tab:PG-1024}). Only nine objects (10\% of the sample) have 12~$\mum$
flux variations larger than 3-$\sigma$ and two objects (2\% of the sample) at
22~$\mum$. Because there may be systematic errors of a few percent between
data obtained from different observatories \citep{carey2010, sloan2015} in
addition to the statistical uncertainties, a 5$\sigma$ significance level is
more convincing for variability detection. With this requirement, only two
quasars, PG 1226+023 (3C273) and PG 0007+106 (a flat-spectrum radio source and
blazar at z=0.089; \citealt{Mao2016}), are left. This is not surprising, since
previous studies of blazars and flat-spectrum radio sources have found them to
vary substantially at 10$\mu$m \citep[e.g.,][]{rieke1972, rieke1974,
Edelson1987, Neugebauer1999}.  In particular, 3C 273 is a well-known IR
variable object \citep[e.g.,][]{Neugebauer1999, Soldi2008}.  It seems that
most, if not all, significant AGN IR variability at $\lambda\gtrsim 10~\mum$
can be associated with non-thermal processes.

\startlongtable
\begin{deluxetable*}{lrrrrrrrrrrl}
    \tabletypesize{\scriptsize}
    \tablewidth{1.0\textwidth}
    \tablecolumns{12}
    \tablecaption{PG quasar variability at 12 and 22 $\mu$m
    \label{tab:PG-1024}
    }
    \tablehead{
	\colhead {name} &
    \colhead {MJD} &
    \colhead {$f_{12\mum}$} &
    \colhead {error} &
    \colhead {stdev}  &
    \colhead {$\Delta M_{W3}$} & 
    \colhead {MJD} &
    \colhead {$f_{22\mum}$} &
    \colhead {error} &
    \colhead {stdev}  &
    \colhead {$\Delta M_{W4}$}  &
    \colhead {comment}    \\
    \colhead {} &
    \colhead {} &
    \colhead {mJy} &
    \colhead {mJy} &
    \colhead {}  &
    \colhead {}  &
    \colhead {}  &
    \colhead {mJy} &
    \colhead {mJy} &
    \colhead {of change}  &
    \colhead {}  &
    \colhead {}
}
\startdata
 PG 0003+158 &  53721.5 &   13.62 &   0.88  &	      &        &  53721.5 &  24.41 &   2.06 &        &        &                \\
             &  55376.1 &   13.96 &   0.34  &  -0.36  &  0.03  &  55376.1 &  25.02 &   1.24 &  -0.25 &   0.03 &\\
 PG 0003+199 &  53559.2 &  175.20 &   3.62  &         &        & $\cdots$ &$\cdots$&$\cdots$&        &        &\\
             &  55377.8 &  191.29 &   2.53  &  -3.65  &  0.10  &  55377.8 & 280.42 &   5.60 &        &        &\\
 PG 0007+106 &  53721.3 &   74.01 &   2.75  &         &        &  53721.3 & 153.34 &   3.90 &        &        &\\
             &  55375.0 &   64.94 &   1.00  &   3.10  &  0.15  &  55375.0 & 126.65 &   2.79 &   5.56 &   0.23 &  variable, FSRQ\\
 PG 0026+129 &  53592.9 &   30.56 &   1.50  &         &        &  53725.8 &  45.84 &   2.34 &        &        &\\
             &  55379.5 &   32.70 &   0.96  &  -1.20  &  0.08  &  55379.5 &  43.63 &   3.55 &   0.52 &   0.05 &\\
 PG 0043+039 &  53381.3 &   14.28 &   1.55  &         &        &  53381.3 &  22.14 &   2.03 &        &        &\\
             &  55383.0 &   16.03 &   0.55  &  -1.06  &  0.13  &  55382.0 &  25.22 &   2.12 &  -1.05 &   0.15 &\\
 PG 0049+171 &  53750.4 &   13.60 &   0.99  &         &        &  53750.4 &  19.02 &   2.40 &        &        &\\
             &  55342.6 &   15.76 &   0.34  &  -2.07  &  0.17  &  55342.6 &  17.96 &   1.17 &   0.40 &   0.06 &\\
  $\cdots$   & $\cdots$ & $\cdots$ & $\cdots$ & $\cdots$ & $\cdots$ &  $\cdots$ &   $\cdots$ &  $\cdots$ &  $\cdots$ &  $\cdots$ & $\cdots$    
\enddata
\tablecomments{
    (This table is available in its entirety in machine-readable form. A
	portion is shown here for guidance regarding its form and content.)
}
\end{deluxetable*}

\subsection{AGN Variability at 22--24~$\mum$ in a Larger Sample}\label{sec:larger-24mum}

\subsubsection{General Variability Behavior}

\begin{figure}[htp]
    \begin{center}
	\includegraphics[width=1.0\hsize]{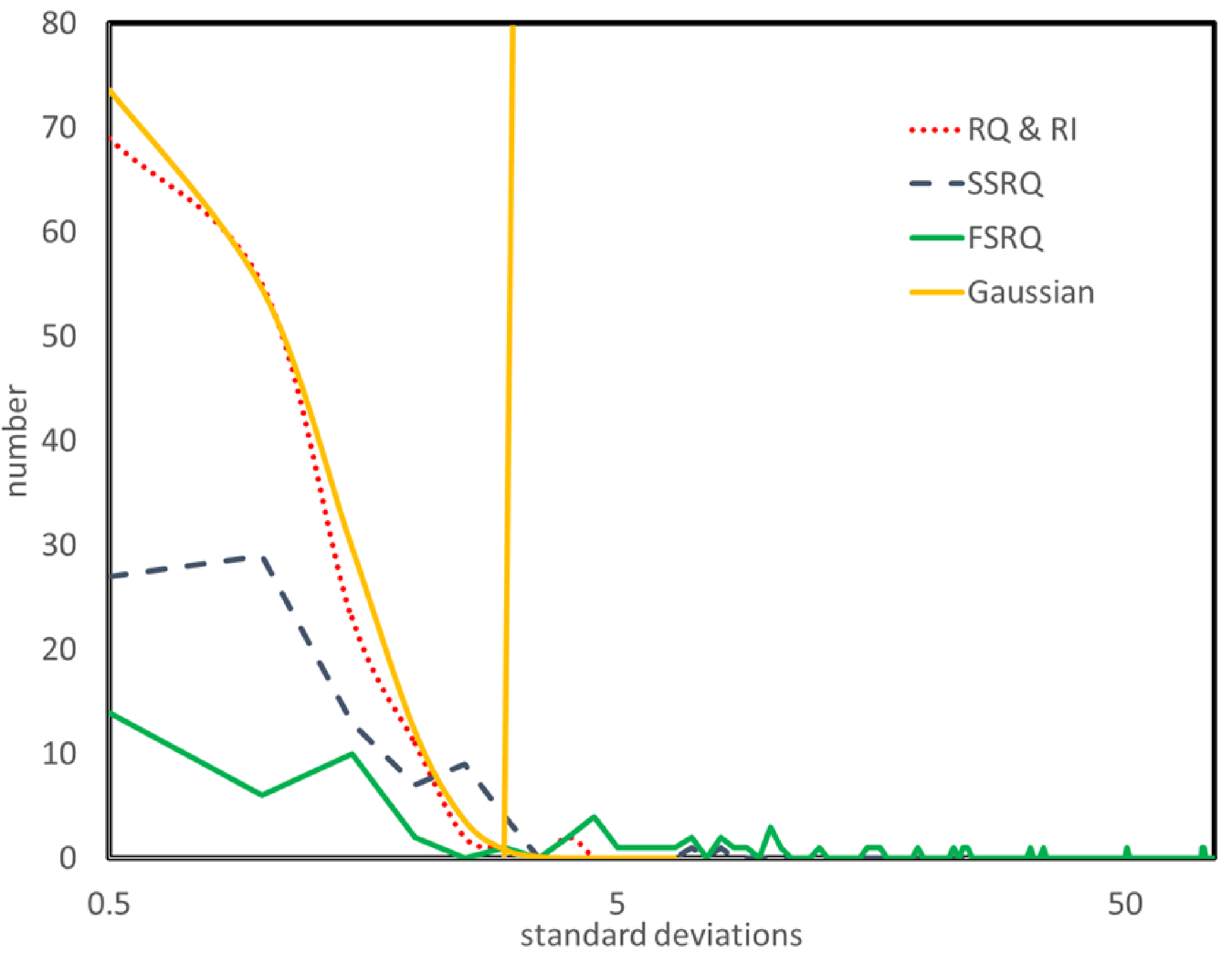}
		\caption{
		    Distribution of variability significance $S_{i,j}$ between
		    different 24~$\mum$ measurements in all 139 quasars.  The
		    calculated $S_{i,j}$ have been linearly binned from 0 with
		    a bin size of 0.5. A Gaussian with $\sigma = 0.9$  has been
		    binned similarly. The vertical orange line indicates
		    $S_{i,j}=3$ used to identify variability.  Two objects,
		    2MASSi~J0918486+211717 and PG~1535+547 stand out as having
		    larger changes and appear to be exceptional radio-quiet
		    variables. Although only one steep-spectrum quasar is
		    identified as being probably variable, their distribution
		    of variability significances is broader than that for the
		    radio-quiet sources suggestive of low level variability in
		    some of them. Compared with step-spectrum radio quasars,
		    the distribution for the flat-spectrum radio quasars is
		    even broader and indicates a strong link between this radio
		    characteristic and variability at 24~$\mu$m.  }
	\label{fig:24var_pattern}
    \end{center}
\end{figure}

With the multi-epoch MIPS and {\it WISE} measurements of 139 objects, we now
further probe AGN variability at 22--24~$\mum$ with a much larger sample. The
measurements and results are also listed in Table~\ref{tab:agn-24-var} and
shown graphically in Figure~\ref{fig:24var_pattern}.  We use $q_{24} =
\log[f_\nu(24~\mum)/f_\nu(1.4~GHz)]$ to determine radio loudness and tag
radio-loud (RL) for $q_{24} < -0.5$, radio-intermediate (RI) for $-0.5 < q_{24}
< 0.5$ and radio-quiet (RQ) for $q_{24} > 0.5$. The radio data were taken from
the NASA Extragalactic Database (NED), the {\it FIRST} survey, \citet[][both
corrected to 1.4 GHz assuming a slope of $-$0.7]{Kellermann1989,
Kellermann2016} , \citet{Hodge2011}, and the Molonglo Galactic Plane Survey
\citep{Murphy2007}.  There is one anomalous case, PG 1309+355, which has a flat
radio spectrum but is radio-intermediate by our definition; by other criteria,
however, it is radio-loud \citep{laor2019}.  We have counted it among the
radio-loud flat-spectrum radio quasars (FSRQs).  Sources lacking radio data are
by default included with the radio-quiet sample. Because there was no
difference in variability behavior, we have also combined the
radio-intermediate sources with this sample. Hereafter this mixed sample is
referred as ``radio-quiet'' for brevity.

As shown in Figure~\ref{fig:24var_pattern}, the flux change significance
distribution of radio-quiet AGNs is best fitted with a Gaussian of width of 0.9
--- that is, this fit indicates that the estimates of individual standard
deviations are high (conservative) by about 10\% under the assumption that the
sample is not variable\footnote{Since PG 1309+355 has not varied in our
observations, classifying it as radio-intermediate would have no effect on this
result.}.  This result confirms that our criterion for identifying variations
is conservative and furthermore emphasizes how rare 24 $\mu$m variability is
for these objects. As indicated by the 10\% smaller width of the Gaussian fit
to the radio-quiet population, the real systematic error might be smaller than
0.7\% as introduced in Section~\ref{sec:data-24}. However, we decided not to
change it since a uniform correction would not apply across the entire sample
(e.g., faint sources probably have larger errors). Given these uncertainties,
two sources, the FSRQ NGC 1275 and the SSRQ 3C 270, are identified as possibly
being variable because they are detected at very high signal to noise and show
flux changes at 2.7--2.8$\sigma$.  Overall, we detect 24~$\mum$ variations in
22 (including the two probable variables) out of 139 targets and set tight
constraints on possible variations for the rest of the sample within the limits
of our observational cadence.

The steep-spectrum radio-loud sources show a distribution of the 24~$\mum$ flux
change significance that is somewhat broader than that for the radio-quiet
sample, suggesting that low levels of infrared variability may apply for some
of them even if it is not detected individually. However, their overall pattern
is not very different from the radio-quiet sample. In contrast, the
distribution of flat-spectrum sources is very broad, showing that variability
is common among them. In fact, we detect variations in 14 out of 19 FSRQs --- a
very high fraction given our small number of observations. 

\subsubsection{Other variability studies}

It seems possible that the measurements by IRAS at 25~$\mu$m could extend the
time scale of our survey. However, the IRAS photometry is of relatively low
accuracy. There are variations at the 5\% level (up to 10\%) in the response
along a pixel, and the simple bulk photoconductors used are susceptible to
responsivity shifts due to exposure to ionizing radiation \citep{beichman1988}.
Evidence for a significant ($> 3 \sigma$) flux change after applying these
systematic errors is found for only three objects: 3C~273, 3C~274, and
PKS~2005-189.  The increased brightness of 3C 274 as seen by IRAS can be
explained by its resolved nature evident in the Spitzer images.  \cite{shi2007}
have shown that the extended 24 $\mu$m emission is the IR synchrotron radiation
from the outer jet and inner radio lobes of this nearby FR I-type radio galaxy.
All of the extended emission would be included within the IRAS beam and
photometry of the Spitzer images using a 1$'$ aperture completely recovers the
excess flux measured by IRAS. However, the IRAS data do demonstrate that 3C273
was nearly twice as bright and PKS~2005-189 about three times as bright when
measured in 1983 compared with our measurements. Both of these sources are FSRQ
blazars, so these results help document the extent of their variability but do
not qualitatively change the findings above. 

\citet{Neugebauer1999} presented a multi-year photometry monitoring for 25 PG
quasars from 1 through 10 $\mu$m (J through N band) and reported 10~$\mum$
variability in 3C 273. They also deduced possible variations at 10 $\mum$ for
PG~1535+547 on the basis that its mid-IR light curve tended to follow the
variations seen at shorter wavelengths even if by itself the measurements did
not have enough signal to noise to make a persuasive detection. They also used
a statistical argument to make a case for variations in other radio-loud
objects. 

\subsubsection{Lack of 24~$\mum$ Variations in Radio-quiet Quasars}\label{sec:radio-quiet-24}

Our study therefore represents a substantial advance over previous work on
variability of AGNs at 10 and 20 $\mu$m. We now use it to evaluate the
overall variability of the radio-quiet sample by computing the weighted
average of the absolute changes in flux between the initial two MIPS
measurements.\footnote{We focus on the initial two MIPS 24~$\mu$m
measurements instead of the {\it WISE} results since they are generally of
higher signal-to-noise ratio and use a smaller beam that reduces
contamination issues. Also, the first two epochs of observations
encompassed the entire sample, whereas the third MIPS observational
epoch only includes radio-loud AGNs.} Although we have only two
measurements per object, the large sample size lets us put interesting
limits on variability using a statistical approach.

The repeatability of the MIPS 24~$\mu$m photometry in the absence of any photon 
noise can be as good as 0.4\% \citep{Engelbracht2007}. If we remove
    2MASSi~J0918486+211717 and PG~1535+547, plus all of the blazars and
    flat-spectrum radio sources, the rms scatter in our quasar measurements is
    $0.0085 \pm 0.0006$ magnitudes ($0.79 \pm 0.06$\%); if we further remove
    all radio-loud objects, the rms scatter reduces to $0.0076 \pm 0.0006$
    magnitudes ($0.70 \pm 0.06$\%).  That is, outside of 2MASSi~J0918486+211717
    and PG~1535+547, plus the flat-spectrum radio sources and blazars, the rms
    scatter in the measurements is consistent with virtually no variability
    over the $\sim$ 3 year period spanned by the MIPS observations.

Now we compare these 24~$\mum$ upper limits to the variability in {\it WISE} W1
and W2 bands, i.e., at 3--5~$\mum$.  For the entire PG sample, the rms scatter
is 0.086 mag and 0.071 mag at W1 and W2, respectively. For the subset for which
we have 24~$\mu$m measurements, the values are the same.  That is, variability
at 3.4~$\mu$m is a factor of ten larger than the upper limits set for 24~$\mu$m. 
As deduced by \citet{Neugebauer1999}, the variability damps out
dramatically going from the near infrared to the mid infrared. The {\it
Spitzer} measurements put this result on much sounder ground than previously:
to illustrate the gain, at 73 mJy, PG 1535+547 is close to the median
brightness of the sample, 86 mJy, yet its variations were just at the detection
limit with the previous ground-based data. 

\subsubsection{Behavior of variable radio-quiet quasars}

PG~1535+547 and 2MASSi~J0918486+211717 are of interest as apparently normal,
radio-quiet, optically-selected quasars that have exceptional mid-infrared
variability signals at 10 and 24~$\mum$. PG 1535+547 (Mrk 486) is of particular
interest because it is very thoroughly studied. \citet{Lyu2017} find its
infrared SED to be fitted very well by their warm dust deficient template,
which makes the assumption that the nuclear non-thermal continuum has dropped
to a negligible level by 24 $\mu$m and that the mid-IR flux comes mostly from
the circumnuclear torus with very little contribution from polar dust.
\citet{Xie2017} show it to have weak silicate emission features, requiring that
some, if not all, of its 24~$\mu$m emission is from heated dust. In the optical
band, PG 1535+547 is moderately obscured with a red continuum and it has the
strongest polarization signals reported in the PG sample ($p\sim2.5\%$;
\citealt{Berriman1990}). Indeed, Hubble Space Telescope ultraviolet and
ground-based spectropolarimetry show that the polarization rises to nearly 8\%
in the UV and that there are complex polarization changes across the H$\alpha$
and H$\beta$ emission-line profiles \citep{Smith1997}. The presence of strong
emission lines in the polarized flux spectrum and the strong rise of the
polarization into the UV suggests that dust scattering of the AGN nuclear
continuum and emission from the BLR is the polarizing mechanism. Despite the
fact that the polarization is from scattering, \cite{Smith1997} found evidence
that the optical polarization varies on time scales as short as a year. In
total, these observations indicate that PG 1535+547 cannot be face-on, but must
be partially-obscured by its circumnuclear torus. In this case, even if a jet
is present, its apparent emission would not be strongly amplified by
relativistic beaming and hence would be weak due to its misalignment with our
line of sight and it would not contribute significantly to the observed flux at
10 and 24~$\mu$m (see more discussion in Section~\ref{sec:disc-origin}).
Instead, we suggest the IR variations of PG 1535+547 at longer wavelengths are
dust reverberation signals.

\begin{figure*}[htp]
    \begin{center}
	\includegraphics[width=0.495\hsize]{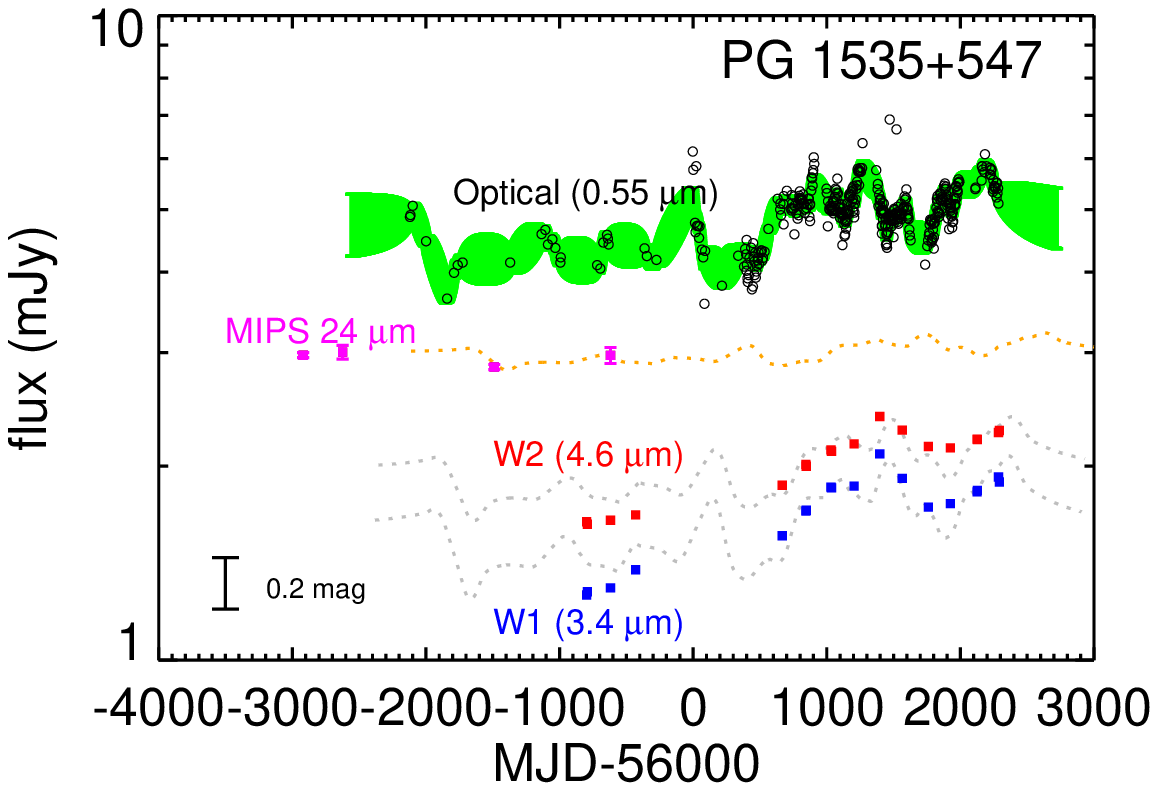}
	\includegraphics[width=0.495\hsize]{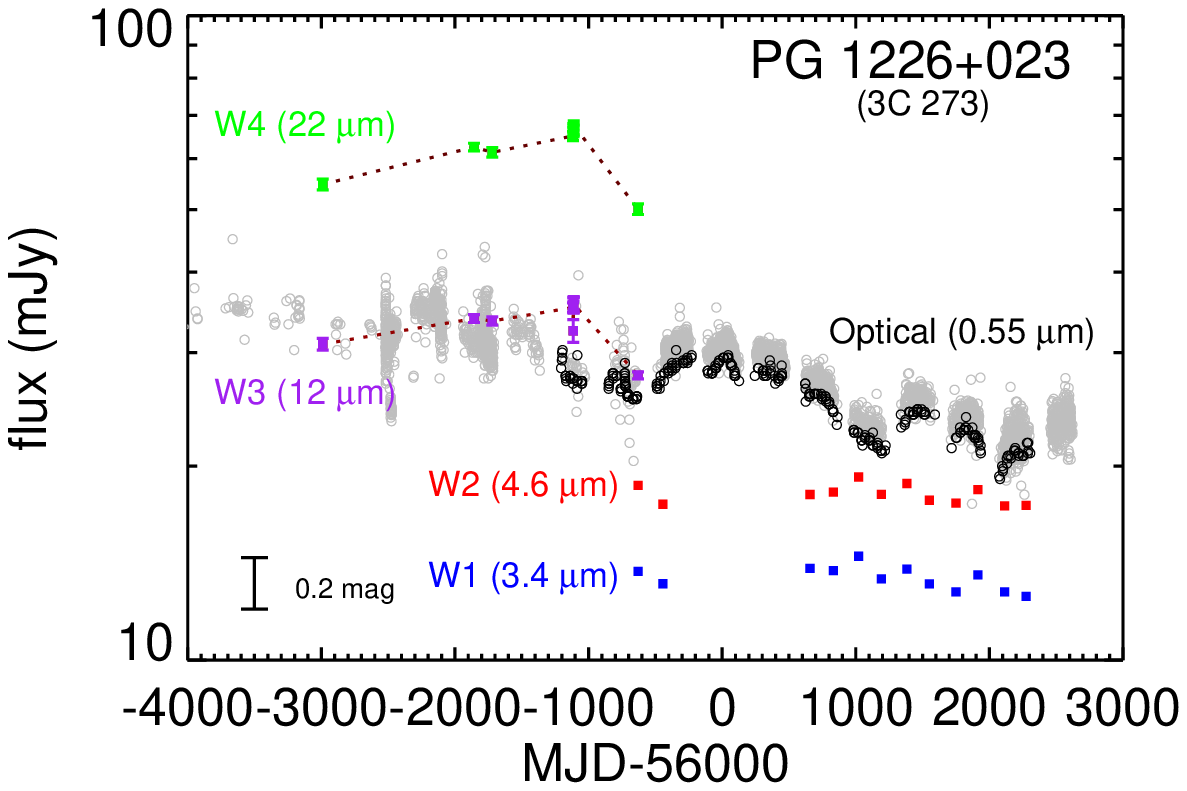}
		\caption{ 
		Multi-band IR and optical light curves of the radio-quiet PG
		1535+547 (left) and the blazar PG 1226+023 (right). See text
		for details.
		    }
	\label{fig:24lc_compare}
    \end{center}
\end{figure*}

To illustrate the different mid-IR variable behaviors between radio-quiet and
radio-loud quasars, we compare the IR light curves of PG~1535+547 and 3C~273 in
Figure~\ref{fig:24lc_compare}. For the blazar 3C 273, there is no obvious
similarity between its optical and mid-IR light curves,
especially given the large variations seen between the neighbouring 
epochs in {\it WISE} W2. The brightness of 3C 273 at W3 and W4  varied by 
$\sim$0.3 mag, about 0.15 mag larger than the variations observed in W1 and W2.  
Such behavior is not expected from the dust emission, since its energy output 
at longer wavelengths must correspond to larger physical scales where the 
amplitude of variability is expected to decrease.

In contrast, the W1 and W2-band light curves of PG 1535+547 can be reproduced
by shifting and scaling its optical light curve, as expected for dust
reverberation signals with time lags of $\sim180$ days for W1 and $\sim205$
days for W2. Assuming a time lag of 2.5 times larger (i.e. $\sim450$ days) than
W1 and an amplitude factor of 0.6, the 24~$\mum$ light curve can be matched by
dust reverberation mapping. The variability amplitude of PG~1535+547 in W1 is
$\gtrsim0.5$~mag, among the 10\% largest W1-band variables in the PG sample. In
comparison, the 24~$\mum$ variability amplitude of PG 1535+547 is about ten
times smaller at only $\sim$0.06 mag. The observed behavior of this object
suggests that a detectable level of mid-IR variability caused by dust
reverberation is plausible given a highly variable quasar, high
signal-to-noise-ratio data, and fortunate time sampling.

\section{Discussion}\label{sec:discussion}

\subsection{The Origin of AGN Mid-IR variability}\label{sec:disc-origin}

\subsubsection{Synchrotron Emission}

The {\it Spitzer} 24~$\mum$ results reaffirm the generally accepted picture
that the smooth UV-millimeter continua of blazars is dominated by the variable,
optically thin synchrotron radiation produced in the core and inner
relativistic jet that also gives rise to the self-absorbed, flat-spectrum radio
emission at centimeter wavelengths (see e.g., \citealt{kellermann1981}).
According to the classical unification scheme \citep{Urry1995}, this type of
object corresponds to an AGN with a powerful jet pointed toward the observer
and its broad-band X-ray to IR SED is dominated by synchrotron emission that is
enhanced by relativistic beaming.  

\begin{figure}[htp]
    \begin{center}
	\includegraphics[width=1.0\hsize]{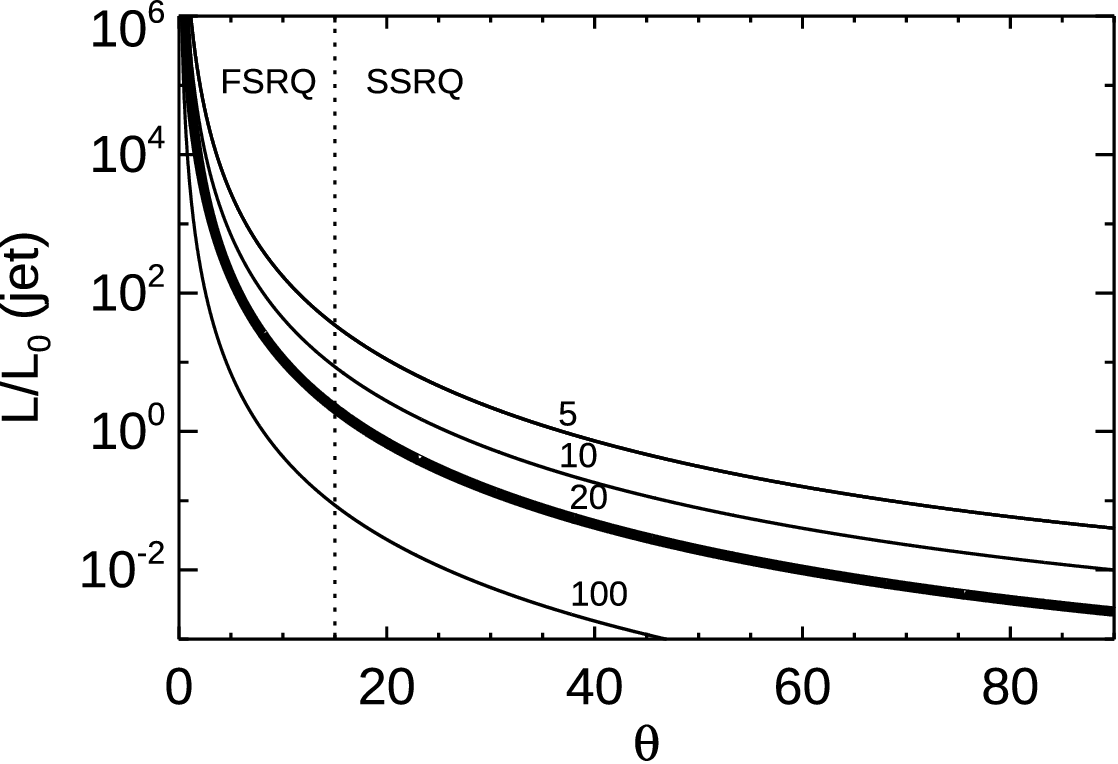}
		\caption{
		  The flux boosting factor as a function of
		  observing angle for different Lorentz factors ($\gamma$=5,
		  10, 20, 100).
		    }
	\label{fig:jet_boost}
    \end{center}
\end{figure}

As shown in Section~\ref{sec:result-24mum}, among the radio-loud AGNs,
steep-spectrum radio quasars (SSRQ) seem to lack strong variability at
24~$\mum$, in contrast to flat-spectrum radio quasars. As suggested by
\citet{Urry1995}, FSRQ are believed to be oriented at relatively small
angles to the line of sight ($\theta\lesssim$15\degree), where SSRQ
have intermediate angles between FSRQ and FR II radio galaxies. For an ideal
relativistic beam, the apparent luminosity can be calculated as $L= L_0
/[\gamma(1-\beta \cos\theta)]^2$, where $\gamma$ is the Lorentz factor, $L_0$
is the intrinsic luminosity, and $\beta=(1-\gamma^{-2})^{1/2}$ is the beam
speed in the AGN frame in units of light speed $c$ \citep{Cohen2007}. For
blazars, Lorenz factors up to 10--20 are common for parsec-scale jets
\citep{Homan2012}.  Assuming $\gamma=20$, the jet variability signal can be
boosted by a factor of 2--1600 for FSRQs (see Figure~\ref{fig:jet_boost}),
while the apparent variability for SSQRs decreases quickly to a few percent or
smaller when $\theta>15$\degree. In other words, given the same level of
intrinsic jet variability, FSRQs are expected to have much higher (synchrotron)
IR variability detection rates compared with SSQRs. In addition, it is possible
that the infrared emission of the SSRQs is no longer dominated by the
relativistic jet but by the emission of a circumnuclear torus. In that case,
their behavior would resemble that of radio-quiet quasars, which do not vary
significantly within our 24~$\mum$ observations. 

Given the arguments above, the different mid-IR variability behavior among
SSRQs, FSRQs, and blazars can be ascribed to first order to different angles of
the radio jets relative to the observer.  This result is consistent with the
unification picture for radio-loud AGN \citep{Urry1995}.

\subsubsection{Dust Infrared Reverberation Signals}

The mid-IR variations of the radio-quiet quasars show little evidence for any
synchrotron component, but are dominated by the dust response to the UV-optical
variations of the central engine. As summarized in
Section~\ref{sec:result-var-summary}, the 3--5~$\mum$ IR emission of $\sim77\%$
of the PG quasars at $z<0.5$ show such dust reverberation signals. If we set
aside the 16 sources where the data were inadequate to look for time lags, 95\%
of the remaining objects behave in this way, with only four objects (5\%)
showing an alternative pattern of variations.  This is a strong evidence for
the common existence of circumnuclear dust outside the black hole accretion
disk.  

The near infrared SEDs of quasars with normal or warm-dust-deficient (WDD)
behavior \citep{Lyu2017b} show direct evidence for this circumnuclear dust
heated by the accretion disk, in the form of a spectral bump at $\sim$3~$\mu$m
that matches the expected emission of dust heated to sublimation temperatures.
Although hot-dust-deficient (HDD) quasars do not show a similar near-IR bump,
as shown in Figure~\ref{fig:tl_sed}, they still follow the $R\propto L^{1/2}$
size-luminosity relation and their dust time lags increase with wavelength.
This is true even for the most HDD quasar --- PG 0049+171, where we detect time
lags $\sim$120 days for its W1 variations and $\sim$157 days for W2. This
behavior is evidence that the emission near 3 $\mu$m is dominated by the
emission of the circumnuclear torus and not by the accretion disk/central
engine. Furthermore, since the IR SEDs of unobscured quasars at $z\sim$0--6
have similar forms as the PG sample \citep{Lyu2017}, the result implies that
there are few, if any, completely dust-free quasars. 

For radio-quiet quasars, the average RMS variability amplitudes at V, W1 and W2
bands are 0.145 mag, 0.093 mag, 0.076 mag. In other words, the W1 and W2 dust
variation signals are typically only $\sim60\%$ and $\sim50\%$ of those in the
optical band. As presented in Section~\ref{sec:result-24mum}, the possible flux
change at 24~$\mum$ is less than 10\% of that at $\sim3~\mum$ (corresponding to
6\% of the variability at V-band). The decrease in variability with increasing
wavelength is likely to be the result of averaging over the variations due to
light travel time to various parts of the extended circumnuclear torus.
Diminished variability at 24~$\mu$m then results from the emission originating
in a substantially larger region, i.e. at a significantly larger radius than
the emission at 3--4~$\mu$m.

\subsubsection{AGN Mid-IR Variability under the Unification Scheme}

Our reverberation mapping study has focused on the PG quasars because their
selection criteria strongly favor cases where we can see the central engine,
accretion disk, and circumnuclear torus all relatively unobscured. We discuss
here the complexities that can be expected where this simple situation is not
the case. 

\begin{figure*}[htp]
    \begin{center}
	\includegraphics[width=1.0\hsize]{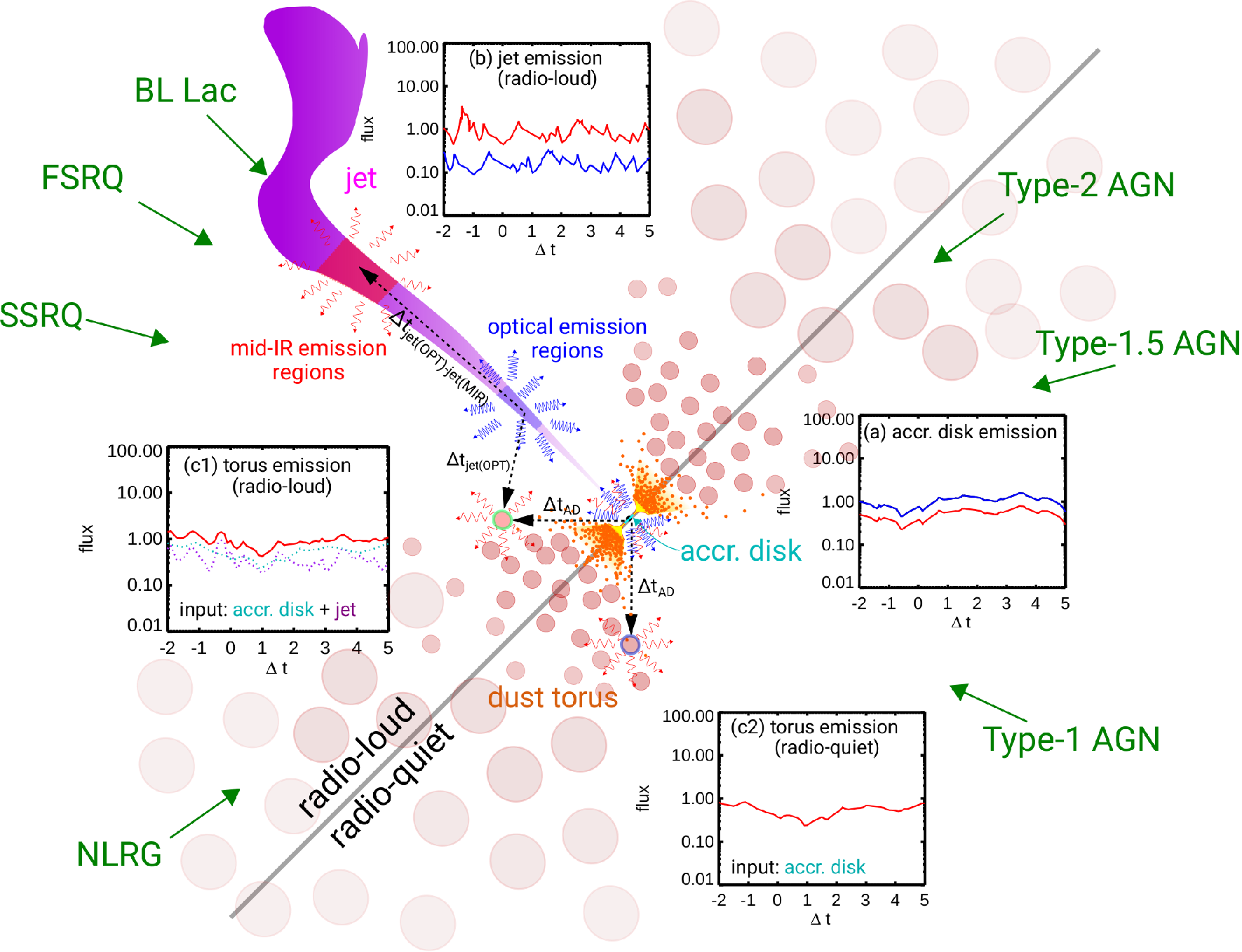}
		\caption{
		    Sketch of the main AGN structures related to the AGN optical
		    and mid-IR continuum variabilities. At the very center is the 
		    BH and its accretion disk (cyan). The dust torus is assumed 
		    to be clumpy and the yellow, orange, red colors represent a 
		    decreasing dust temperature. A strong jet (purple) exists for 
		    the radio-loud case and we highlight its optical emission (blue) 
		    and mid-IR emission (red) regions. For the accretion
		    disk and the jet component, we plot their optical (blue solid line) 
		    and mid-IR (red solid line) light curves in panels (a) and (b). 
		    The mid-IR light curves (red solid line) of the torus in 
		    the radio-loud and radio-quiet cases are plotted in panels (c1) and
		    (c2). For the radio-loud case, we show the contribution from
		    the dust reverberation signals to the accretion disk
		    and the jet in dashed cyan and purple lines. According to the 
		    AGN unification model \citep{Antonucci1993a, Urry1995}, we denote 
		    different AGN types at different inclination angles for radio-loud 
		    and radio-quite cases with green arrows.
		    See text for detailed explanations.
		    }
	\label{fig:var_cartoon}
    \end{center}
\end{figure*}

In general, both synchrotron emission and torus-reprocessed emission will
contribute to the mid-IR variability of a radio-loud AGN; their relative
importance depends on the inclination angle and radio-loudness. In addition,
the obscuration of the central engine will substantially affect the apparent
optical variability, making the interpretation of the mid-IR light curves
difficult.  Figure~\ref{fig:var_cartoon} provides an illustration of the
classical unification model \citep{Antonucci1993a, Urry1995} and shows the
optical and mid-IR continuum light curves of various AGN components.

In the radio-loud case, the integrated optical continuum emission of the AGN 
is a combination of the jet component ($f_{\rm jet, opt}(t)$) and the 
accretion disk ($f_{\rm AD, opt}(t)$),
\begin{equation}
F_{\rm radio-loud, opt}\approx C_{\rm torus} (\theta) f_{\rm
AD, opt}(t) + C_{\rm beam}(\theta) f_{\rm jet, opt}(t)~~,
\end{equation}
where $C_{\rm torus}(\theta)$  reflects the obscuration level of the accretion
disk light by the torus ($C_{\rm torus}\sim1$ for type-1 AGNs and $\sim0$ for
type-2 AGNs) and $C_{\rm beam}(\theta)$ is the flux boosting factor due to
realistic beaming. As the presumed inclination angle of the jet to our line of
sight increases from BL Lac, FSRQ to SSRQ, the relative contribution of $f_{\rm
AD, opt}$ increases quickly.  For Narrow-Line Radio Galaxies (NLRG), although
the absolute strength of the jet variability is weaker than for the SSRQ case,
the optical light curve is dominated by the jet since the accretion disk is
obscured ($C_{\rm torus}\sim0$).  In principle, both the jet component and the
accretion disk can provide UV-optical energy to heat the torus and produce the
mid-IR emission. As they locate in different regions, the mid-IR dust
reverberation signals of the torus might be mixed with the two separated time
lags ($\Delta t_{\rm jet(opt)}$ and $\Delta t_{\rm AD}$).  The relativistic
beaming effects also apply to the jet IR emission, so the observed integrated
mid-IR emission can be written as
\begin{equation}
\begin{split}
    F_{\rm radio-loud,
IR}& \approx  f_{\rm AD, IR}(t) + C_{\rm beam}(\theta) f_{\rm jet, IR}(t) + \\
      & ~~~~~~~~~\mathcal{F}_{\rm AD,
opt}(t-\Delta t_{\rm AD}) +\mathcal{F}_{\rm jet, opt}(t-\Delta t_{\rm jet(opt)}),
\end{split}
\end{equation}
where we use $\mathcal{F}(t-\Delta t)$ to denote the dust-reprocessed emission
of the optical light $f(t)$ with a time lag $\Delta t$. Since the jet is highly
beamed and perpendicularly distributed, its illumination to the torus could
often be ignored. As a result, the observed total IR emission can be further
simplied into
\begin{equation}
    F_{\rm radio-loud,
IR}\approx  f_{\rm AD, IR}(t) + C_{\rm beam}(\theta) f_{\rm jet, IR}(t) + \mathcal{F}_{\rm AD,
opt}(t-\Delta t_{\rm AD}) ~.
\end{equation}
Comparing it with the form of $F_{\rm radio-loud, opt}$, if dominated by
different components, the optical and mid-IR light curves could be uncorrelated
(the jet component and accretion disk have different variability patterns).
This could explain the behavior of 3C 273 (as seen in
Figure~\ref{fig:24lc_compare}).

The situation for radio-quiet AGNs is much simpler; their optical light curve
is $F_{\rm radio-quiet, opt}\approx C_{\rm torus} (\theta) f_{\rm AD, opt}(t)$
and their mid-IR light curve $F_{\rm radio-quiet, IR}\approx f_{\rm AD,
IR}(t)+\mathcal{F}_{\rm AD, opt}(t-\Delta t_{AD})$. When the nucleus is
unobscured ($C_{\rm torus}\sim1$), we readily observe the time lag $\Delta
t_{AD}$, which reflects the light travel time from the accretion disk to the
torus, as demonstrated by our dust reverberation mapping. However, if the
nucleus is obscured, the optical band could have a significant contribution
from the host galaxy. Given the large beams used for much of our visible data,
in this situation it would be difficult to extract the nuclear component
accurately. To illustrate this issue, consider the full sample observed at 24
$\mu$m, excluding the radio loud sources. 50\% of the remaining sources are
2MASS quasars, which show varying amounts of AGN obscuration (e.g.,
\citealt{Cutri2002}). Of the remaining sources, 40\% are of types 1.8, 1.9, or
2 (from the NED database), again indicating substantial obscuration.

\subsubsection{Other Causes of IR Variations}

Besides the radio-loud PG 1226+023 (3C 273), four other PG objects --- PG
0003+199, PG 0923+201, PG 1534+580, and PG 1612+261 --- have uncorrelated
optical and mid-IR variability. None of these objects are blazars or even
radio-loud, so it is unlikely that non-thermal emission contributes to their IR
variations. In addition, since all of these four objects show blue optical
continua and strong broad Balmer emission lines, the optical light curve should
delineate the variations of the accretion disk.  Consequently, their IR
variations cannot be the dust reverberation signals from a classical torus. We
cannot rule out the possibility that these objects have very complicated dust
structures (e.g., there are several polar dust clouds or filaments) that cause
reverberation signals with a wide range of time lags with diverse luminosity
weights --- for some configurations, the integrated IR light curve might not
show any variation pattern similar to the input optical signals. Nevertheless,
there is no new mechanism involved in this possibility -- it still belongs to
the dust reverberation signals.

Dynamical motion of the torus material could be another possibility to cause
AGN IR variability.  Assuming the dusty environment of an AGN can be described
as a radiation-driven fountain, \cite{Schartmann2014} explored the
time-resolved IR SEDs and argued for possible variations, but over a timescale
of 0.1 Myr.  As their simulation indicates, there could be significant but
temporary dust in the line-of-sight that is heated by the AGN, causing changes
in the IR (by dust emission) and optical flux (by dust extinction).  On yearly
timescales, however, the turbulence/fountain argument cannot explain the
observed mid-IR variations of these four quasars by merely moving dust clouds.
Taking the typical velocities of order 300--400 km/sec, the moving dusty
clouds/filaments could only travel a few $10^{-4}$ parsec, which would not
change the exposed solid angle significantly at a distance of a parsec from the
heating source.

However, there is a chance that dust grains could be evaporated when the dusty
clouds/filaments move to an environment with strong radiation heating by the
AGN or where an outburst (which could be at a non-visible wavelength, e.g.,
X-rays) by the central engine evaporates dust in a formerly `safe' environment.
As a result, their IR emission could be quickly reduced.  In fact, the dust
destruction timescale around the sublimation region is estimated to be order of
10 days \citep{Kishimoto2013}, quick enough to produce the mid-IR variation we
see. Observed optical polarization variability seen in objects like PG0050+124,
PG1535+547 \citep{Smith1997}, and Mrk231 \citep{Gallagher2005} on timescales of
less than a year may also be related to rapid sublimation events.  In another
paper, we will provide a detailed study on how to explain these peculiar mid-IR
variability features with this possibility and discuss its broad implications.

\vspace{10pt}

In conclusion, (1) observed mid-IR variability on monthly to yearly time-scales
of most quasars is caused by dust reverberation signals that are correlated
with the variations of the accretion disk emission. The amplitude of the dust
IR variability decreases quickly with increasing wavelength.  (2) If a jet is
present and the system is viewed close to the line of sight to the jet axis
(e.g., FSRQ or blazar), non-thermal processes dominate the IR variations with
larger variation amplitudes over shorter timescales. (3) Only a small subset of
quasars appears to have IR variability that does not fall into these two
categories, which might be explained by reverberation signals off structures
other than the circumnuclear torus, or by the motions of dusty clumps combined
with the heating and destruction of the constituent dust grains.

\subsection{Constraints on the Torus Properties}

\subsubsection{Wavelength-dependent Torus Size and Issues with the Classical
Clumpy Model}

To compare the torus sizes constrained through dust reverberation signals at
different wavelengths, we fix the slope of the time-lag vs. AGN luminosity
relation to be 0.5, and repeat the fitting for W1 and W2-band measurements of
the PG quasars. The results can be found in Table~\ref{tab:fit_fix_slope}.  For
the 67 objects with convincing dust time lags, we find that the dust emission
size ratios follow $R_{W1}:R_{W2}=1:1.2$. As suggested by \cite{Lyu2017},
quasars having different intrinsic IR SED variations may be a reflection of
different torus structures. Therefore, we also compute separately the
corresponding $\Delta t$---$L_{AGN}$ relations for normal, warm-dust-deficient
(WDD) and hot-dust-deficient (HDD) populations. Discussion of how the torus
size depends on SED type will be expanded in Section~\ref{sec:disc-torus-type}.
Here we simply assume all AGNs have a similar torus structure.

\capstartfalse
\begin{deluxetable*}{cccccccc}
    \tabletypesize{\footnotesize}
    \tablewidth{1.0\hsize}
    \tablecolumns{8}
    \tablecaption{Results of Linear Regression of Lag-Luminosity Relation \label{tab:fit_fix_slope}
    }
    \tablehead{
        \colhead{Sample} &
        \colhead{Observed Band} &
	\colhead{<z>} &
	\colhead{$\lambda_{\it rest}$} &
	\colhead{Sample Size} &
	\multicolumn{2}{c}{$\alpha$} &
	\colhead{$\beta$}  \\
	&
	&
	&
	($\mum$) &
	&
	\colhead{($\Delta L_\text{AGN}=$ 0.3 dex)} & 
	\colhead{($\Delta L_\text{AGN}=$ 0.025 dex)} & \\
	\colhead{(1)} &
	\colhead{(2)} &
	\colhead{(3)} &
	\colhead{(4)} &
	\colhead{(5)} &
	\colhead{(6)} &
	\colhead{(7)} &
	\colhead{(8)} 
}
\startdata
    Normal QSOs  &  W1  & 0.145 & 2.95 &  44   &  2.13$\pm$0.04 &   2.21$\pm$0.01 & 0.5  \\
                 &  W2  & 0.145 & 4.04 &  44   &  2.21$\pm$0.04 &   2.28$\pm$0.01 & 0.5  \\
    WDD QSOs     &  W1  & 0.220 & 2.78 &  8    &  2.03$\pm$0.09 &   2.04$\pm$0.05 & 0.5  \\
                 &  W2  & 0.241 & 3.72 &  7    &  2.10$\pm$0.09 &   2.12$\pm$0.05 & 0.5  \\
    HDD QSOs     &  W1  & 0.122 & 3.01 &  15   &  1.97$\pm$0.07 &   1.98$\pm$0.04 & 0.5  \\ 
                 &  W2  & 0.122 & 4.12 &  15   &  2.08$\pm$0.07 &   2.08$\pm$0.04 & 0.5  \\
    All QSOs     &  W1  & 0.149 & 2.94 &  67   &  2.08$\pm$0.03 &   2.19$\pm$0.01 & 0.5  \\
                 &  W2  & 0.149 & 4.03 &  66   &  2.16$\pm$0.03 &   2.27$\pm$0.01 & 0.5  \\
    Seyfert-1    &   K  & 0.038 & 2.11 &  17   &  1.88$\pm$0.05 &   1.93$\pm$0.02 & 0.5  
\enddata
    \tablenotetext{}{The fitted model is $\log(\Delta t) = \alpha + \beta \log(L_{\rm AGN})$.}
    \tablecomments{Col. (1): sample information; Col. (2): the studied
	photometry band; Col. (3): the average redshift of the sample; Col.
	(4): the effecive rest-frame wavelength that the observations are
	probing; (5): number of sample objects; Col. (6): derived $\alpha$
	assuming the
	same 0.025 dex uncertainty of $L_\text{AGN}$ for all objects; Col.
	(7): similar to Col. (5) but the uncertainty is assumed to be 0.3 dex;
    Col.  (8): $\beta$ is fixed to 0.5.}
\end{deluxetable*}
\capstarttrue

We also make a similar correlation analysis for the K-band reverberation
signals of the 17 Seyfert-1 sample in \cite{Koshida2014}, finding a size
ratio between K band and W1 band of $R_{K}:R_{W1}=0.6:1$. In
Appendix~\ref{app:koshida-sed}, we present the SED decomposition results of
these Seyfert-1 nuclei and estimate that half of the sample should have
intrinsic IR SEDs best-described by the normal AGN template.  This value is
similar to the $60-70\%$ normal AGN fraction in the PG quasar sample
\citep{Lyu2017}. As a result, the AGN dust emission size differences at K and
W1 bands are unlikely to be related to the different AGN IR SED types. To test
whether the results depend strongly on the uncertainties in the AGN luminosity
estimate, we changed the uncertainty of $L_\text{AGN}$ from 0.3 dex to 0.025
dex (the latter value corresponds to the typical V-band magnitude uncertainty
of Seyfert-1 nuclei in \citealt{Koshida2014}) and repeated the fits. The
$\Delta t$ -- $L_{\rm AGN}$ correlations shift by 0.05-0.1 dex, but the size
ratios between K, W1 and W2 are not significantly affected, i.e.,
$R_{K}:R_{W1}:R_{W2}=0.6:1:1.2$.

The poor time sampling of the 24~$\mum$ observations prevents us from deriving
any torus size through reverberation mapping. On the other hand, the lack of
variability at this wavelength is most readily explained if the torus zone
dominating the 24~$\mum$ emission is much larger than the zone producing the
bulk of the flux observed at W1 and W2.  In this case, the time lags have a
large range so that the variability is smoothed out.

These observational results provide some challenges to the predictions based on
the classical "clumpy" torus picture. \cite{Almeyda2017} simulated the IR
reverberation response of such a torus model and suggested that the dust time
lags at 3--5~$\mum$ have very limited wavelength dependence (see their section
4.1). In addition, their model also predicted substantial reverberation signals
at 10 and 30 $\mu$m (see their Figure 5).  These conclusions are expected with
their basic model assumption: from the illuminated face to the dark side, the
same optically-thick clouds can have a broad temperature distribution that
results in IR emission over a wide wavelength range. However, this picture is
not favored by our observations.  Given the typical redshift of the sources we
have observed at 24 $\mu$m, the rest wavelength is $\sim$ 20 $\mu$m.
Interpolating their results at 10 and 30 $\mu$m for an extended clumpy torus,
the same model would predict a peak response at the wavelength of our
observations of about 35\% of the response at rest 3.6 and 4.5 $\mu$m. In
contrast, we have found that the response at 24 $\mu$m is an order of magnitude
smaller than that at (observed) 3.4 and 4.5 $\mu$m. 

Rather than being purely clumpy, the torus is likely to be a mixture of
optically-thick dusty clouds and some diffuse distribution of low-density
dust. Such a structure is likely because clumps at the inner edge of the torus
are not likely to be completely stable against gravitational shearing and hence
must be sustained as a result of turbulence or similar effects. In fact, it is
known that the clumpy torus model \citep[e.g.,][]{Nenkova2008a} alone can not
reasonably fit the $\sim3~\mum$ hot dust emission peak among quasars and
another blackbody with $T\sim1300$K needs to be added \citep[e.g.,][]{Mor2009,
Leipski2013}. In contrast, the two-phase (clumps + diffuse dust) torus model
developed by \citet{Stalevski2012, Stalevski2016} does not have this problem
and appears to reproduce the AGN hot dust emission features.

Previous work has often argued that a single black body plus a power-law SED is
enough to reproduce the quasar continuum at $\lambda\sim$0.5--3~$\mum$
\citep[e.g.,][]{Glikman2006, Kim2015, Hernan-Caballero2016}. Since dust
temperature is a strong function of distance, the dust grains responsible for
the $\lambda\sim$1--3~$\mum$ emission should be located at similar radii.
However, our study finds contradictory evidence with $R_{K}:R_{W1}=0.6:1$ and
indicates a complicated picture. Interestingly, under the assumption of similar
grain sizes, the sublimation radii for graphite and silicate dust grains have
$R_\text{sub,C}/R_\text{sub, S}\sim0.5$ (see Section~\ref{sec:lag-lum}), very
close to the reported $R_{K}:R_{W1}\sim0.6$. Future work should address whether
this possibility can be supported by other evidence, as well as folding in the
constraints on torus structure provided by the relative lack of variability at
24~$\mu$m. 

\subsubsection{Surface Density Profiles of the Hot Dust Emission}

Previously, the radial structures of AGN tori have been explored at 8--13
$\mum$ only by long-baseline infrared interferometry observations of $<10$ Seyfert
nuclei \citep{Kishimoto2009b, Kishimoto2011b}. Assuming a power-law radial
surface density distribution of heated dust, the profile ranges from $\sim r^0$
to $\sim r^{-1}$, and might be dependent on AGN luminosity
\citep{Kishimoto2011b}. However, it is likely that most of these objects have
some contribution of extended polar dust emission in the mid-IR ($f_{\rm pol,
10\mum}/f_{\rm total, 10\mum}\sim0.3-0.8$ estimated from SED analysis in
\citealt{Lyu2018}), so the real density profile of the compact torus at these
wavelengths is still highly uncertain.

With the measurements of variability amplitude differences between the optical
and IR light curves, crude constraints on the dust surface density profile can
be calculated with some simple approximations. 

For simplicity, we assume the dust sublimation zone is smooth (i.e., not
clumpy) and that its surface density profile can be described by a power-law
$\Sigma(r)\propto r^{\alpha}$. In addition, the dust grains in the
    sublimation zone will have only a modest range of temperature given that
    the AGN hot dust emission feature is well-matched by a single black body
    spectrum \citep[e.g.,][]{Mor2009}. Furthermore, the dust grains will be hot
    enough that the temperature-dependence of their emission will be modest
    (i.e., they approach the Rayleigh-Jeans regime). Therefore, we can ignore
    the radial dependence of the temperature and express the IR emission as:
\begin{equation}
    f_{\rm IR} \propto \int \Sigma(r) r d r \propto r^{\alpha+2} ~.
\end{equation}
Since the hot dust time lag has been found to be linearly correlated with
the square root of AGN luminosity, we can argue that the dust distance $r$
    is correlated to the AGN optical luminosity by the inverse square law, $r
    \propto L_\text{opt}^{0.5}$, so
\begin{equation}
    f_{\rm IR} \propto f_\text{opt}^{(\alpha+2)/2} ~.
\end{equation}
When the optical flux $f_\text{opt}$ is changed to $f_\text{opt}'$ by a factor
of $\xi$, the corresponding IR flux changes from $f_\text{IR}$ to
$f_\text{IR}'$ and
\begin{equation}\label{eqn:ir_surface}
    \frac{f_\text{IR}'}{f_\text{IR}} = \left(\frac{f_\text{opt}'}{f_\text{opt}}\right)^{(\alpha+2)/2} = \xi^{(\alpha+2)/2}~.
\end{equation}
With our RM model, the absolute flux change in the IR is correlated with that
in the optical according to Equation~\ref{eqn:opt_ir_cor},
\begin{equation}
    \Delta f_\text{IR} = f_\text{IR}'-f_\text{IR} = AMP\times(\xi -1)f_\text{opt}~.
\end{equation}
We can rewrite Equation~\ref{eqn:ir_surface} as
\begin{equation}
    1+AMP(\xi-1)\frac{f_\text{opt}}{f_\text{IR}}=\xi^{(\alpha+2)/2}~.
\end{equation}
The IR variability can be characterized by magnitude change $\Delta M_\text{IR}$,
\begin{equation}
    \Delta M_\text{IR} = -2.5 \log\left(\frac{f_\text{IR}'}{f_\text{IR}}\right)~.
\end{equation}
Finally we have
\begin{equation}
    \alpha = 2\left(\frac{\log\left(10^{-\Delta M_\text{IR}/2.5}\right)}{\log\left(1+(10^{-\Delta M_\text{IR}/2.5}-1)(f_\text{IR}/f_\text{opt})AMP^{-1}\right)}-1\right)~,
\end{equation} 
where $f_\text{IR}/f_\text{opt}$ is the infrared to optical color of the AGN
SED.

We summarize the mean values of $\Delta M$ and $AMP$ of Seyfert-1 AGN and PG
quasar samples in Table~\ref{tab:torus-profile}. Assuming a normal AGN
template, the characteristic dust surface density profiles are found to be
$r^{0.3}$ in K, $r^{-0.7}$ in W1 and $r^{-1.0}$ in W2. Such a result suggests
that the surface density gradient of sublimating dust increases gradually with radius 
at 2.0~$\mum$, reaches a peak after that, and drops quickly at longer
wavelengths.


\begin{deluxetable}{cccccc}
    \tabletypesize{\footnotesize}
    \tablewidth{1.0\hsize}
    \tablecolumns{4}
    \tablecaption{Parameters related to the torus radial density profiles\label{tab:torus-profile}}
    \tablehead{
	\colhead{Parameter} &
	\colhead{K} &
	\colhead{W1} &
	\colhead{W2} 
	}
    \startdata
    $<\lambda_\text{rest}>/\mum$  & 2.1   &  3.0  & 4.1 \\
    $\Delta M_\text{IR} $ &  0.48$\pm$0.33    &  0.19$\pm$0.11   &  0.16$\pm$0.09 \\
    $ AMP $    &  4.65$\pm$3.07    &  4.65$\pm$3.51   &  4.68$\pm$3.47 \\
    $f_\text{IR}/f_\text{opt}$ &   4.16  &   6.90  & 9.49 \\  
    $\alpha$   &  0.3  & -0.7 & -1
    \enddata
\end{deluxetable}

\subsubsection{Normal Quasars versus Dust-deficient Quasars}\label{sec:disc-torus-type}

\cite{Lyu2017} showed that the intrinsic IR SEDs of PG quasars can be grouped
into normal, WDD and HDD populations that are likely associated with different
torus structures. Now we discuss if such arguments are supported by the dust
reverberation results. Considering the relatively small numbers of dust-deficient
quasars in our PG sample, we reduce the number of free parameters by assuming
that the time lag goes exactly as the square root of the luminosity (i.e.,
$\beta = 0.5$) to facilitate comparisons of the time lag - AGN luminosity
relations for the different quasar types.  The fitted parameters can be found
in Table~\ref{tab:fit_fix_slope}. At the same AGN luminosity, WDD and HDD
quasars appear to have smaller time lags (see also Figure~\ref{fig:tl_sed}),
suggesting a more compact hot dust emission zone.

\begin{figure}[htp]
    \begin{center}
	\includegraphics[width=1.0\hsize]{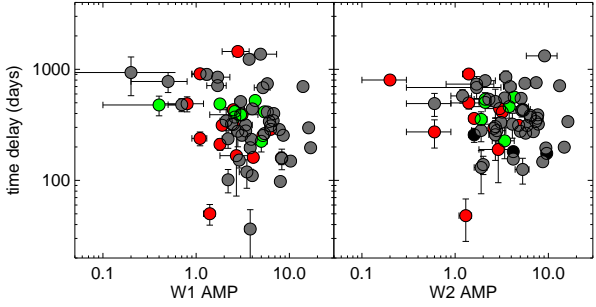}
		\caption{
		    The distribution of variation amplitudes and time lags for
		    normal (grey), WDD (green) and HDD (red) quasars.
		    }
	\label{fig:tl_amp}
    \end{center}
\end{figure}

In Figure~\ref{fig:tl_amp}, we compare the variation amplitudes and time lags
for different populations of quasars. First, there is no strong correlation
between these two properties. The HDD quasars have relatively smaller variation
amplitudes compared with normal quasar population, as expected from their
different SED features. However, the WDD quasars do not show much of a
difference. This is possibly related to their similar SEDs to normal quasars at
these wavelengths.

We have carried out Kolmogorov-Smirnov (K-S) tests to check if the distribution
of the variation amplitudes and time lags among dust-deficient quasars and
normal quasars are different at a significant level. The K-S probabilities that
describe the likelihood that the two samples do not differ significantly can be
found in Table~\ref{tab:K-S_prob}. As demonstrated in \cite{Lyu2017}, the
distribution of AGN luminosities among HDD quasars is not significantly
different from that for normal quasars. For the mid-IR time lags, the K-S
probabilities of the HDD against normal quasars are close to unity
($p\sim0.95$). In other words, the time lag distributions of the W1 and W2
bands between the HDD population and the normal population are similar (but
their dependences on AGN luminosity are different, as shown in
Table~\ref{tab:fit_fix_slope}).  However, the variation amplitude distribution
between HDD and normal quasars are significantly different ($p\sim0.009$ for W1
and $p\sim0.005$ for W2), which can be expected if HDD quasars have fewer hot
dust grains heated by the AGN, as previously proposed based on SED features in
\cite{Lyu2017}.  For WDD quasars, although the K-S test results do not
support any significant differences with normal quasars, the situation is
uncertain since WDD quasars have higher redshifts and higher luminosities.
These two features would impact the observed torus size in the opposite
directions, hindering meaningful comparisons.


\begin{deluxetable}{cccc}
    \tablewidth{1.0\hsize}
    \tablecolumns{4}
    \tablecaption{K-S probabilities of the HDD and WDD quasars against normal quasars\label{tab:K-S_prob}
    }
    \tablehead{
        \colhead{Measurement} & \colhead{HDD}  & \colhead{WDD} & \colhead{HDD+WDD}
    }
    \startdata
    W1 AMP      &     {\bf 0.009}\tablenotemark{*}     & 0.315 &   0.014 \\
    W1 lag    &       0.947                          & 0.046 &   0.439 \\
    W2 AMP      &     {\bf 0.005 }                     & 0.227 &   {\bf 0.005} \\
    W2 lag    &       0.949                          & 0.351 &   0.847
    \enddata
    \tablenotetext{*}{We indicate significant differences ($p<0.01$) in bold.}
\end{deluxetable}

\subsection{Relation between Dusty Torus and Broad-line Regions}

\begin{deluxetable*}{ccccccc}
    \tablewidth{1.0\hsize}
    \tablecolumns{7}
    \tablecaption{Time lag comparison of BLR and mid-IR dust for 12 PG quasars \label{tab:BLR-torus-size}
    }
    \tablehead{
	\colhead{Name} & 
	\colhead{Type}  & 
	\colhead{$\log(L_\text{AGN, bol}/L_\odot)$}  & 
	\colhead{$\Delta t_{H\alpha}$}  & 
	\colhead{$\Delta t_{H\beta}$} & 
	\colhead{$\Delta t_\text{W1}$} &
	\colhead{$\Delta t_\text{BLR}/\Delta t_\text{TOR, W1}$}
    }
    \startdata
    PG 0026+129  & HDD  &  12.44 & $132^{+29}_{-31}$  &  $125^{+29}_{-31}$      & 577.0$\pm$20.4  & 0.47 \\
    PG 0052+251  & HDD  &  12.56 & $211^{+66}_{-44}$  &   $99^{+30}_{-31}$      & 360.0$\pm$75.1  & 0.58 \\
    PG 0804+761  & WDD  &  12.55 & $193^{+20}_{-17}$  &  $151^{+26}_{-24}$ 	& 659.9$\pm$40.4   & 0.30 \\
    PG 8444+349  & HDD? &  11.87 &  $39^{+16}_{-16}$  &   $13^{+14}_{-11}$      & 224.0$\pm$26.4  & 0.17 \\
    PG 0953+414  & WDD  &  12.93 &                    &  $187^{+27}_{-33}$   	& 913.0$\pm$68.7   & 0.20 \\
    PG 1211+143  & NORM &  12.13 & $116^{+38}_{-46}$  &  $103^{+32}_{-44}$      & 365.6$\pm$22.4  & 0.32 \\
    PG 1229+204  & NORM &  11.53 &  $71^{+39}_{-46}$  &   $36^{+32}_{-18}$      & 266.3$\pm$2.3  & 0.27 \\
    PG 1307+085  & HDD  &  12.16 & $179^{+94}_{-145}$ & $108^{+46}_{-115}$      & 357.6$\pm$25.2  & 0.50 \\
    PG 1411+442  & NORM &  12.02 & $103^{+40}_{-37}$  &  $118^{+72}_{-71}$      & 441.7$\pm$26.2  & 0.27 \\
    PG 1426+015  & NORM &  12.09 &  $90^{+46}_{-68}$  &  $115^{+49}_{-68}$      & 264.0$\pm$34.3  & 0.44 \\
    PG 1613+658  & NORM &  12.42 &  $43^{+40}_{-22}$  &   $44^{+20}_{-23}$      & 371.9$\pm$86.3  & 0.12 \\
    PG 2130+099  & NORM &  11.90 & $237^{+53}_{-28}$  &         		        &  525.3$\pm$18.5  & 0.45
    \enddata
\end{deluxetable*}

In the classical AGN unification scheme, the broad-line region (BLR) is
dust-free and relatively small, well-separated from the outer dusty torus
\citep{Antonucci1993a, Urry1995}. In recent years, proposals have been made
that the BLR is a failed dusty wind from the outer accretion disc
\citep{Czerny2011}. Under this picture, \cite{Baskin2018} explored the expected
dust properties and corresponding BLR structure, and argued for the presence of
large graphite grains ($a\gtrsim0.3\mum$) down to the observed size of the BLR.
By comparing the reverberation mapping analysis of the BLR and dusty torus
sizes, these statements can be tested.

Previous comparisons of K-band and the UV-optical emission-line time lags
relative to the optical continuum  have suggested the torus inner radius is
larger than the BLR region, with $R_\text{BLR}\sim0.5 R_\text{TOR, K}$ for
Seyfert-1 nuclei \citep{Suganuma2006, Koshida2014}. This result was obtained by
comparing the time lag - AGN luminosity correlation for the BLR and torus from
different samples. A very small number of objects have size estimates of both
the torus (in the near-IR band) and the BLR from dust and emission line
reverberation mappings at similar observing epochs.  \cite{Clavel1989} reported
$R_{\rm BLR, MgII}/R_\text{TOR, K}\sim0.4$ for Fairall~9, a bright quasar with
$L_\text{AGN, bol}\sim3\times10^{12} L_\odot$.  \cite{Pozo2015} found $R_{\rm
BLR, H\alpha}/R_\text{TOR, K}\sim0.4$ for PGC~50424, a Seyfert-1 nucleus with
$L_\text{AGN, bol}\sim3\times10^{10}~L_\odot$. \cite{Ramolla2018} studied the
optical and near-IR time lags of 3C~120, a type-1 AGN with $L_\text{AGN,
bol}\sim 4\times 10^{11}~L_\odot$, and argued for $R_{\rm BLR,
H\alpha}/R_\text{TOR, K}\sim0.7$.

With our mid-IR reverberation analysis and previous BLR size measurements of
the PG sample, we can explore the BLR and torus relation in a large sample of
bright quasars. \cite{Kaspi2000} studied the variations in the optical Balmer
emission lines and the continuum emission of 28 PG quasars, and reported time
lag measurements for 17 objects. Among these 17 objects, we have detected
mid-IR time lags in 15. The results are summarized in
Table~\ref{tab:BLR-torus-size}. The mean value of $\Delta t_{\rm BLR}/\Delta
t_{\rm TOR, 3.0~\mum}\sim0.23\pm0.10$, with the maximum value 0.44, minimum
value at 0.06 and the median at 0.23.

\begin{figure}[htp]
    \begin{center}
	\includegraphics[width=1.0\hsize]{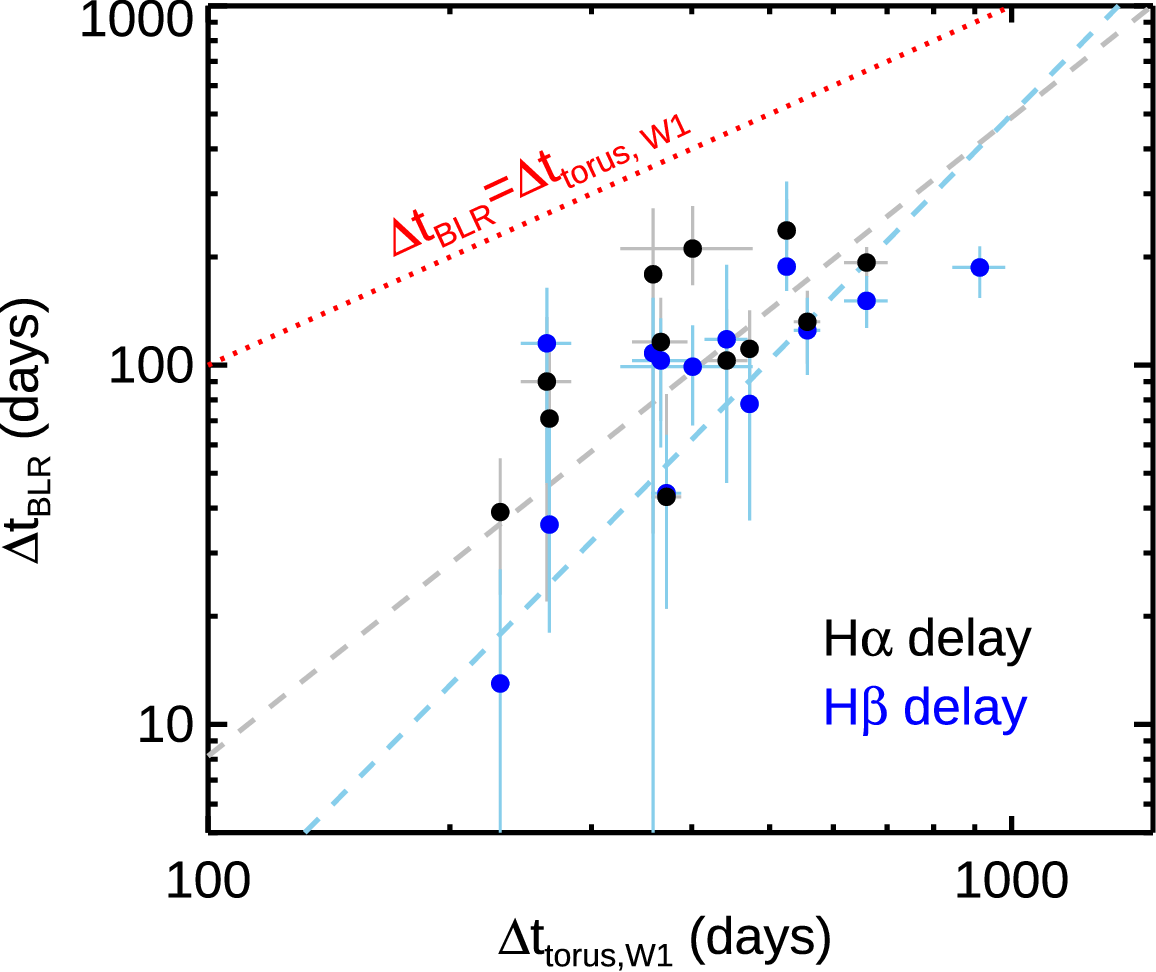}
		\caption{
		    Comparison of the infrared light lags in the {\it WISE} W1
		    band ($\sim3.4~\mum$) from this work and the broad emission
		    line lags of 17 PG quasars from \cite{Kaspi2000}.  We fit
		    linear functions for H$\beta$ lag vs. W1 lag (blue dots)
		    and H$\alpha$ lag vs. W1 lag (black dots) separately, shown
		    as blue and black dashed lines (see text for details).
		    The red dotted line represents where the torus IR time lags
		    equal the BLR time lags.
		    }
	\label{fig:BLR}
    \end{center}
\end{figure}

In Figure~\ref{fig:BLR}, we explore if there are correlations between $\Delta
t_\text{BLR}$ and $\Delta t_\text{TOR, W1}$. Linear fits result in:
\begin{equation}
    \Delta t_{\rm BLR, H\alpha} = (0.37\pm0.06)\Delta t_{\rm TOR, W1} + (-43.00\pm26.49)~,
\end{equation}
and
\begin{equation}
    \Delta t_{\rm BLR, H\beta} = (0.28\pm0.05)\Delta t_{\rm TOR, W1} + (-44.18\pm21.98)~.
\end{equation}
In log-log space, we find
\begin{equation}
    \Delta t_{\rm BLR, H\alpha} = 10^{-2.64\pm0.29} (\Delta t_{\rm TOR, W1})^{1.77\pm 0.11}~,
\end{equation}
and
\begin{equation}
    \Delta t_{\rm BLR, H\beta} = 10^{-4.12\pm0.26} (\Delta t_{\rm TOR, W1})^{2.27\pm 0.09}
\end{equation}.

Given the mean value of $\Delta t_{\rm BLR}/\Delta t_{\rm TOR, 3\mum}\sim0.23$
for the PG quasars and $\Delta t_{\rm TOR, K}/\Delta t_{\rm TOR, W1}\sim$0.6
from our fits of \cite{Koshida2014} measurements (Section~\ref{sec:lag-lum}),
we find  $R_\text{BLR}/R_\text{TOR, K}\sim0.4$. This result statistically
confirms the values obtained in the three cases discussed above. It suggests
that the BLR region is only slightly smaller than the size of the K-band dust
emission, providing strong support for the picture proposed by
\cite{Baskin2018} that large graphite dust grains might survive down to
$R_\text{BLR}$ and that the very hot dust emission comes from regions very
close to the BLR.

\section{Summary}\label{sec:summary}

We present the first statistical mid-IR dust reverberation mapping study of the
quasar torus with an innovative usage of the long-term time-series data from
the mid-IR \textit{WISE}/\textit{NEOWISE} mission and several optical transient
surveys (CRTS, ASAS-SN and PTF) over a timescale of about 8 yr. Compared with
previous dedicated targeted observations, the data in these public archives
have various complications and uncertainties. We have developed procedures to
maximize the usefulness of these data and characterize the dust reverberation
signals by comparing the mid-IR and optical light curves with a simple linear
model. The success of this approach has been demonstrated by our detection of
mid-IR time dust lags to the optical variation signals in 67 out of 87 $z<0.5$
PG quasars. Most of the remaining quasars have data quality issues or
featureless mid-IR light curves, making analysis ambiguous. Our key results for
AGN variability at 1--3~$\mum$ are as follows.
\begin{enumerate}[label=\arabic*.]
    \item The majority of PG quasars ($\sim77\%$) have convincing dust
	    reverberation signals with time-lags that follow the expected
		$\Delta t\propto L_{\rm AGN}^{0.5}$ relation. For the {\it
		WISE} W1 and W2 filter bandpasses, we find
	\begin{equation*}
           \Delta t_{\rm torus, W1}/{\rm day} = 10^{2.10\pm0.06} (L_{\rm AGN, SED}/10^{11}L_\odot) ^{0.47\pm0.06}
	\end{equation*}
		and
	\begin{equation*}
           \Delta t_{\rm torus, W2}/{\rm day} = 10^{2.20\pm0.06} (L_{\rm AGN, SED}/10^{11}L_\odot) ^{0.45\pm0.05}~.
	\end{equation*}
	Combined with previous studies in the near-IR, the AGN IR time lags share
	the same scaling relation with AGN luminosity over four orders
	of magnitude, indicating that similar circumnuclear dust
	structures are common to AGNs;

    \item By combining our mid-IR results of PG quasars and previous K-band
	analysis of Seyfert-1 nuclei by \cite{Koshida2014}, we provide the
	first multiwavelength torus size constraints. Assuming the same $\Delta
		t\propto L_{\rm AGN}^{0.5}$ relation for different bands, we
		find average time lag ratios of $\Delta t_{\rm K}: \Delta
		t_{\rm W1}: \Delta t_{\rm W2}\sim0.6:1.0:1.2$ ($\lambda_{\rm
		rest}\sim$2.1, 2.9, 4.0~$\mu$m);

    \item With the variability amplitudes derived from
	reverberation model fitting, it is possible to put some crude
	constraints on surface density profiles of AGN-heated hot dust grains.
	We find $\Sigma(r) \propto r^{0.3}$ at K-band, $\Sigma(r) \propto
	r^{-0.7}$ at {\it WISE} W1-band and $\Sigma(r) \propto r^{-1.0}$ at
	W2-band, indicating the concentration of the hottest dust grains in the
	innermost regions of the torus;

    \item For the same AGN luminosity, the mid-IR emission region sizes of
	dust-deficient quasars are only 60--70\% of those found for normal
	quasars. In addition, their relative size differences in the {\it WISE}
	W1 and W2 bands are smaller, possibly indicating compact dust
	structures. Meanwhile, the difference between the optical and mid-IR
	variability amplitudes are smaller for hot-dust-deficient quasars than
	for normal quasars, indicating a smaller amount of dust reprocessing
	the accretion disk emission. On the other hand, there is no difference
	in the ratio of mid-IR to optical variability amplitude between
	warm-dust-deficient and normal quasars. These results are roughly
	consistent with their SED features and support their different torus
	structures as argued by \cite{Lyu2017};

    \item Using previous measurements of BLR size measurements in the
	literature, the mean value of $R_{\rm BLR}/R_{\rm torus,
	W1}\sim0.23\pm0.10$. Given that the relative IR time lags between W1
	and K bands,  $R_{\rm BLR}/R_{\rm torus, K}\sim0.4$, indicating that the
	dust torus is located just outside of AGN BLRs. 
\end{enumerate}

With the multi-epoch infrared data from {\it Spitzer} and WISE, we also studied
the AGN variability behavior at 10--24~$\mum$ and explored its relationship to
that at 3--5$\mum$. The most important results are:
\begin{enumerate}[label=\arabic*.]
	  \setcounter{enumi}{5}
    \item With very few exceptions, significant AGN IR variability at
	$\lambda\gtrsim10~\mum$ is only found among blazars and/or
	flat-spectrum radio sources (FSRQs).  Considering the limited number of
	24~$\mum$ observations, it is likely that all of these sources vary in
	the mid-IR.  In contrast with blazars and FSRQ, we have found only one
	steep-spectrum radio quasar (SSRQ) to probably be variable at
	24~$\mum$. Such a difference can be explained under the radio-loud AGN
	unification scheme. Since SSRQs have larger inclination angles to our
	line of sight compared with FSRQs, realistic beaming effects greatly
	enhance the variability of the latter population.

  \item Besides blazars and FSRQs, the vast majority of AGNs do not show
      variability at 24$\mum$. That is the IR variability amplitude decreases
      quickly as a function of wavelength. Compared with W1 band, the W2 band
      variability is reduced to 90\% and the 24$\mum$ is less than 10\%. Given
      the fact that typical W1-band variation RMS$\lesssim$0.1 mag for most
      quasars, the corresponding flux change at $\sim24~\mum$ is no more than
      0.01 mag. Only in very rare cases is a possible dust reverberation signal
      detected (e.g., PG 1535+547).

\end{enumerate}

\acknowledgements

We thank the referee, Makoto Kishimoto, for his constructive report.

This work was supported by NASA grants NNX13AD82G and 1255094.

This publication has made use of data products from the {\it Wide-field
Infrared Survey Explorer}, which is a joint project of the University of
California, Los Angeles, and the Jet Propulsion Laboratory/California Institute
of Technology, funded by the National Aeronautics and Space Administration.
This publication also makes use of data products from NEOWISE, which is a
project of the Jet Propulsion Laboratory/California Institute of Technology,
funded by the Planetary Science Division of the National Aeronautics and Space
Administration. This work is based in part on archival data obtained with the
Spitzer Space Telescope, which is operated by the Jet Propulsion Laboratory,
California Institute of Technology under a contract with NASA. Support for this
work was provided by an award issued by JPL/Caltech.

The CSS survey is funded by the National Aeronautics and Space Administration
under Grant No. NNG05GF22G issued through the Science Mission Directorate
Near-Earth Objects Observations Program.  The CRTS survey is supported by the
U.S.~National Science Foundation under grants AST-0909182 and AST-1313422.
ASAS-SN is supported by the Gordon and Betty Moore Foundation through grant
GBMF5490 to the Ohio  State University  and NSF grant AST-1515927.  Development
of ASAS-SN has been supported by NSF grant AST-0908816, the Mt.  Cuba
Astronomical Foundation, the Center for Cosmology and Astro Particle Physics at
the Ohio State University, the Chinese Academy of Sciences South America Center
for Astronomy (CASSACA), the Villum Foundation, and George Skestos.  Data from
the Steward Observatory spectro-polarimetric monitoring project were used. This
program is supported by Fermi Guest Investigator grants NNX08AW56G, NNX09AU10G,
NNX12AO93G, and NNX15AU81G.  We also acknowledge with thanks the 3C 273
observations from the {\it AAVSO International Database} contributed by
observers worldwide and used in this research.

This work is based in part on observations made with the Spitzer Space
Telescope, which is operated by the Jet Propulsion Laboratory, California
Institute of Technology under a contract with NASA. The Combined Atlas of
Sources with Spitzer IRS Spectra (CASSIS) is a product of the IRS instrument
team, supported by NASA and JPL.

\software{SExtractor \citep{SExtractor}, DAOPHOT package \citep{stetson1987},
IRAF \citep{Tody1986, Tody1993}, MPFIT \citep{MPFIT}, JAVELIN \citep{Zu2013},
Matplotlib \citep{matplotlib}}

\appendix

\twocolumngrid

\section{Photometric Stability}
\label{app:pho_stable}

None of the CRTS, ASAS-SN and \textit{WISE}/\textit{NEOWISE} missions has put AGN
variability study as its main science objective; thus, we need to assess the
noise characteristics of their photometric measurements. For our time-series
analysis, a clear idea of the instrument photometric stability is of paramount
importance. With careful selections of photometric standard stars in the
optical and mid-IR bands, we construct the light curves of non-variable sources
and explore the systematics and noise characteristics in this appendix.

\subsection{Optical Data from CRTS and ASAS-SN}

\begin{figure*}[htp]
    \begin{center}
	\includegraphics[width=1.0\hsize]{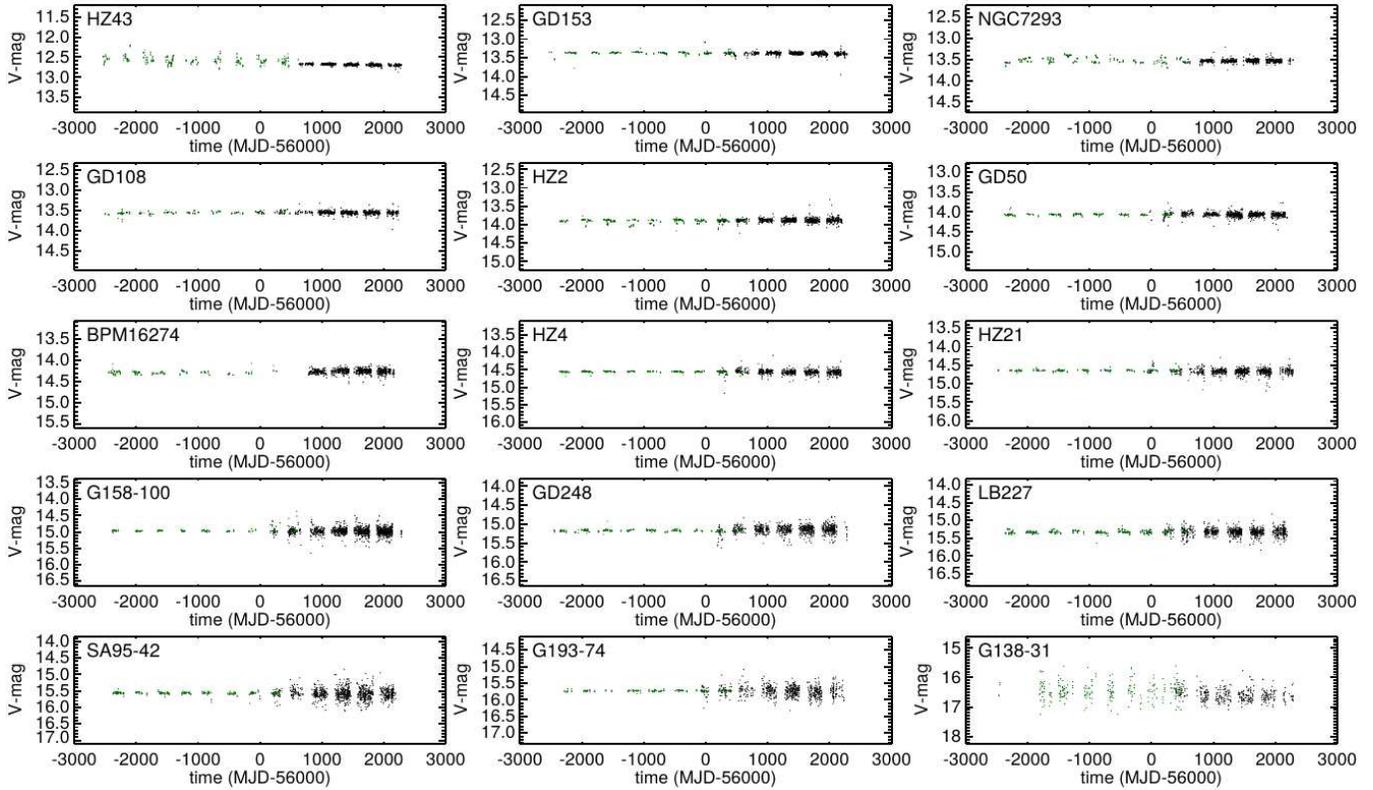}
		\caption{ Optical light curves of 15 optical spectrophotometric
		standard stars. The CRTS data are in green and the ASAS-SN data
		are in black.}
	\label{fig:opt_standard}
    \end{center}
\end{figure*}

From the Optical and UV Spectrophotometric Standard Stars web page at
ESO,\footnote{\url{http://www.eso.org/sci/observing/tools/standards/spectra/stanlis.html}}
we have selected 15 stars to study the optical photometric stability of the
CRTS and ASAS-SN data. Besides GD 108, a sub-dwarf, all of these stars are
white dwarfs. We have compiled their optical measurements from the CRTS and
ASAS-SN archives and constructed the corresponding optical light curves. Since
the CRTS data were taken without a photometric filter, we matched the CRTS
light curve to those derived from ASAS-SN by introducing a constant offset. We
also rejected photometry outliers that showed a discrepancy of $>0.5$ mag from
the average values. The final light curves are plotted in
Figure~\ref{fig:opt_standard}.

We calculate the mean magnitudes and the root-mean-square (RMS) offsets of the
CRTS and ASAS-SN light curves separately and summarize the results in
Table~\ref{tab:opt-standard}. Besides the uncertainties caused by standard
photometry measurements, any issues with system stability can further increase
the observed magnitude RMS offset, and these values could be related as
\begin{equation}\label{eqn:error}
    RMS_{M-\bar M}^2 = RMS_{\Delta M}^2+ \sigma_{s.s.}^2 ~~.
\end{equation}
On average, we get $\sigma_{s.s.}\sim0.018$ mag for ASA-SN data and
$\sigma_{s.s.}\sim$0.023 mag for CRTS data.

\begin{deluxetable*}{lccccccc}
    \tablewidth{1.0\hsize}
    \tablecolumns{8}
    \tablecaption{Measurements of Optical Standard Stars\label{tab:opt-standard}
    }
    \tablehead{
	\colhead{Name} & 
	\colhead{$M_V$}  & 
	\multicolumn{3}{c}{CRTS}  & 
	\multicolumn{3}{c}{ASAS-SN}  \\
	 &
	 &
	\colhead{$\bar M$}  & 
	\colhead{RMS$_{M-\bar M}$}  & 
	\colhead{RMS$_{\Delta M}$}  & 
	\colhead{$\bar M$}  & 
	\colhead{RMS$_{M-\bar M}$}  & 
	\colhead{RMS$_{\Delta M}$}  \\
	\colhead{(1)} &
	\colhead{(2)} &
	\colhead{(3)} &
	\colhead{(4)} &
	\colhead{(5)} &
	\colhead{(6)} &
	\colhead{(7)} &
	\colhead{(8)}
    }
    \startdata
HZ43        &	12.91 & 12.577  & 0.092 &  0.009  &  12.687  & 0.023 &  0.015 \\
GD153       &	13.35 & 13.371  & 0.036 &  0.011  &  13.384  & 0.041 &  0.022 \\
NGC7293     &	13.51 & 13.527  & 0.065 &  0.013  &  13.534  & 0.035 &  0.030 \\
GD108       &	13.56 & 13.551  & 0.034 &  0.012  &  13.548  & 0.039 &  0.027 \\
HZ2         &	13.86 & 13.902  & 0.051 &  0.012  &  13.882  & 0.053 &  0.034 \\
GD50        &	14.06 & 14.070  & 0.021 &  0.013  &  14.068  & 0.052 &  0.038 \\
BPM16274    &	14.20 & 14.285  & 0.040 &  0.018  &  14.257  & 0.051 &  0.047 \\
HZ4         &	14.52 & 14.555  & 0.016 &  0.017  &  14.568  & 0.066 &  0.048 \\
HZ21        &	14.68 & 14.646  & 0.022 &  0.019  &  14.670  & 0.077 &  0.054 \\
G158-100    &	14.89 & 14.968  & 0.022 &  0.018  &  14.980  & 0.109 &  0.076 \\
GD248       &	15.09 & 15.170  & 0.025 &  0.020  &  15.146  & 0.102 &  0.083 \\
LB227       &	15.34 & 15.326  & 0.038 &  0.019  &  15.324  & 0.101 &  0.083 \\
SA95-42     &	15.61 & 15.567  & 0.040 &  0.023  &  15.581  & 0.156 &  0.123 \\
G193-74     &	15.70 & 15.732  & 0.019 &  0.023  &  15.743  & 0.156 &  0.130 \\
G138-31     &   16.14 & 16.461  & 0.297 &  0.245  &  16.594  & 0.195 &  0.177 
\enddata
    \tablecomments{Col. (1): star name; Col. (2): V-band magnitude given in the
    ESO standard star list; Col. (3): mean magnitude of all CRTS measurements,
    an offset is included to match the CRTS light curve; Col. (4): RMS value
    for all the magnitude deviations relative to the mean magnitude for CRTS
    data; Col. (5): RMS value for all magnitude errors reported in the CRTS
    catalog; Col. (6)-(8): similar to Col. (3)-(5) but for ASAS-SN data.
    }
\end{deluxetable*}

\subsection{Mid-IR from {\it WISE}/NEOWISE}

\begin{figure*}[htp]
    \begin{center}
	\includegraphics[width=1.0\hsize]{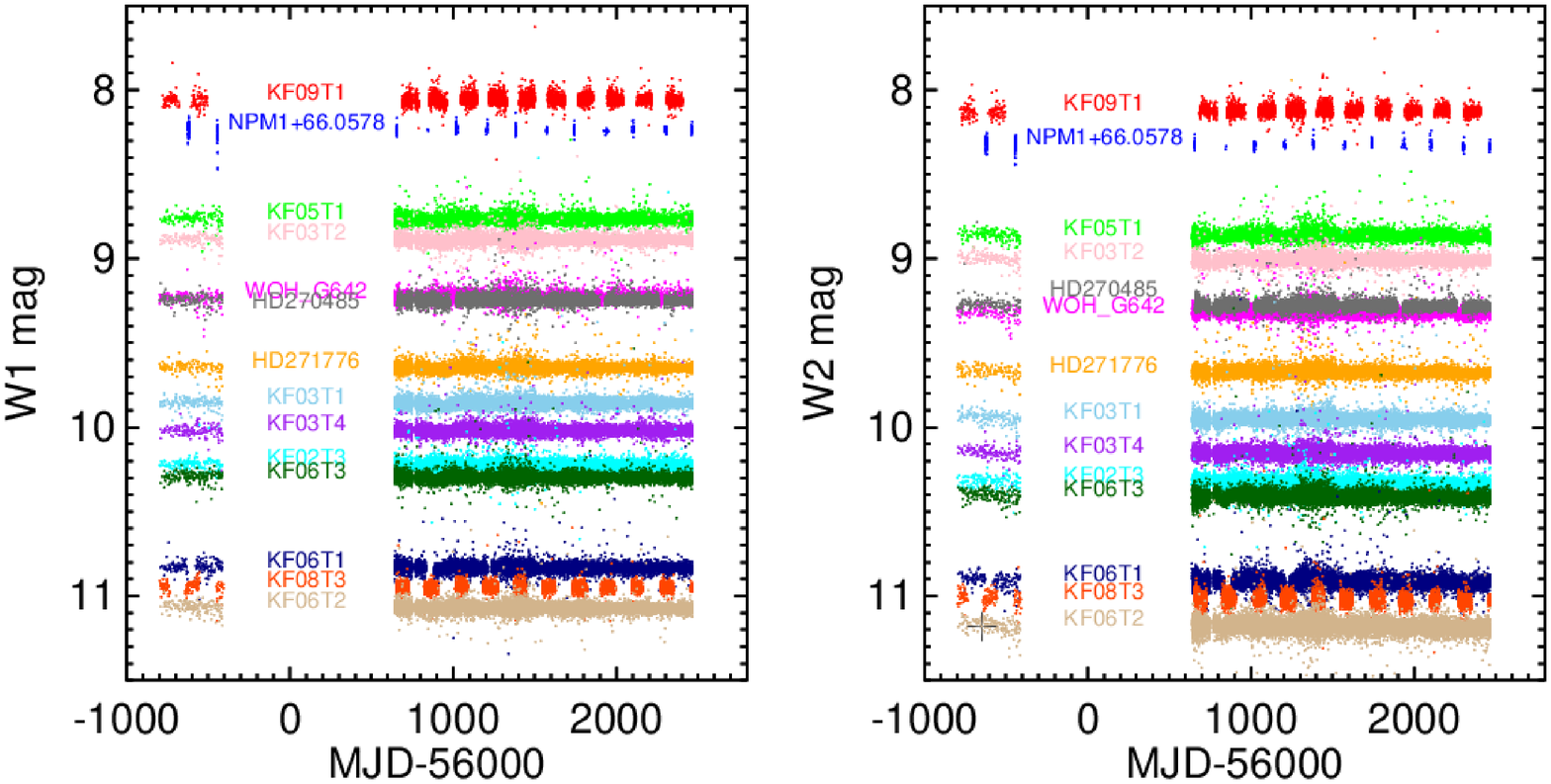}
		\caption{{\it WISE} W1 (left) and W2 (right) light curves of 14 IR
		calibration standard stars }
	\label{fig:wise_standard}
    \end{center}
\end{figure*}

With the {\it WISE}/NEOWISE catalogs, we examined the photometric
stability of W1 and W2 bands as a function of time. Fourteen mid-IR standard
stars from the {\it Spitzer}/IRAC primary
calibrators\footnote{\url{https://irsa.ipac.caltech.edu/data/SPITZER/docs/irac/iracinstrumenthandbook/17/}}
and {\it WISE} (W1, W2) Calibration
stars\footnote{\url{http://wise2.ipac.caltech.edu/docs/release/allsky/expsup/sec4_4ht7.html}}
were chosen for this purpose. The constructed mid-IR light curves can be seen
in Figure~\ref{fig:wise_standard}. We summarize the RMS magnitude offsets and
RMS errors of W1 and W2 bands during the {\it WISE} and NEOWISE mission
in
Table~\ref{tab:mid-standard}.

With equation~\ref{eqn:error}, we get $\sigma_{s.s.}\sim0.029$ mag for {\it
WISE} and 0.016 mag for {NEOWISE}. Since we have averaged the 10$\sim$20
observations for the mid-IR light curve construction
(Section~\ref{sec:wise-lc-construction}), the contribution to the measurement
uncertainties from the system instability is expected to be $<$0.009 mag for
{\it WISE} and $<0.005$ mag for NEOWISE.

\begin{deluxetable*}{l|rcccc|rcccc}
    \tablewidth{1.0\hsize}
    \tablecolumns{11}
    \tablecaption{Measurements of Mid-IR Standard Stars\label{tab:mid-standard}
    }
    \tablehead{
	\colhead{Name} & 
	\multicolumn{5}{c}{W1 band}  & 
	\multicolumn{5}{c}{W2 band}  \\
	  & 
	  &
	\multicolumn{2}{c}{ALLWISE} &
	\multicolumn{2}{c|}{NEOWISE} &
	  &
	\multicolumn{2}{c}{ALLWISE} &
	\multicolumn{2}{c}{NEOWISE} \\
	 & 
	\colhead{$\bar M$}  & 
	\colhead{RMS$_{M-\bar M}$}  & 
	\colhead{RMS$_{\Delta M}$} &
	\colhead{RMS$_{M-\bar M}$}  & 
	\multicolumn{1}{c|}{RMS$_{\Delta M}$}  &
	\colhead{$\bar M$}  & 
	\colhead{RMS$_{M-\bar M}$}  & 
	\colhead{RMS$_{\Delta M}$} &
	\colhead{RMS$_{M-\bar M}$}  & 
	\colhead{RMS$_{\Delta M}$}  \\
	\colhead{(1)} &
	\colhead{(2)} &
	\colhead{(3)} &
	\colhead{(4)} &
	\colhead{(5)} &
	\colhead{(6)} &
	\colhead{(7)} &
	\colhead{(8)} &
	\colhead{(9)} &
	\colhead{(10)} &
	\colhead{(11)} 
    }
    \startdata
   KF09T1           &   8.055   &  0.038 & 0.024 &  0.022 &0.016 &    8.123  &   0.037 & 0.021 &  0.022 & 0.016 \\
   NPM1+66.0578     &   8.237   &  0.041 & 0.024 &  0.020 &0.016 &    8.323  &   0.032 & 0.021 &  0.020 & 0.016 \\
   KF05T1           &   8.767   &  0.035 & 0.024 &  0.019 &0.016 &    8.863  &   0.027 & 0.022 &  0.019 & 0.016 \\
   KF03T2           &   8.891   &  0.029 & 0.024 &  0.020 &0.016 &    9.007  &   0.033 & 0.022 &  0.020 & 0.016 \\
   HD271776         &   9.647   &  0.032 & 0.024 &  0.023 &0.016 &    9.671  &   0.034 & 0.022 &  0.023 & 0.016 \\
   WOH\_G642        &   9.229   &  0.042 & 0.024 &  0.021 &0.016 &    9.319  &   0.036 & 0.022 &  0.021 & 0.016 \\
   HD270485         &   9.248   &  0.030 & 0.025 &  0.031 &0.016 &    9.277  &   0.052 & 0.022 &  0.031 & 0.016 \\
   KF03T1           &   9.858   &  0.033 & 0.025 &  0.022 &0.016 &    9.952  &   0.049 & 0.023 &  0.022 & 0.016 \\
   KF03T4           &  10.019   &  0.034 & 0.025 &  0.021 &0.016 &   10.155  &   0.035 & 0.023 &  0.021 & 0.016 \\
   KF02T3           &  10.223   &  0.037 & 0.025 &  0.028 &0.017 &   10.332  &   0.035 & 0.025 &  0.028 & 0.017 \\
   KF06T3           &  10.298   &  0.030 & 0.025 &  0.022 &0.017 &   10.410  &   0.032 & 0.025 &  0.022 & 0.017 \\
   KF06T1           &  10.832   &  0.036 & 0.026 &  0.023 &0.018 &   10.912  &   0.039 & 0.027 &  0.023 & 0.018 \\
   KF08T3           &  10.946   &  0.033 & 0.027 &  0.021 &0.018 &   11.020  &   0.039 & 0.028 &  0.021 & 0.018 \\
   KF06T2           &  11.072   &  0.031 & 0.026 &  0.031 &0.018 &   11.179  &   0.042 & 0.029 &  0.031 & 0.018
\enddata
    \tablecomments{Col. (1): star name; Col. (2): average W1 magnitude of all
    {\it WISE}/NEOWISE measurements; Col. (3): RMS magnitude offset relatively to the
    mean W1 magnitude for measurements taken during the ALLWISE mission; Col.
    (4): RMS value of all the W1 magnitude errors reported in the ALLWISE
    Multiepoch Photometry Table; Col. (5)-(6): similar to Col. (3)-(4) but for
    W1 measurements taken during the NEOWISE mission; Col. (7)-(11): similar to
    Col. (2)-(6), but for W2 band measurements.
    }
\end{deluxetable*}

\section{A simple model for relating time lags to the torus structures}
\label{app:model}

\begin{figure}[htp]
    \begin{center}
	\includegraphics[width=1.0\hsize]{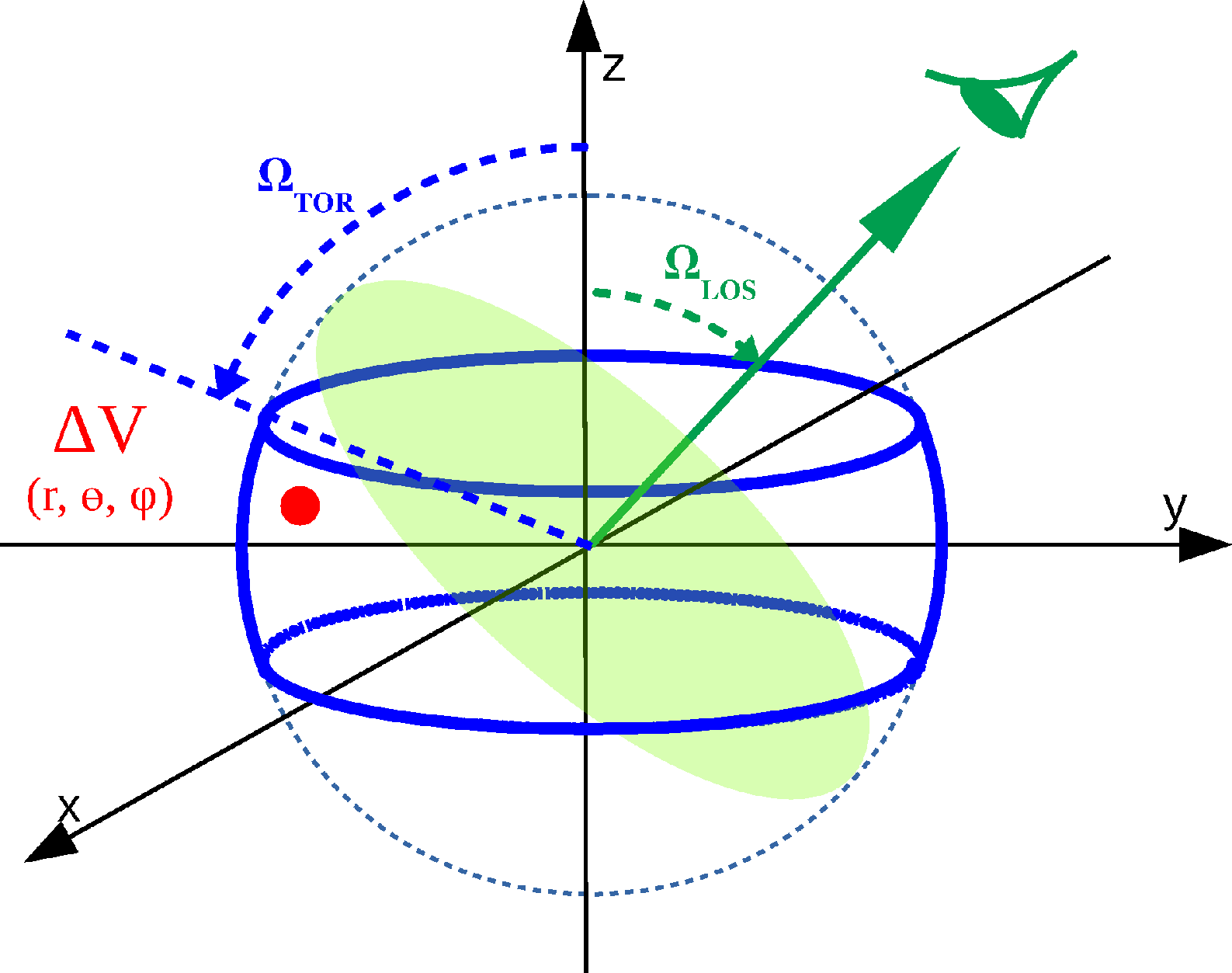}
		\caption{
		     Geometry of a dusty shell with a radius $r$
		    from the origin (accretion disk) and a half opening angle
		    $\Omega_{\rm TOR}$.  The observer has a viewing angle
		    $\Omega_{\rm LOS}$.
		    }
	\label{fig:rm_toy_model}
    \end{center}
\end{figure}

As shown in Figure~\ref{fig:rm_toy_model}, we consider a thin dust shell with a
half opening angle $\Omega_{\rm TOR}$ that surrounds the accretion disk. We
also assume that a single {\it WISE} band traces the IR emission from dust at
some similar distance, $R$, to the accretion disk. Placing the axis of symmetry
along the z-axis, we have a dusty layer with $r=R$, $\theta = [\Omega_{\rm
TOR}, \pi - \Omega_{\rm TOR}]$ and $\phi = [0, 2\pi)$. The observer is located
at an observing angle of $\Omega_{\rm LOS}$ to the z-axis and parallel to the
$z$-$y$ plane.

Now imagine a plane that is perpendicular to the observing angle and also
tangent to the dusty sphere, which can be described by the function
\begin{equation}
    [\sin\Omega_{\rm LOS}]y + [\cos \Omega_{\rm LOS}] z - R=0 ~.
\end{equation}
\noindent It takes time, $R/c$ for emission from the accretion disk to reach this plane.

For a point on the dusty sphere with coordinates $(x, y, z) =
R(\sin\theta\cos\phi,  \sin\theta\sin\phi, \cos\theta)$, the light travel
distance would be
\begin{align}
    l & = R + | R[\sin\Omega_{\rm LOS} \sin\theta\sin\phi + \cos\Omega_{\rm LOS} \cos\theta -1] | \\
      & = R [ 2 - \sin\Omega_{\rm LOS} \sin\theta\sin\phi - \cos\Omega_{\rm LOS} \cos\theta] ~.
\end{align}
An average time lag of all elements on that sphere is given by
\begin{align}
    <t> & = \frac{\int w l ds} { c \int ds} \\
        & = \frac{\int^{2\pi}_{0} \int^{\pi - \Omega_{\rm TOR}}_{\Omega_{\rm TOR}} w l R \sin\theta d\theta d\phi}{ c \int^{2\pi}_{0} \int^{\pi - \Omega_{\rm TOR}}_{\Omega_{\rm TOR}} R \sin\theta d\theta d\phi} ~,
\end{align}
where $w$ is the weight of each element, which is dependent on the dust optical
depth in the IR band along the LOS and the local strength of dust emission. If
the emission is optically thin and homogeneously distributed, we have $w=1$, so
\begin{align}
    <t> & = \text{\footnotesize  $\frac{\int^{2\pi}_{0} \int^{\pi - \Omega_{\rm TOR}}_{\Omega_{\rm TOR}} R [ 2 - \sin\Omega_{\rm LOS} \sin\theta\sin\phi - \cos\Omega_{\rm LOS} \cos\theta]
    R\sin\theta d\theta d\phi }{ c \int^{2\pi}_{0} \int^{\pi - \Omega_{\rm TOR}}_{\Omega_{\rm TOR}} R\sin\theta d\theta d\phi}$} \\
       & = \frac{2R}{c}.
\end{align}
Consequently, the average time lag between the surrounding dusty torus shell
and the accretion disk is
\begin{equation}
    \Delta t = \frac{R}{c} ~,
\end{equation}
This means the time lag is not dependent on the viewing angle in the optically
thin case.

If the dusty torus is extremely optically thick, we can only see flux from the
edge of the structure that is exposed to the observer. When the observer is
looking through the torus opening angle (type-1; $\Omega_{\rm LOS} <
\Omega_{\rm TOR}$),  the time lag can be easily calculated to be
\begin{equation} 
    \Delta t = \frac{R}{c} (1-\cos\Omega_{\rm LOS} \cos\Omega_{\rm TOR}).
\end{equation} 
It is straightforward to prove the time lag in this situation will always be
shorter than the optically-thin case for the same $R$.

In true physical situations where the torus dust is not perfectly optically
thin or optically thick in the mid-IR, we should have
\begin{equation}
    \frac{R}{c} > \Delta t > \frac{R}{c} (1-\cos\Omega_{\rm LOS} \cos \Omega_{\rm TOR})
\end{equation}
Knowing the fraction of type-2 objects in the quasar population
($\sim$0.5--0.6; \citet{Reyes2008}), we have roughly
$\Omega_{\rm TOR}= \pi/3$ and
\begin{equation}
    \frac{R}{c} > \Delta t \gtrsim 0.5 \frac{R}{c}
\end{equation}

In other words, a naive size estimation from $c\Delta t$ could slightly
underestimate the true physical size up to a factor of two.  For a single ring
with $\theta=\theta_0$, the range of time lags, $\Delta t$, is given by
\begin{equation}
    max(\Delta t) - min(\Delta t) = \frac{2R}{c}\sin\Omega_{\rm LOS} \sin\theta_0~.
\end{equation}
For a complete shell, 
\begin{equation}
    max(\Delta t) - min(\Delta t) = \frac{2R}{c}\cos (\Omega_{\rm TOR} -\Omega_{\rm LOS} ) ~.
\end{equation}
Finally, the standard deviation of the time lag $\Delta t$ for a dusty shell
\begin{equation}
    \sigma (\Delta t) = \frac{R}{c}\sqrt{\frac{1}{3} \cos ^2 \Omega_{\rm TOR} + \frac{1}{2} \sin ^2\Omega_{\rm LOS} \sin ^2 \Omega_{\rm TOR}}~.
\end{equation}
This can introduce a smoothing effect on the integrated IR reverberation light curves, whose
strength is proportional to the torus size (or time lag $\Delta t$)

\section{A Revisit of the Koshida et al. Sample}\label{app:koshida}

\subsection{SED Analysis}\label{app:koshida-sed}

For consistency with the luminosity estimates for the PG sample, we examined
the broad-band IR SEDs of the Seyfert-1 sample presented by \cite{Koshida2014}
with the decomposition model introduced in \cite{Lyu2018}.  The total IR
emission is assumed to be a linear combination of the contributions from three
components: AGN-heated dust, near-IR starlight and mid-to-far IR emission from
HII regions within the host galaxy. To reduce the ambiguity of interpretation, we
have used well-tested empirical templates to describe each component. For the
AGN templates, we first used the three types of intrinsic AGN SEDs proposed in
\cite{Lyu2017} to represent the intrinsic variations of the torus. We also
allow obscuration by an IR optically-thin extended dust distribution of large
grains and the corresponding IR reprocessed emission. The latter component has
been shown to be a valid explanation for the AGN polar dust emission seen by
mid-IR interferometry analysis (see details in \citealt{Lyu2018}). We used
$\chi^2$ minimization to determine the final best-fit model. The dust-deficient
AGN template was selected only when it improved the $\chi^2$ value by a factor
greater than two compared with the normal AGN template.

\begin{figure*}[htp]
    \begin{center}
	\includegraphics[width=1.0\hsize]{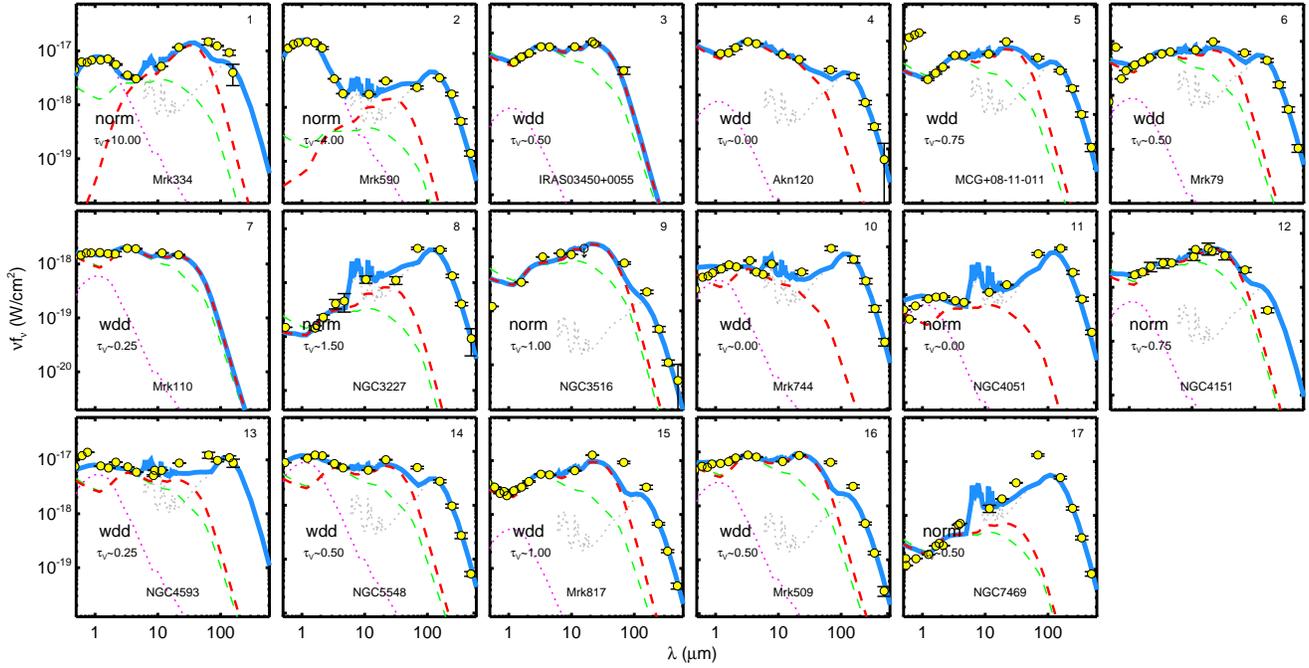}
		\caption{Best-fit results for the Seyfert-1 nuclei studied in
		\cite{Koshida2014}. Photometric data points are shown as yellow
		dots. The SED model (blue thick solid lines) is composed of the
		AGN component (red dashed lines), the stellar component
		(magenta dotted lines), and the far-IR star formation component
		(grey dotted lines). We also plot the suggested intrinsic AGN
		template (green dashed line) for each object from our SED
		fittings to compare with the observed SED.
		    }
	\label{fig:koshida-sed}
    \end{center}
\end{figure*}

\begin{figure}[htp]
    \begin{center}
	\includegraphics[width=1.0\hsize]{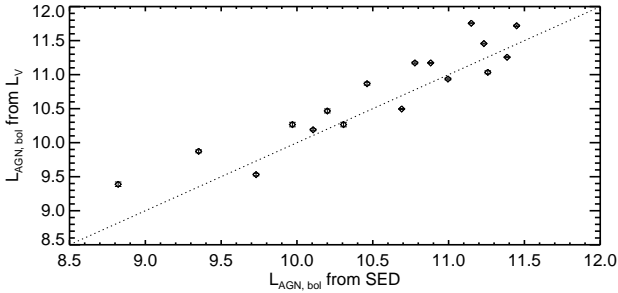}
		\caption{
		    AGN luminosity estimation comparison between SED fitting
		    and V-band photometry from \cite{Koshida2014}. The dotted line
		    is the 1:1 relation.
		    }
	\label{fig:koshida-lum-compare}
    \end{center}
\end{figure}

The results can be seen in Figure~\ref{fig:koshida-sed}.  About half of the
\cite{Koshida2014} sample have been identified as WDD AGNs and the other half
are normal AGNs. With the derived intrinsic AGN SEDs, we calculated the AGN
bolometric luminosities following Section~\ref{sec:lum-estimate} and summarized
the results in Table~\ref{tab:koshida-new}. As shown in
Figure~\ref{fig:koshida-lum-compare}, the results are consistent with those
converted from the observed $V$-band measurements provided in \cite{Koshida2014},
yielding an average $L_\text{AGN bol, SED}/L_\text{AGN bol, V}=0.98\pm0.03$.  

\subsection{Time Lag Measurements}

\begin{figure}[htp]
    \begin{center}
	\includegraphics[width=1.0\hsize]{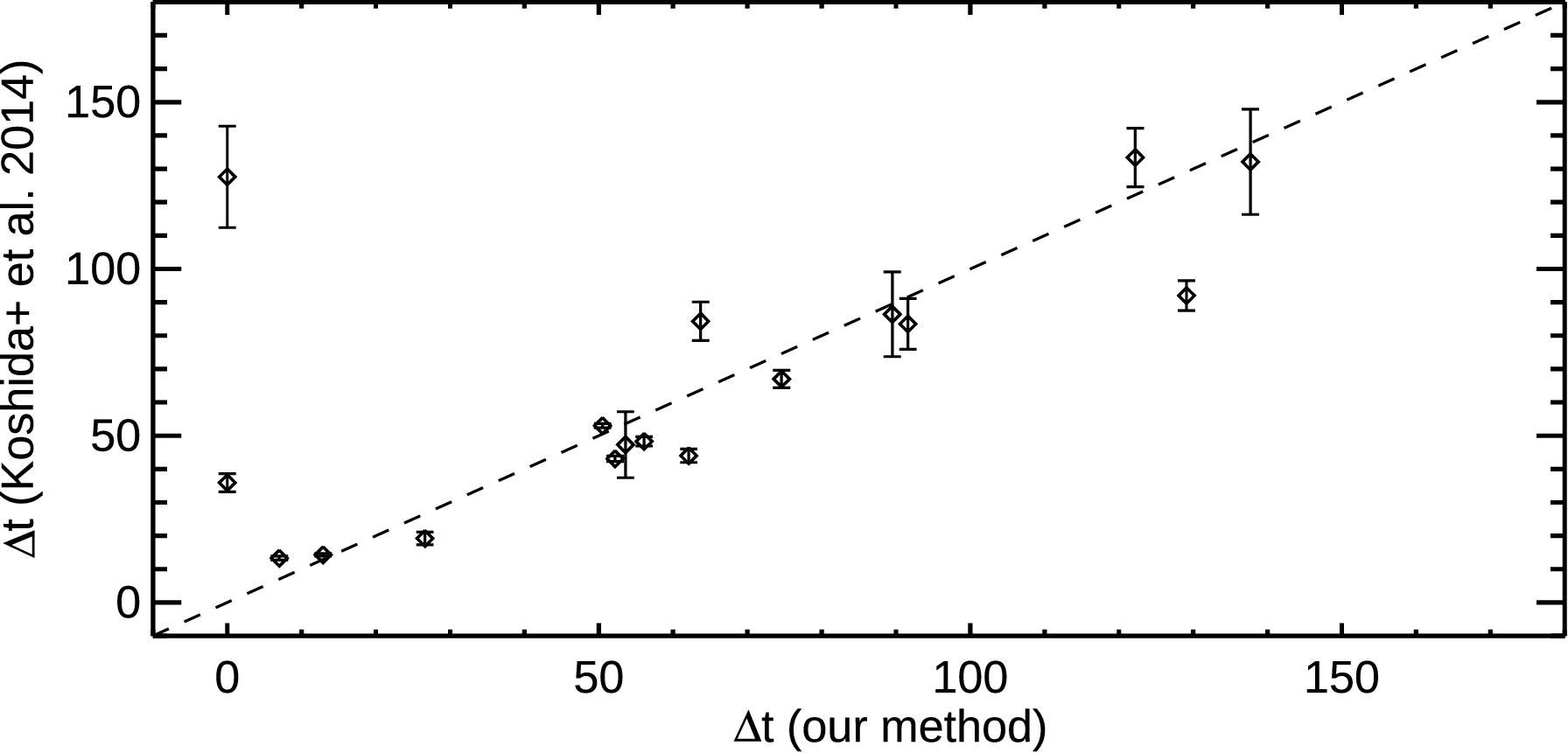}
		\caption{
		    K-band time lag comparison between our model and the \cite{Koshida2014} method.
		    The dotted line is the 1:1 relation.
		    }
	\label{fig:koshida-lag-compare}
    \end{center}
\end{figure}

To determine if our time-lag measurements are significantly different from
those determined by traditional CCF analysis, we apply the model introduced in
Section~\ref{sec:model} to fit the high-cadence light curves of 17 Seyfert-1
nuclei in \cite{Koshida2014} and present our measurements of the $K$-band time
lags in Table~\ref{tab:koshida-new}. In Figure~\ref{fig:koshida-lag-compare},
we compare the results from these two approaches. Except for Mrk 590 and Akn
120, most measurements are consistent with each other, and the average
difference is $<10\%$.  This suggests that the $R_\text{K}/R_\text{W1}\sim0.6$
is not caused by the systematics from the different measurement approaches.

\begin{deluxetable}{llcccc}
    \tablewidth{1.0\hsize}
    \tablecolumns{6}
    \tablecaption{New Time Lag Measurements of the \cite{Koshida2014} Seyfert-1 Sample \label{tab:koshida-new}
    }
    \tablehead{
	\colhead{Name} & 
	\colhead{Type}  & 
	\multicolumn{2}{c}{$\log(L_\text{AGN, bol}/L_\odot)$}  & 
	\colhead{$\Delta t $/day}  & 
	\colhead{AMP}  \\
	&
	&
	\colhead{(V-band)} &
	\colhead{(SED)} &
	&
    }
    \startdata
    Mrk 334 		& Norm 	&  11.17 & 10.88  & 137.7 &  5.1  \\
    Mrk 590  		& Norm  &  10.46 & 10.20  &   0   &  6.2  \\
    IRAS 03450+0055  	& WDD  	&  11.46 & 11.23  & 129.1 &  6.2  \\
    Akn 120  		& WDD 	&  11.87 & 11.14  &   0   &  4.0  \\
    MCG+08-11-011  	& WDD 	&  11.76 & 11.25  &  89.5 &  9.5  \\
    Mrk 79    		& WDD 	&  10.93 & 10.99  &  74.6 &  2.1  \\
    Mrk 110  		& Norm  &  11.17 & 10.77  &  63.7 &  2.8  \\
    NGC 3227  		& Norm 	&   9.85 &  9.35  &   7.0 &  5.3  \\
    NGC 3516  		& Norm	&  10.26 & 10.30  &  53.6 &  6.8  \\
    Mrk 744    		& WDD 	&   9.48 &  9.73  &  26.6 & 19.9  \\
    NGC 4051  		& NORM 	&   9.30 &  8.82  &  12.9 &  6.4  \\
    NGC 4151  		& NORM 	&  10.26 &  9.97  &  52.2 &  4.2  \\
    NGC 4593  		& WDD 	&  10.17 & 10.10  &  62.1 &  6.9  \\
    NGC 5548  		& WDD 	&  10.49 & 10.69  &  50.5 &  4.9  \\
    Mrk 817  		& NORM 	&  11.26 & 11.38  &  91.6 &  6.7  \\
    Mrk 590  		& WDD 	&  11.71 & 11.44  & 122.2 &  2.4  \\
    NGC 7469 		& NORM 	&  10.87 & 10.46  &  56.1 &  5.1     
    \enddata
\end{deluxetable}

\bibliographystyle{apj.bst}

\begin{thebibliography}{91}
\expandafter\ifx\csname natexlab\endcsname\relax\def\natexlab#1{#1}\fi

\bibitem[{{Almeyda} {et~al.}(2017){Almeyda}, {Robinson}, {Richmond}, {Vazquez},
  \& {Nikutta}}]{Almeyda2017}
{Almeyda}, T., {Robinson}, A., {Richmond}, M., {Vazquez}, B., \& {Nikutta}, R.
  2017, \apj, 843, 3

\bibitem[{{Antonucci}(1993)}]{Antonucci1993a}
{Antonucci}, R. 1993, \araa, 31, 473

\bibitem[{{Barvainis}(1987)}]{Barvainis1987}
{Barvainis}, R. 1987, \apj, 320, 537

\bibitem[{{Barvainis}(1992)}]{Barvainis1992}
---. 1992, \apj, 400, 502

\bibitem[{{Baskin} \& {Laor}(2018)}]{Baskin2018}
{Baskin}, A., \& {Laor}, A. 2018, \mnras, 474, 1970

\bibitem[Beichman et al. (1988)]{beichman1988} Beichman, C. A., Neugebauer, G., Habing, H. J., Clegg, P. E., \& 
Chester, T. J., IRAS Explanatory Supplement, NASA

\bibitem[Bertin, \& Arnouts(1996)]{SExtractor} Bertin, E., \& Arnouts, S.\ 1996, \aaps, 117, 393


\bibitem[{{Berriman} {et~al.}(1990){Berriman}, {Schmidt}, {West}, \&
  {Stockman}}]{Berriman1990}
{Berriman}, G., {Schmidt}, G.~D., {West}, S.~C., \& {Stockman}, H.~S. 1990,
  \apjs, 74, 869

\bibitem[{{Blandford} \& {McKee}(1982)}]{Blandford1982}
{Blandford}, R.~D., \& {McKee}, C.~F. 1982, \apj, 255, 419


\bibitem[Boroson, \& Green(1992)]{Boroson1992} Boroson, T.~A., \& Green, R.~F.\ 1992, \apjs, 80, 109


\bibitem[Carey (2010)]{carey2010} Carey, S. J. 2010, "Some thoughts on cross-calibration in the mid-infrared,"
on 2010 STScI Calibration Workshop, ed. S. Deustua \& C. Oliveira, page 37 

\bibitem[{{Clavel} {et~al.}(1989){Clavel}, {Wamsteker}, \&
  {Glass}}]{Clavel1989}
{Clavel}, J., {Wamsteker}, W., \& {Glass}, I.~S. 1989, \apj, 337, 236

\bibitem[{{Cohen} {et~al.}(2007){Cohen}, {Lister}, {Homan}, {Kadler},
  {Kellermann}, {Kovalev}, \& {Vermeulen}}]{Cohen2007}
{Cohen}, M.~H., {Lister}, M.~L., {Homan}, D.~C., {et~al.} 2007, \apj, 658, 232

\bibitem[Cutri et al.(2002)]{Cutri2002} Cutri, R.~M., Nelson, B.~O., Francis, P.~J., et al.\ 2002, IAU Colloq. 184: AGN Surveys, 127



\bibitem[{{Czerny} \& {Hryniewicz}(2011)}]{Czerny2011}
{Czerny}, B., \& {Hryniewicz}, K. 2011, \aap, 525, L8

\bibitem[{{Diamond-Stanic} {et~al.}(2009){Diamond-Stanic}, {Rieke}, \&
  {Rigby}}]{Diamond-Stanic2009}
{Diamond-Stanic}, A.~M., {Rieke}, G.~H., \& {Rigby}, J.~R. 2009, \apj, 698, 623

\bibitem[{{Draine} \& {Lee}(1984)}]{Draine1984}
{Draine}, B.~T., \& {Lee}, H.~M. 1984, \apj, 285, 89

\bibitem[{{Drake} {et~al.}(2009){Drake}, {Djorgovski}, {Mahabal}, {Beshore},
  {Larson}, {Graham}, {Williams}, {Christensen}, {Catelan}, {Boattini},
  {Gibbs}, {Hill}, \& {Kowalski}}]{CRTS2009}
{Drake}, A.~J., {Djorgovski}, S.~G., {Mahabal}, A., {et~al.} 2009, \apj, 696,
  870

\bibitem[{{Edelson} \& {Malkan}(1987)}]{Edelson1987}
{Edelson}, R.~A., \& {Malkan}, M.~A. 1987, \apj, 323, 516

\bibitem[{{Elvis} {et~al.}(1994){Elvis}, {Wilkes}, {McDowell}, {Green},
  {Bechtold}, {Willner}, {Oey}, {Polomski}, \& {Cutri}}]{Elvis1994}
{Elvis}, M., {Wilkes}, B.~J., {McDowell}, J.~C., {et~al.} 1994, \apjs, 95, 1

\bibitem[{{Engelbracht} {et~al.}(2007){Engelbracht}, {Blaylock}, {Su}, {Rho},
  {Rieke}, {Muzerolle}, {Padgett}, {Hines}, {Gordon}, {Fadda},
  {Noriega-Crespo}, {Kelly}, {Latter}, {Hinz}, {Misselt}, {Morrison},
  {Stansberry}, {Shupe}, {Stolovy}, {Wheaton}, {Young}, {Neugebauer},
  {Wachter}, {P{\'e}rez-Gonz{\'a}lez}, {Frayer}, \&
  {Marleau}}]{Engelbracht2007}
{Engelbracht}, C.~W., {Blaylock}, M., {Su}, K.~Y.~L., {et~al.} 2007, \pasp,
  119, 994

\bibitem[Gallagher et al.(2005)]{Gallagher2005} Gallagher, S.~C., Schmidt, G.~D., Smith, P.~S., et al.\ 2005, \apj, 633, 71

\bibitem[Glass(2004)]{Glass2004} Glass, I.~S.\ 2004, \mnras, 350, 1049


\bibitem[{{Glikman} {et~al.}(2006){Glikman}, {Helfand}, \&
  {White}}]{Glikman2006}
{Glikman}, E., {Helfand}, D.~J., \& {White}, R.~L. 2006, \apj, 640, 579

\bibitem[{{Gordon} {et~al.}(2005){Gordon}, {Rieke}, {Engelbracht}, {Muzerolle},
  {Stansberry}, {Misselt}, {Morrison}, {Cadien}, {Young}, {Dole}, {Kelly},
  {Alonso-Herrero}, {Egami}, {Su}, {Papovich}, {Smith}, {Hines}, {Rieke},
  {Blaylock}, {P{\'e}rez-Gonz{\'a}lez}, {Le Floc'h}, {Hinz}, {Latter},
  {Hesselroth}, {Frayer}, {Noriega-Crespo}, {Masci}, {Padgett}, {Smylie}, \&
  {Haegel}}]{Gordon2005}
{Gordon}, K.~D., {Rieke}, G.~H., {Engelbracht}, C.~W., {et~al.} 2005, \pasp,
  117, 503

\bibitem[{{Graham} {et~al.}(2017){Graham}, {Djorgovski}, {Drake}, {Stern},
  {Mahabal}, {Glikman}, {Larson}, \& {Christensen}}]{Graham2017}
{Graham}, M.~J., {Djorgovski}, S.~G., {Drake}, A.~J., {et~al.} 2017, \mnras,
  470, 4112

\bibitem[Graham et al.(2014)]{Graham2014} Graham, M.~J., Djorgovski, S.~G., Drake, A.~J., et al.\ 2014, \mnras, 439, 703


\bibitem[Hern{\'a}n-Caballero et al.(2016)]{Hernan-Caballero2016}
    Hern{\'a}n-Caballero, A., Hatziminaoglou, E., Alonso-Herrero, A., et al.\
    2016, \mnras, 463, 2064

\bibitem[Hodge et al.(2011)]{Hodge2011} Hodge, J.~A., Becker, R.~H., White, R.~L., et al.\ 2011, \aj, 142, 3

\bibitem[{{Homan}(2012)}]{Homan2012}
{Homan}, D.~C. 2012, in International Journal of Modern Physics Conference
  Series, Vol.~8, International Journal of Modern Physics Conference Series,
  163--171

\bibitem[{{H{\"o}nig} {et~al.}(2010){H{\"o}nig}, {Kishimoto}, {Gandhi},
  {Smette}, {Asmus}, {Duschl}, {Polletta}, \& {Weigelt}}]{Honig2010}
{H{\"o}nig}, S.~F., {Kishimoto}, M., {Gandhi}, P., {et~al.} 2010, \aap, 515,
  A23

\bibitem[{{Houck} {et~al.}(2004){Houck}, {Roellig}, {van Cleve}, {Forrest},
  {Herter}, {Lawrence}, {Matthews}, {Reitsema}, {Soifer}, {Watson}, {Weedman},
  {Huisjen}, {Troeltzsch}, {Barry}, {Bernard-Salas}, {Blacken}, {Brandl},
  {Charmandaris}, {Devost}, {Gull}, {Hall}, {Henderson}, {Higdon}, {Pirger},
  {Schoenwald}, {Sloan}, {Uchida}, {Appleton}, {Armus}, {Burgdorf},
  {Fajardo-Acosta}, {Grillmair}, {Ingalls}, {Morris}, \&
  {Teplitz}}]{spitzerirs}
{Houck}, J.~R., {Roellig}, T.~L., {van Cleve}, J., {et~al.} 2004, \apjs, 154,
  18

\bibitem[Hunter(2007)]{matplotlib} Hunter, J.~D.\ 2007, Computing in Science and Engineering, 9, 90

\bibitem[{{Jiang} {et~al.}(2017){Jiang}, {Green}, {Greene}, {Morganson},
  {Shen}, {Pancoast}, {MacLeod}, {Anderson}, {Brandt}, {Grier}, {Rix}, {Ruan},
  {Protopapas}, {Scott}, {Burgett}, {Hodapp}, {Huber}, {Kaiser}, {Kudritzki},
  {Magnier}, {Metcalfe}, {Tonry}, {Wainscoat}, \& {Waters}}]{Jiang2017}
{Jiang}, Y.-F., {Green}, P.~J., {Greene}, J.~E., {et~al.} 2017, \apj, 836, 186

\bibitem[{{Kaspi} {et~al.}(2000){Kaspi}, {Smith}, {Netzer}, {Maoz}, {Jannuzi},
  \& {Giveon}}]{Kaspi2000}
{Kaspi}, S., {Smith}, P.~S., {Netzer}, H., {et~al.} 2000, \apj, 533, 631


\bibitem[Kasliwal et al.(2017)]{Kasliwal2017} Kasliwal, V.~P., Vogeley, M.~S., \& Richards, G.~T.\ 2017, \mnras, 470, 3027


\bibitem[{{Kawaguchi} \& {Mori}(2011)}]{Kawaguchi2011}
{Kawaguchi}, T., \& {Mori}, M. 2011, \apj, 737, 105

\bibitem[Kellermann \& Pauline-Toth (1981)]{kellermann1981} Kellermann, K. I., \& Pauliny-Toth, I. I. K. 1981, ARA\&A, 19, 373

\bibitem[Kellermann et al.(1989)]{Kellermann1989} Kellermann, K.~I., Sramek, R., Schmidt, M., et al.\ 1989, \aj, 98, 1195

\bibitem[Kellermann et al.(2016)]{Kellermann2016} Kellermann, K.~I., Condon, J.~J., Kimball, A.~E., et al.\ 2016, \apj, 831, 168

\bibitem[{{Kelly} {et~al.}(2009){Kelly}, {Bechtold}, \&
  {Siemiginowska}}]{Kelly2009}
{Kelly}, B.~C., {Bechtold}, J., \& {Siemiginowska}, A. 2009, \apj, 698, 895

\bibitem[Kelly et al.(2014)]{Kelly2014} Kelly, B.~C., Becker, A.~C., Sobolewska, M., et al.\ 2014, \apj, 788, 33


\bibitem[{{Kim} {et~al.}(2015){Kim}, {Im}, {Kim}, {Jun}, {Woo}, {Lee}, {Lee},
  {Nakagawa}, {Matsuhara}, {Wada}, {Oyabu}, {Takagi}, {Ohyama}, \&
  {Lee}}]{Kim2015}
{Kim}, D., {Im}, M., {Kim}, J.~H., {et~al.} 2015, \apjs, 216, 17

\bibitem[{{Kishimoto} {et~al.}(2008){Kishimoto}, {Antonucci}, {Blaes},
  {Lawrence}, {Boisson}, {Albrecht}, \& {Leipski}}]{Kishimoto2008}
{Kishimoto}, M., {Antonucci}, R., {Blaes}, O., {et~al.} 2008, \nat, 454, 492

\bibitem[{{Kishimoto} {et~al.}(2011){Kishimoto}, {H{\"o}nig}, {Antonucci},
  {Millour}, {Tristram}, \& {Weigelt}}]{Kishimoto2011b}
{Kishimoto}, M., {H{\"o}nig}, S.~F., {Antonucci}, R., {et~al.} 2011, \aap, 536,
  A78

\bibitem[{{Kishimoto} {et~al.}(2007){Kishimoto}, {H{\"o}nig}, {Beckert}, \&
  {Weigelt}}]{Kishimoto2007}
{Kishimoto}, M., {H{\"o}nig}, S.~F., {Beckert}, T., \& {Weigelt}, G. 2007,
  \aap, 476, 713

\bibitem[{{Kishimoto} {et~al.}(2009){Kishimoto}, {H{\"o}nig}, {Tristram}, \&
  {Weigelt}}]{Kishimoto2009b}
{Kishimoto}, M., {H{\"o}nig}, S.~F., {Tristram}, K.~R.~W., \& {Weigelt}, G.
  2009, \aap, 493, L57

\bibitem[Kishimoto et al.(2013)]{Kishimoto2013} Kishimoto, M.,
    H{\"o}nig, S.~F., Antonucci, R., et al.\ 2013, \apjl, 775, L36

\bibitem[{{Kochanek} {et~al.}(2017){Kochanek}, {Shappee}, {Stanek}, {Holoien},
  {Thompson}, {Prieto}, {Dong}, {Shields}, {Will}, {Britt}, {Perzanowski}, \&
  {Pojma{\'n}ski}}]{Kochanek2017}
{Kochanek}, C.~S., {Shappee}, B.~J., {Stanek}, K.~Z., {et~al.} 2017, \pasp,
  129, 104502

\bibitem[{{Koshida} {et~al.}(2014){Koshida}, {Minezaki}, {Yoshii}, {Kobayashi},
  {Sakata}, {Sugawara}, {Enya}, {Suganuma}, {Tomita}, {Aoki}, \&
  {Peterson}}]{Koshida2014}
{Koshida}, S., {Minezaki}, T., {Yoshii}, Y., {et~al.} 2014, \apj, 788, 159

\bibitem[{{Koz{\l}owski} {et~al.}(2010){Koz{\l}owski}, {Kochanek}, {Udalski},
  {Wyrzykowski}, {Soszy{\'n}ski}, {Szyma{\'n}ski}, {Kubiak}, {Pietrzy{\'n}ski},
  {Szewczyk}, {Ulaczyk}, {Poleski}, \& {OGLE Collaboration}}]{Kozlowski2010}
{Koz{\l}owski}, S., {Kochanek}, C.~S., {Udalski}, A., {et~al.} 2010, \apj, 708,
  927

\bibitem[{{Laor} \& {Draine}(1993)}]{Laor1993}
{Laor}, A., \& {Draine}, B.~T. 1993, \apj, 402, 441

\bibitem[Laor et al (2019)]{laor2019} Laor, A., Baldi, R. D., \& Behar, E. 2019, MNRAS, 482, 5513

\bibitem[{{Law} {et~al.}(2009){Law}, {Kulkarni}, {Dekany}, {Ofek}, {Quimby},
  {Nugent}, {Surace}, {Grillmair}, {Bloom}, {Kasliwal}, {Bildsten}, {Brown},
  {Cenko}, {Ciardi}, {Croner}, {Djorgovski}, {van Eyken}, {Filippenko}, {Fox},
  {Gal-Yam}, {Hale}, {Hamam}, {Helou}, {Henning}, {Howell}, {Jacobsen},
  {Laher}, {Mattingly}, {McKenna}, {Pickles}, {Poznanski}, {Rahmer}, {Rau},
  {Rosing}, {Shara}, {Smith}, {Starr}, {Sullivan}, {Velur}, {Walters}, \&
  {Zolkower}}]{PTF}
{Law}, N.~M., {Kulkarni}, S.~R., {Dekany}, R.~G., {et~al.} 2009, \pasp, 121,
  1395

\bibitem[{{Lebouteiller} {et~al.}(2011){Lebouteiller}, {Barry}, {Spoon},
  {Bernard-Salas}, {Sloan}, {Houck}, \& {Weedman}}]{cassis}
{Lebouteiller}, V., {Barry}, D.~J., {Spoon}, H.~W.~W., {et~al.} 2011, \apjs,
  196, 8

\bibitem[{{Leipski} {et~al.}(2013){Leipski}, {Meisenheimer}, {Walter}, {Besel},
  {Dannerbauer}, {Fan}, {Haas}, {Klaas}, {Krause}, \& {Rix}}]{Leipski2013}
{Leipski}, C., {Meisenheimer}, K., {Walter}, F., {et~al.} 2013, \apj, 772, 103

\bibitem[{{Lira} {et~al.}(2011){Lira}, {Ar{\'e}valo}, {Uttley}, {McHardy}, \&
  {Breedt}}]{Lira2011}
{Lira}, P., {Ar{\'e}valo}, P., {Uttley}, P., {McHardy}, I., \& {Breedt}, E.
  2011, \mnras, 415, 1290

\bibitem[{{Lyu} \& {Rieke}(2017)}]{Lyu2017b}
{Lyu}, J., \& {Rieke}, G.~H. 2017, \apj, 841, 76

\bibitem[{{Lyu} \& {Rieke}(2018)}]{Lyu2018}
---. 2018, \apj, 866, 92

\bibitem[{{Lyu} {et~al.}(2017){Lyu}, {Rieke}, \& {Shi}}]{Lyu2017}
{Lyu}, J., {Rieke}, G.~H., \& {Shi}, Y. 2017, \apj, 835, 257

\bibitem[{{MacLeod} {et~al.}(2010){MacLeod}, {Ivezi{\'c}}, {Kochanek},
  {Koz{\l}owski}, {Kelly}, {Bullock}, {Kimball}, {Sesar}, {Westman}, {Brooks},
  {Gibson}, {Becker}, \& {de Vries}}]{MacLeod2010}
{MacLeod}, C.~L., {Ivezi{\'c}}, {\v Z}., {Kochanek}, C.~S., {et~al.} 2010,
  \apj, 721, 1014

\bibitem[{{Mainzer} {et~al.}(2014){Mainzer}, {Bauer}, {Cutri}, {Grav},
  {Masiero}, {Beck}, {Clarkson}, {Conrow}, {Dailey}, {Eisenhardt}, {Fabinsky},
  {Fajardo-Acosta}, {Fowler}, {Gelino}, {Grillmair}, {Heinrichsen}, {Kendall},
  {Kirkpatrick}, {Liu}, {Masci}, {McCallon}, {Nugent}, {Papin}, {Rice},
  {Royer}, {Ryan}, {Sevilla}, {Sonnett}, {Stevenson}, {Thompson}, {Wheelock},
  {Wiemer}, {Wittman}, {Wright}, \& {Yan}}]{neowise}
{Mainzer}, A., {Bauer}, J., {Cutri}, R.~M., {et~al.} 2014, \apj, 792, 30

\bibitem[{{Mandal} {et~al.}(2018){Mandal}, {Rakshit}, {Kurian}, {Stalin},
  {Mathew}, {Hoenig}, {Gandhi}, {Sagar}, \& {Pandge}}]{Mandal2018}
{Mandal}, A.~K., {Rakshit}, S., {Kurian}, K.~S., {et~al.} 2018, \mnras, 475,
  5330

\bibitem[{{Mao} {et~al.}(2016){Mao}, {Urry}, {Massaro}, {Paggi}, {Cauteruccio},
  \& {K{\"u}nzel}}]{Mao2016}
{Mao}, P., {Urry}, C.~M., {Massaro}, F., {et~al.} 2016, \apjs, 224, 26

\bibitem[Markwardt(2009)]{MPFIT} Markwardt, C.~B.\ 2009,
Astronomical Data Analysis Software and Systems XVIII, 411, 251




\bibitem[{{Mel{\'e}ndez} {et~al.}(2008){Mel{\'e}ndez}, {Kraemer}, {Armentrout},
  {Deo}, {Crenshaw}, {Schmitt}, {Mushotzky}, {Tueller}, {Markwardt}, \&
  {Winter}}]{Melendez2008}
{Mel{\'e}ndez}, M., {Kraemer}, S.~B., {Armentrout}, B.~K., {et~al.} 2008, \apj,
  682, 94

\bibitem[{{Mor} {et~al.}(2009){Mor}, {Netzer}, \& {Elitzur}}]{Mor2009}
{Mor}, R., {Netzer}, H., \& {Elitzur}, M. 2009, \apj, 705, 298

\bibitem[Murphy et al.(2007)]{Murphy2007} Murphy, T., Mauch, T., Green, A., et al.\ 2007, \mnras, 382, 382

\bibitem[Mushotzky et al.(2011)]{Mushotzky2011} Mushotzky, R.~F., Edelson, R., Baumgartner, W., et al.\ 2011, \apjl, 743, L12

\bibitem[{{Nenkova} {et~al.}(2008){Nenkova}, {Sirocky}, {Ivezi{\'c}}, \&
  {Elitzur}}]{Nenkova2008a}
{Nenkova}, M., {Sirocky}, M.~M., {Ivezi{\'c}}, {\v Z}., \& {Elitzur}, M. 2008,
  \apj, 685, 147

\bibitem[{{Netzer}(2015)}]{Netzer2015}
{Netzer}, H. 2015, \araa, 53, 365

\bibitem[{{Netzer} {et~al.}(2004){Netzer}, {Shemmer}, {Maiolino}, {Oliva},
  {Croom}, {Corbett}, \& {di Fabrizio}}]{Netzer2004}
{Netzer}, H., {Shemmer}, O., {Maiolino}, R., {et~al.} 2004, \apj, 614, 558

\bibitem[{{Neugebauer} \& {Matthews}(1999)}]{Neugebauer1999}
{Neugebauer}, G., \& {Matthews}, K. 1999, \aj, 118, 35

\bibitem[{{Oknyanskij} \& {Horne}(2001)}]{Oknyanskij2001}
{Oknyanskij}, V.~L., \& {Horne}, K. 2001, in Astronomical Society of the
  Pacific Conference Series, Vol. 224, Probing the Physics of Active Galactic
  Nuclei, ed. B.~M. {Peterson}, R.~W. {Pogge}, \& R.~S. {Polidan}, 149

\bibitem[{{Peterson} {et~al.}(2004){Peterson}, {Ferrarese}, {Gilbert}, {Kaspi},
  {Malkan}, {Maoz}, {Merritt}, {Netzer}, {Onken}, {Pogge}, {Vestergaard}, \&
  {Wandel}}]{Peterson2004}
{Peterson}, B.~M., {Ferrarese}, L., {Gilbert}, K.~M., {et~al.} 2004, \apj, 613,
  682

\bibitem[{{Pozo Nu{\~n}ez} {et~al.}(2014){Pozo Nu{\~n}ez}, {Haas}, {Chini},
  {Ramolla}, {Westhues}, {Steenbrugge}, {Kaderhandt}, {Drass}, {Lemke}, \&
  {Murphy}}]{PozoNunez2014}
{Pozo Nu{\~n}ez}, F., {Haas}, M., {Chini}, R., {et~al.} 2014, \aap, 561, L8

\bibitem[{{Pozo Nu{\~n}ez} {et~al.}(2015){Pozo Nu{\~n}ez}, {Ramolla},
  {Westhues}, {Haas}, {Chini}, {Steenbrugge}, {Barr Dom{\'{\i}}nguez},
  {Kaderhandt}, {Hackstein}, {Kollatschny}, {Zetzl}, {Hodapp}, \&
  {Murphy}}]{Pozo2015}
{Pozo Nu{\~n}ez}, F., {Ramolla}, M., {Westhues}, C., {et~al.} 2015, \aap, 576,
  A73

\bibitem[{{Ramolla} {et~al.}(2018){Ramolla}, {Haas}, {Westhues}, {Pozo
  Nu{\~n}ez}, {Sobrino Figaredo}, {Blex}, {Zetzl}, {Kollatschny}, {Hodapp},
  {Chini}, \& {Murphy}}]{Ramolla2018}
{Ramolla}, M., {Haas}, M., {Westhues}, C., {et~al.} 2018, \aap, 620, A137

\bibitem[{{Reyes} {et~al.}(2008){Reyes}, {Zakamska}, {Strauss}, {Green},
  {Krolik}, {Shen}, {Richards}, {Anderson}, \& {Schneider}}]{Reyes2008}
{Reyes}, R., {Zakamska}, N.~L., {Strauss}, M.~A., {et~al.} 2008, \aj, 136, 2373


\bibitem[Rieke (1972)]{rieke1972} Rieke, G. H. 1972, ApJL, 176, L61

\bibitem[Rieke \& Kinman (1974)]{rieke1974} Rieke, G. H., \& Kinman, T. D. 1974, ApJL, 192, L115

\bibitem[{{Rigby} {et~al.}(2009){Rigby}, {Diamond-Stanic}, \&
  {Aniano}}]{Rigby2009}
{Rigby}, J.~R., {Diamond-Stanic}, A.~M., \& {Aniano}, G. 2009, \apj, 700, 1878

\bibitem[{{Risaliti} \& {Elvis}(2004)}]{Risaliti2004}
{Risaliti}, G., \& {Elvis}, M. 2004, in Astrophysics and Space Science Library,
  Vol. 308, Supermassive Black Holes in the Distant Universe, ed. A.~J.
  {Barger}, 187

\bibitem[{{Runnoe} {et~al.}(2012){Runnoe}, {Brotherton}, \&
  {Shang}}]{Runnoe2012}
{Runnoe}, J.~C., {Brotherton}, M.~S., \& {Shang}, Z. 2012, \mnras, 426, 2677

\bibitem[{{Schartmann} {et~al.}(2014){Schartmann}, {Wada}, {Prieto}, {Burkert},
  \& {Tristram}}]{Schartmann2014}
{Schartmann}, M., {Wada}, K., {Prieto}, M.~A., {Burkert}, A., \& {Tristram},
  K.~R.~W. 2014, \mnras, 445, 3878

\bibitem[Schmidt, \& Green(1983)]{Schmidt1983} Schmidt, M., \& Green, R.~F.\ 1983, \apj, 269, 352

\bibitem[Schmidt et al.(1992)]{Schmidt1992} Schmidt, G.~D., Stockman, H.~S., \& Smith, P.~S.\ 1992, \apjl, 398, L57

\bibitem[{{Shappee} {et~al.}(2014){Shappee}, {Prieto}, {Grupe}, {Kochanek},
  {Stanek}, {De Rosa}, {Mathur}, {Zu}, {Peterson}, {Pogge}, {Komossa}, {Im},
  {Jencson}, {Holoien}, {Basu}, {Beacom}, {Szczygie{\l}}, {Brimacombe},
  {Adams}, {Campillay}, {Choi}, {Contreras}, {Dietrich}, {Dubberley},
  {Elphick}, {Foale}, {Giustini}, {Gonzalez}, {Hawkins}, {Howell}, {Hsiao},
  {Koss}, {Leighly}, {Morrell}, {Mudd}, {Mullins}, {Nugent}, {Parrent},
  {Phillips}, {Pojmanski}, {Rosing}, {Ross}, {Sand}, {Terndrup}, {Valenti},
  {Walker}, \& {Yoon}}]{asassn}
{Shappee}, B.~J., {Prieto}, J.~L., {Grupe}, D., {et~al.} 2014, \apj, 788, 48

\bibitem[Shi et al. (2007)]{shi2007} Shi, Y., Rieke, G. H., Hines, D. C., Gordon, K. D., \& Egami, E. 2007, ApJ, 655, 781

\bibitem[{{Shi} {et~al.}(2014){Shi}, {Rieke}, {Ogle}, {Su}, \&
  {Balog}}]{Shi2014}
{Shi}, Y., {Rieke}, G.~H., {Ogle}, P.~M., {Su}, K.~Y.~L., \& {Balog}, Z. 2014,
  \apjs, 214, 23

\bibitem[Sloan et al. (2015)]{sloan2015} Sloan, G. C., Herter, T. L., Charmandaris, V., Seth, K., Burgdorf, M., \& Houck, J. R. 2015, AJ, 149, 11

\bibitem[Smith et al.(1997)]{Smith1997} Smith, P.~S., Schmidt, G.~D., Allen, R.~G., et al.\ 1997, \apj, 488, 202

\bibitem[Smith(2016)]{Smith2016} Smith, P.~S.\ 2016, Galaxies, 4, 27



\bibitem[{{Soldi} {et~al.}(2008){Soldi}, {T{\"u}rler}, {Paltani}, {Aller},
  {Aller}, {Burki}, {Chernyakova}, {L{\"a}hteenm{\"a}ki}, {McHardy}, {Robson},
  {Staubert}, {Tornikoski}, {Walter}, \& {Courvoisier}}]{Soldi2008}
{Soldi}, S., {T{\"u}rler}, M., {Paltani}, S., {et~al.} 2008, \aap, 486, 411

\bibitem[{{Stalevski} {et~al.}(2012){Stalevski}, {Fritz}, {Baes}, {Nakos}, \&
  {Popovi{\'c}}}]{Stalevski2012}
{Stalevski}, M., {Fritz}, J., {Baes}, M., {Nakos}, T., \& {Popovi{\'c}}, L.~{\v
  C}. 2012, \mnras, 420, 2756

\bibitem[{{Stalevski} {et~al.}(2016){Stalevski}, {Ricci}, {Ueda}, {Lira},
  {Fritz}, \& {Baes}}]{Stalevski2016}
{Stalevski}, M., {Ricci}, C., {Ueda}, Y., {et~al.} 2016, \mnras, 458, 2288

\bibitem[Stetson (1987)]{stetson1987} Stetson, P. B. 1987, PASP, 99, 191

\bibitem[{{Suganuma} {et~al.}(2006){Suganuma}, {Yoshii}, {Kobayashi},
  {Minezaki}, {Enya}, {Tomita}, {Aoki}, {Koshida}, \&
  {Peterson}}]{Suganuma2006}
{Suganuma}, M., {Yoshii}, Y., {Kobayashi}, Y., {et~al.} 2006, \apj, 639, 46

\bibitem[Tody(1993)]{Tody1993} Tody, D.\ 1993, Astronomical Data Analysis Software and Systems II, 173

\bibitem[Tody(1986)]{Tody1986} Tody, D.\ 1986, \procspie, 733

\bibitem[{{Urry} \& {Padovani}(1995)}]{Urry1995}
{Urry}, C.~M., \& {Padovani}, P. 1995, \pasp, 107, 803

\bibitem[{{Vazquez} {et~al.}(2015){Vazquez}, {Galianni}, {Richmond},
  {Robinson}, {Axon}, {Horne}, {Almeyda}, {Fausnaugh}, {Peterson}, {Bottorff},
  {Gallimore}, {Eltizur}, {Netzer}, {Storchi-Bergmann}, {Marconi}, {Capetti},
  {Batcheldor}, {Buchanan}, {Stirpe}, {Kishimoto}, {Packham}, {Perez},
  {Tadhunter}, {Upton}, \& {Estrada-Carpenter}}]{Vazquez2015}
{Vazquez}, B., {Galianni}, P., {Richmond}, M., {et~al.} 2015, \apj, 801, 127

\bibitem[{{Wright} {et~al.}(2010){Wright}, {Eisenhardt}, {Mainzer}, {Ressler},
  {Cutri}, {Jarrett}, {Kirkpatrick}, {Padgett}, {McMillan}, {Skrutskie},
  {Stanford}, {Cohen}, {Walker}, {Mather}, {Leisawitz}, {Gautier}, {McLean},
  {Benford}, {Lonsdale}, {Blain}, {Mendez}, {Irace}, {Duval}, {Liu}, {Royer},
  {Heinrichsen}, {Howard}, {Shannon}, {Kendall}, {Walsh}, {Larsen}, {Cardon},
  {Schick}, {Schwalm}, {Abid}, {Fabinsky}, {Naes}, \& {Tsai}}]{WISE}
{Wright}, E.~L., {Eisenhardt}, P.~R.~M., {Mainzer}, A.~K., {et~al.} 2010, \aj,
  140, 1868

\bibitem[{{Xie} {et~al.}(2017){Xie}, {Li}, \& {Hao}}]{Xie2017}
{Xie}, Y., {Li}, A., \& {Hao}, L. 2017, \apjs, 228, 6

\bibitem[{{Zu} {et~al.}(2013){Zu}, {Kochanek}, {Koz{\l}owski}, \&
  {Udalski}}]{Zu2013}
{Zu}, Y., {Kochanek}, C.~S., {Koz{\l}owski}, S., \& {Udalski}, A. 2013, \apj,
  765, 106

\bibitem[{{Zu} {et~al.}(2011){Zu}, {Kochanek}, \& {Peterson}}]{Zu2011}
{Zu}, Y., {Kochanek}, C.~S., \& {Peterson}, B.~M. 2011, \apj, 735, 80

\end{thebibliography}

\end{document}